\documentclass[11pt,xcolor=dvipsnames]{article}
\pdfoutput=1 

\usepackage[colorlinks,urlcolor=black,linkcolor = blue,citecolor = black,]{hyperref}
\usepackage{subfigure}
\usepackage{latexsym}
\usepackage{epsfig}
\usepackage[mathscr]{eucal}
\usepackage{amsfonts}
\usepackage{amscd}
\usepackage{cite}
\usepackage{array}
\usepackage{amssymb}
\usepackage{colordvi}
\usepackage[centertags]{amsmath}
\usepackage{enumerate}
\usepackage{graphicx}
\usepackage{booktabs}
\usepackage{theorem}
\usepackage[footnotesize]{caption}
\usepackage{soul}
\usepackage{mcite}
\usepackage{slashed}
\usepackage{braket}
\usepackage{units}

\usepackage{xcolor}
\usepackage{ulem}
\usepackage{bbm}
\usepackage[utf8]{inputenc}
\usepackage{fancyvrb}
\usepackage{framed}
\usepackage{xspace}
\usepackage{todonotes}

\newcommand{\newc}{\newcommand}
\newc{\be}{\begin{equation}}
\newc{\bea}{\begin{eqnarray}}
\newc{\eea}{\end{eqnarray}}
\newc{\ol}{\overline}
\newc{\wt}{\widetilde}
\newc{\bs}{\boldsymbol}
\newc{\m}{\mathcal}
\newc{\la}{\langle}
\newc{\ra}{\rangle}

\newcommand{\beq}{\begin{eqnarray}}
\newcommand{\eeq}{\end{eqnarray}}
\newcommand{\bpmatrix}{\begin{pmatrix}}
\newcommand{\epmatrix}{\end{pmatrix}}
\newcommand{\ba}{\begin{array}}
\newcommand{\ea}{\end{array}}

\newcommand{\mdeg}{\mathcal{M}_{\text{deg}}}
\newcommand{\msep}{\mathcal{M}_{\text{sep}}}

\newcommand{\br}{\mathrm{BR}}

\renewcommand{\ol}{\text{1l}}
\renewcommand{\Re}{\text{Re}\!}
\renewcommand{\Im}{\text{Im}\!}
\renewcommand{\eqref}[1]{Eq.~(\ref{#1})}

\newcommand{\cbrak}[1]{\left(#1\right)}
\newcommand{\sbrak}[1]{\left[#1\right]}

\allowdisplaybreaks

\usepackage[printonlyused]{acronym}

\renewcommand{\ol}{\text{1l}}

\renewcommand{\Re}{\text{Re}\!}
\renewcommand{\Im}{\text{Im}\!}

\renewcommand{\eqref}[1]{Eq.~(\ref{#1})}

\newcommand{\MSb}{\overline{\text{MS}}}

\newcommand{\bc}{\begin{center}}
\newcommand{\ec}{\end{center}}

\newcommand{\gev}{~\text{GeV}}

\newcommand{\tev}{~\text{TeV}}
\newcommand{\pb}{{~\text{pb}}}
\newcommand{\fb}{{~\text{fb}}}

\newcommand{\s}{\newline \vspace*{-3.5mm}}

\newcommand{\nullel}{\mathcal{O}}

\newcommand{\ii}{\mathrm{i}}

\newcommand{\figref}[2][{}]{\hyperref[#2]{\figurename~\ref{#2}#1}}

\newenvironment{kasten*}[1]
{
\hspace{0.05\linewidth}
\begin{minipage}{0.95\linewidth}
\setlength{\fboxsep}{10pt}
\definecolor{shadecolor}{gray}{0.9}
\definecolor{framecolor}{gray}{0}

\MakeFramed {\FrameRestore}
\subsection*{#1}
}
{
\endMakeFramed
\end{minipage}
\vspace{1em}
}

\newcommand{\mHc}{m_{H^{\pm}}}
\newcommand{\mdown}{m_{H_\downarrow}}
\newcommand{\mup}{m_{H_\uparrow}}
\newcommand{\hup}{H_{\uparrow}}
\newcommand{\hdown}{H_{\downarrow}}

\newcommand{\hHc}{H^{\pm}}
\renewcommand{\Re}{\operatorname{Re}}
\renewcommand{\Im}{\operatorname{Im}}

\newcommand{\xic}{\ensuremath{\xi_c}\xspace}

\allowdisplaybreaks

\usepackage[printonlyused]{acronym}

\usepackage{cleveref}
\crefname{chapter}{Chapter}{Chapter}
\crefname{section}{Sec.}{Secs.}
\crefname{table}{Tab.}{Tabs.}
\crefname{figure}{Fig.}{Figs.}
\crefname{equation}{Eq.}{Eqs.}
\crefname{appendix}{Appendix\ }{Appendix\ }

\begin{document}
\title{
\vspace*{-3.7cm}
\phantom{h} \hfill\mbox{\small KA-TP-27-2019}\\[-1.1cm]
\vspace*{2.7cm}
\textbf{Electroweak Phase Transition \\ in Non-Minimal Higgs Sectors \\[4mm]}}

\date{}
\author{
Philipp Basler$^{1\,}$\footnote{E-mail:
  \texttt{philipp.basler@kit.edu}} ,
Margarete M\"{u}hlleitner$^{1\,}$\footnote{E-mail:
\texttt{margarete.muehlleitner@kit.edu}} ,
Jonas M\"{u}ller$^{1\,}$\footnote{E-mail:
\texttt{jonas.mueller@kit.edu}}
\\[9mm]
{\small\it
$^1$Institute for Theoretical Physics, Karlsruhe Institute of Technology,} \\
{\small\it 76128 Karlsruhe, Germany}}
\maketitle

\begin{abstract}
Higgs sector extensions beyond the Standard Model (BSM) provide
additional sources of CP violation and further scalar states
that help to trigger a strong first order electroweak phase transition
(SFOEWPT) 
required to generate the observed baryon asymmetry of the Universe
through electroweak baryogenesis. We investigate the CP-violating
2-Higgs-Doublet Model (C2HDM) and the Next-to-Minimal 2-Higgs-Doublet
Model (N2HDM) with respect to their potential to generate an SFOEWPT
while being compatible with all relevant and recent theoretical and
experimental constraints. The implications of an SFOEWPT on the collider
phenomenology of the two models are analysed in detail in particular with respect to
Higgs pair production. We provide benchmark points for parameter
points that are compatible with an SFOEWPT and provide distinct
di-Higgs signatures. 
\end{abstract}
\thispagestyle{empty}
\vfill
\newpage
\setcounter{page}{1}

\maketitle

\section{Introduction}
The discovery of the Higgs boson by the LHC experiments ATLAS \cite{Aad:2012tfa}
and CMS \cite{Chatrchyan:2012xdj} has been a great success for particle physics. While
the Higgs boson behaves very Standard Model (SM)-like \cite{Aad:2015mxa,Khachatryan:2014kca,Aad:2015gba,Khachatryan:2014jba} it is clear that
there must be physics beyond the SM (BSM) in order to solve the remaining
puzzles that cannot be explained within the SM. Thus the observed baryon
asymmetry of the Universe \cite{Bennett:2012zja} calls for new physics
extensions. Electroweak baryogenesis \cite{Kuzmin:1985mm,Cohen:1990it,Cohen:1993nk,Quiros:1994dr,Rubakov:1996vz,Funakubo:1996dw,Trodden:1998ym,Bernreuther:2002uj,Morrissey:2012db} provides a mechanism for its
generation if all three Sakharov conditions \cite{Sakharov:1967dj} are
fulfilled. These are baryon number violation, C and CP violation and
departure from the thermal equilibrium. The asymmetry can be generated
if the electroweak phase transition (EWPT) is of strong first order
\cite{Trodden:1998ym,Morrissey:2012db}. The strong first order phase
transition proceeds through bubble
formation and suppresses the baryon number violating sphaleron
transitions in the false vacuum
\cite{Manton:1983nd,Klinkhamer:1984di}. While the SM in principle
fulfills all three Sakharov conditions the EWPT is not of strong first
order \cite{firstorder}. This would 
require an SM Higgs boson mass of around 70-80 GeV \cite{smmass} in contrast to the
mass value of 125~GeV of the discovered Higgs boson
\cite{Aad:2015zhl}. Moreover, the SM CP violation arising from the
Cabibbo-Kobayashi-Maskawa (CKM) matrix is not large enough
\cite{Morrissey:2012db,ckmsize}. Extended Higgs sectors beyond the SM
provide additional sources of CP violation and further scalar states
that help to trigger a strong first order EWPT (SFOEWPT). Thus
previous studies within the 2-Higgs-Doublet Model (2HDM)
\cite{Lee:1973iz,Branco:2011iw}, which belongs to the simplest BSM
extensions, have shown that it provides a framework where an SFOEWPT
can be realized in accordance with the relevant theoretical and
experimental constraints, both in the CP-conserving 
\cite{previous2hdm,Basler:2016obg,Laine:2017hdk,Dorsch:2017nza,Andersen:2017ika,Bernon:2017jgv,Wang:2018hnw,Kainulainen:2019kyp}
and in the CP-violating case \cite{previousc2hdm,C2HDM,Wang:2019pet}. \s

With the increasing precision in the Higgs property measurements new
physics extensions get more and more constrained and also models
being so far successful in generating the observed baryon-antibaryon
asymmetry get more and more under tension. In view of this situation,
we first of all revisit our results of Ref.~\cite{C2HDM}, where we
investigated the CP-violating 2HDM (C2HDM) \cite{Ginzburg:2002wt} with
respect to an SFOEWPT, by taking into account the newest collider
constraints. We then move on to the Next-to-Minimal 2HDM (N2HDM)
\cite{Chen:2013jvg,Muhlleitner:2016mzt} in order to investigate the
question if it can provide an SFOEWPT and how this connects with
collider phenomenology. The N2HDM is based on the extension of the
CP-conserving 2HDM by a real
singlet scalar field, inducing a Higgs sector consisting of three
scalar, one pseudoscalar and two charged Higgs boson fields. The C2HDM
and N2HDM resemble each other in the sense that they both provide at
least three neutral Higgs bosons. While in the C2HDM their
phenomenology is determined by the amount of CP admixture to the mass
eigenstates, in the N2HDM it is their singlet admixture that governs
phenomenology. Moreover, the N2HDM features more input parameters
that can be tuned to render the model compatible with all theoretical
and experimental constraints, so that it provides more flexibility in
finding parameter points that are both compatible with these constraints
and provide an SFOEWPT. We compare both models, by taking into account
all relevant and recent theoretical and experimental constraints, with
respect to their ability to provide an SFOEWPT. We investigate the
potential to distinguish between both models in view of these requirements. In
particular, the size of the trilinear Higgs self-couplings that is
necessary for an SFOEWPT is analysed and their implication for Higgs
pair production. We provide benchmark points that connect the
requirement of an SFOEWPT with specific features in the collider
phenomenology, in particular Higgs pair production. Both models are
interesting here also because their non-minimal Higgs sectors allow
for the production of mixed Higgs pair final states. With our analysis
we hope to provide a link between collider phenomenology and
cosmology. \s

The paper is organized as follows. In Sec.~\ref{sec:effpot} we
introduce the computation of the effective potential in the C2HDM and
N2HDM and set our notation. In Sec.~\ref{sec:renormalisation}, we
describe the renormalisation of the N2HDM that is new. The one for
the C2HDM has been provided previously in \cite{C2HDM}. The basics of
our numerical analysis are introduced in
Sec.~\ref{sec:numerical}. Section~\ref{sec:results} is devoted to the
presentation of our results, where we first give in
Subsec.~\ref{subsec:c2hdm} an update of the
C2HDM by including the newest constraints. We then move on to the
detailed presentation of the N2HDM phenomenology of the SFOEWPT in
Subsec.~\ref{subsec:n2hdmpheno}, both discussing the related mass spectra
and trilinear Higgs self-couplings and providing benchmark points. In
Subsec.~\ref{subsec:comparison} we compare the C2HDM and N2HDM rates
for Higgs pair production induced by the requirement of an
SFOEWPT. Our conclusions are given in Sec.~\ref{sec:concl}.

\section{The Effective Potential \label{sec:effpot}}
In order to decide if the EWPT is of strong
first order, we have to determine the value $v_c$ of the VEV at the critical
temperature $T_c$. The VEV at the temperature $T$ is given by the
minimum of the one-loop corrected effective potential at non-zero
temperature $T$. In this section we provide the loop-corrected
effective potentials at finite temperature for our two models under
investigation, the C2HDM and the N2HDM. We start with the tree-level
potentials and thereby set our notation.

\subsection{The Tree-Level C2HDM Potential}
We briefly introduce the C2HDM Higgs sector and refer to
\cite{Fontes:2017zfn,C2HDM} for a more detailed introduction.  
In 2HDMs \cite{Lee:1973iz,Branco:2011iw} the SM Higgs potential is
extended by an additional 
$\mathrm{SU(2)}_L$ scalar doublet yielding the Higgs potential,
\begin{align}
V_{\text{C2HDM}} =~& m_{11}^2 \Phi_1^{\dagger}\Phi_1 + m_{22}^2
              \Phi_2^{\dagger}\Phi_2 + \frac{\lambda_1}{2}
              \left(\Phi_1^{\dagger}\Phi_1\right)^2 +
              \frac{\lambda_2}{2}
              \left(\Phi_1^{\dagger}\Phi_1\right)^2+ \lambda_3
              \left(\Phi_1^{\dagger}\Phi_1\right)\left(
              \Phi_2^{\dagger}\Phi_2\right) \nonumber\\ 
&+ \lambda_4 \left( \Phi_1^\dagger \Phi_2\right)^2  + \left[
  \frac{\lambda_5}{2} \left(\Phi_1^\dagger\Phi_2\right)^2 - m_{12}^2
  \left(\Phi_1^\dagger \Phi_2\right) + h.c. \right] \,, \label{2HDM:Pot}
\end{align}
with a softly broken discrete $\mathbb{Z}_2$ symmetry under which 
$\Phi_1\rightarrow\Phi_1$, $\Phi_2\rightarrow-\Phi_2$, which
ensures the absence of tree-level flavour-changing neutral currents
(FCNC) upon extension to the Yukawa sector. The hermiticity of the
Lagrangian requires all couplings to be 
real except for $m_{12}^2$ and $\lambda_5$. If their complex phases
are unrelated the model is CP-violating and called C2HDM
\cite{Ginzburg:2002wt}. In the following, we will adopt the
conventions of Refs.~\cite{Fontes:2014xva,Fontes:2017zfn}.
We denote the VEVs of the EW minimum by
$\omega_i\in\mathbb{R}$ ($i=1,2$) and write the two Higgs doublets as
expansions around the VEVs in terms of the charged field components
$\rho_i$ and $\eta_i$ and the neutral CP-even and CP-odd fields
$\zeta_i$ and $\psi_i$. At tree level, the general vacuum structure of
the 2HDM allows for three different possible vacua that are given by the
normal EW-breaking vacuum, a CP-breaking and a charge-breaking (CB)
vacuum. It has been shown that vacua breaking different symmetries
cannot coexist at tree level in the 2HDM
\cite{Ivanov:2007de,Barroso:2007rr,Ferreira:2004yd}. Since this
statement might not hold at higher orders, or be broken by finite
temperature effects, we allow for a more general vacuum structure. We
therefore include the possibility of a CB- and CP-breaking VEV in the
field expansion, denoted by $\omega_{\text{CB}}$ and $\omega_{\text{CP}}$, respectively,
\begin{align}
\Phi_1 =& \frac{1}{\sqrt{2}} \begin{pmatrix}
 \rho_1 + \ii \eta_1 \\ \zeta_1 + \omega_1 + \ii \psi_1
\end{pmatrix} \qquad \Phi_2 = \frac{1}{\sqrt{2}} \begin{pmatrix}
\rho_2 + \omega_{\text{CB}} +\ii \eta_2 \\ \zeta_2 + \omega_2 + \ii
\left(\psi_2 + \omega_{\text{CP}} \right) 
\end{pmatrix}  \;,
\end{align}
with
\begin{align}
\braket{\Phi_1} =& \frac{1}{\sqrt{2}}\begin{pmatrix}
0 \\ \omega_1
\end{pmatrix}	\quad \mbox{and} \quad \braket{\Phi_2} =
  \frac{1}{\sqrt{2}} \begin{pmatrix} 
\omega_{\text{CB}} \\ \omega_2 + \ii \omega_{\text{CP}}
\end{pmatrix} \,,
\label{eq:2hdmdoublets1}
\end{align}
where the brackets around the doublets stand for their vacuum state. 
The VEVs of our present vacuum at zero temperature are denoted by
\begin{equation}
  v_i \equiv \omega_i\big\vert_{T=0}\,  \qquad i =
  1,2,\mathrm{CP},\mathrm{CB} \;,
\label{eq:2hdmdoublets2}
\end{equation}
with
\begin{equation}
  v_{\text{CP}}=v_{\text{CB}} \equiv 0\,.
\end{equation}
A non-zero CB VEV would break the conservation of electric charge
and introduce massive photons so that we will neglect in our analysis parameter points
evolving such unphysical vacuum structures. The VEVs
of the normal EW minimum are related to the SM VEV $v\approx 246\gev$
by 
\begin{equation}
  v_1^2+v_2^2 \equiv v^2\,.
\end{equation}
The angle $\beta$ is defined by the ratio of $v_1$ and $v_2$,
\begin{equation}
  \tan\beta = \frac{v_2}{v_1}\,.
\end{equation}
Using the minimum condition of the potential
\begin{equation}
  \frac{\partial V_{\text{tree}}}{\partial\Phi_i^{\dagger}}\big\vert_{\Phi_j=\braket{\Phi_j}} \overset{!}{=} 0 \,,\quad   i,j\in\lbrace 1,2\rbrace\,,
\end{equation}
yields
\begin{subequations}
\begin{align}
	m_{11}^2 =& \Re m_{12}^2 \frac{v_2}{v_1} - \frac{\lambda_1}{2} v_1^2 - \frac{\lambda_3+\lambda_4+\Re\lambda_5}{2} v_2^2 \label{C2HDM::tadpol1}\\
	m_{22}^2 =& \Re m_{12}^2 \frac{v_1}{v_2} - \frac{\lambda_2}{2} v_2^2 - \frac{\lambda_3+\lambda_4+\Re\lambda_5}{2} v_1^2 \label{C2HDM::tadpol2}\\
	\Im m_{12}^2 =& \Im \lambda_5 \frac{v_1v_2}{2}\label{C2HDM::tadpol3} \,.
\end{align} 
\end{subequations}
We use \eqref{C2HDM::tadpol1} and \eqref{C2HDM::tadpol2} to trade
$m_{11}^2$ and $m_{22}^2$ for $v_1$ and $v_2$. The two possible
sources of CP violation are related to each other through
\eqref{C2HDM::tadpol3} so that one independent CP-violating phase
remains in the C2HDM. Rotating the fields $\psi_1$ and $\psi_2$ with
$\beta$ yields 
\begin{equation}
	\begin{pmatrix}
		G^0 \\ \zeta_3 
	\end{pmatrix}
	 =
	  \begin{pmatrix}
	\cos\beta & \sin\beta \\ -\sin\beta & \cos\beta
	\end{pmatrix}
	\begin{pmatrix}
	\psi_1 \\ \psi_2
	\end{pmatrix} \,,
\end{equation}
where $G^0$ denotes the neutral Goldstone boson. The three neutral
mass eigenstates $H_k$ ($k=1,2,3$) of the C2HDM are obtained by the
rotation of the gauge eigenstates $\zeta_k$ ($k=1,2,3$) to the mass basis, 
\begin{equation}
\label{C2HDM::rotation}
\begin{pmatrix}
H_1\\H_2\\H_3
\end{pmatrix}
=R \begin{pmatrix} \zeta_1\\\zeta_2\\\zeta_3 \end{pmatrix} \,,
\end{equation}
with the rotation matrix $R$ parametrized in terms of the mixing
angles $\alpha_k$ as ($\cos\alpha_k\equiv c_k,$ $\sin\alpha_k \equiv s_k$)
\begin{equation}
R = \begin{pmatrix}
c_1c_2 & s_1 c_2 & s_s\\
-\left(c_1 s_2s_3+ s_1 c_3\right) & c_1 c_3 - s_1 s_2 s_3 & c_2 s_3\\
-c_1s_2c_3+s_1 s_3 & -\left(c_1 s_3+s_1s_2c_3\right) & c_2 c_3
\end{pmatrix}\,.
\label{eq:RotationMatrix}
\end{equation}
Without loss of generality, the rotation angles $\alpha_{1,2,3}$ can
be chosen in the interval
\begin{equation}
  -\frac{\pi}{2}\leq \alpha_i < \frac{\pi}{2}\,.
\end{equation}
The mass eigenvalues are given by
\begin{equation}
R M^2_{\text{Scalar}} R^T = \operatorname{diag}(m_{H_1}^2,m_{H_2}^2,m_{H_3}^2)\,.
\end{equation}
The mass eigenstates are ordered by ascending masses as 
\begin{equation}
  m_{H_1}\leq m_{H_2}\leq m_{H_3}\,.
\end{equation}
Using the minimum conditions and the rotation to the mass eigenstates,
the following set of nine independent parameters of the C2HDM remains
\cite{ElKaffas:2007rq}, 
\begin{equation}
v\,,\quad \tan\beta\,,\quad \alpha_{1,2,3}\,,\quad m_{H_i}\,,\quad
m_{H_j}\,,\quad m_{H^{\pm}}\,\quad \text{and}\,\quad \Re(m_{12}^2)\,. 
\end{equation}
The $m_{H_i}$ and $m_{H_j}$ denote any of the masses of two among the
three neutral Higgs bosons. The third neutral Higgs boson mass is determined through 
\begin{align}
	\sum\limits_{k=1}^{3} m_{H_k}^2 R_{k3} \left( R_{k2} \tan\beta - R_{k1} \right) = 0 \,.
\end{align}
One of the three Higgs bosons is identified with the SM-like Higgs boson
with the measured mass value of $m_h=125.09\gev$ \cite{Aad:2015zhl}. As
mentioned above, extending the
$\mathbb{Z}_2$ symmetry to the Yukawa sector ensures that each type of
the up- and down-type quarks and charged leptons can couple to
only one of the two Higgs doublets so that FCNCs are avoided at tree
level. In \cref{tab::TypeTable} the different types of the 2HDM, type
I, type II, lepton-specific and flipped, are listed. In this analysis
we will concentrate on type I and type II. 
\begin{table}
\center
\begin{tabular}{c c c c | c c c c c }
\toprule
	 		& 	$u$-type &	$d$-type 	&	leptons	 	&	Q	&	$u_R$	&	$d_R$	&	L	& 	$l_R$\\\midrule
Type I  & $\Phi_2$ & $\Phi_2$ & $\Phi_2$ & +& $-$ &$-$&+&$-$\\
Type II & $\Phi_2$ & $\Phi_1$& $\Phi_1$ & +&$-$&+&+&$-$\\
lepton-specific & $\Phi_2$&$\Phi_2$&$\Phi_1$&+&$-$&+&+&$-$\\
flipped & $\Phi_2$&$\Phi_1$&$\Phi_2$&+&$-$&$-$&+&+\\\bottomrule
\end{tabular}
\caption{Left: Definition of the 2HDM types through the allowed couplings among
  fermions and Higgs doublets. Right: Corresponding $\mathbb{Z}_2$
  parity assignments to the left-handed quark and lepton doublets, $Q$,
  $L$, and the right-handed singlets of the up-type and down-type
  quarks, $u_R$ and $d_R$, and right-handed leptons $l_R$.}
\label{tab::TypeTable}
\end{table}

\subsection{The Tree-Level N2HDM Potential}
We give a brief introduction in the N2HDM and refer for more details
to \cite{Muhlleitner:2016mzt}. The tree-level potential of the N2HDM
consists of a CP-conserving 2HDM which is extended by a real singlet field
$\Phi_S$. The potential is invariant under two discrete $\mathbb{Z}_2$
symmetries. We call the one given by the generalisation of the 2HDM
symmetry to avoid tree-level FCNCs $\mathbb{Z}_2$ under which
\beq  
\Phi_1 \to \Phi_1 \;, \quad \Phi_2 \to -\Phi_2 \;, \quad \Phi_S \to  
\Phi_S \;.  
\eeq  
We call the second one $\mathbb{Z}_2^\prime$. It is defined as 
\beq 
\Phi_1 \to \Phi_1 \;, \quad \Phi_2 \to \Phi_2 \;, \quad \Phi_S \to 
- \Phi_S \;.
\eeq 
The most general tree-level potential invariant under these
transformations apart from a term proportional to $m_{12}^2$ that softly
breaks $\mathbb{Z}_2$, reads \cite{Muhlleitner:2016mzt}
\begin{align}
\label{N2HDM_Pot}
V_{\text{N2HDM}}=\, &
m_{11}^2\Phi_1^{\dag}\Phi_1+m_{22}^2\Phi_2^{\dag}\Phi_2-m_{12}^2\left(\Phi_1^{\dag}\Phi_2+\text{h.c.}\right)+\frac{\lambda_1}{2}\left(\Phi_1^{\dag}\Phi_1\right)^2+\frac{\lambda_2}{2}\left(\Phi_2^{\dag}\Phi_2\right)^2\\\notag  
&+\lambda_3\Phi_1^{\dag}\Phi_1\Phi_2^{\dag}\Phi_2+\lambda_4\Phi_1^{\dag}\Phi_2\Phi_2^{\dag}\Phi_1+\frac{\lambda_5}{2}\left((\Phi_1^{\dag}\Phi_2)^2+\text{h.c.}\right)\\\notag
&+\frac{1}{2}m_S^2\Phi_S^2+\frac{\lambda_6}{8}\Phi_S^4+\lambda_7\left(\Phi_1^{\dag}\Phi_1\right)\Phi_S^2+\lambda_8\left(\Phi_2^{\dag}\Phi_2\right)\Phi_S^2\,,
\end{align}
where all parameters are real due to CP-conservation. After
electroweak symmetry breaking the two Higgs doublets and the real 
singlet acquire VEVs $\omega_i\in\mathbb{R}$. As a first analysis of
the N2HDM vacuum structure performed in
Ref.~\cite{Muhlleitner:2016mzt} has shown, the N2HDM exhibits a
different vacuum structure than the 2HDM. The impact of the N2HDM
vacuum structure has been further studied in \cite{Ferreira:2019iqb}
by applying the method of Ref.~\cite{Hollik:2018wrr}. Additionally,
loop corrections and finite temperature effects may change the vacuum
texture, so that we expand the doublet and singlet fields including
the most general vacuum structure, as
\begin{equation}
\label{N2HDM::EWMini1}
\Phi_1 = \frac{1}{\sqrt{2}} \begin{pmatrix}
\rho_1 + \ii \eta_1 \\ \zeta_1 + \omega_1 + \ii \psi_1
\end{pmatrix} \;,\qquad \Phi_2 = \frac{1}{\sqrt{2}} \begin{pmatrix}
\rho_2 + \omega_{\text{CB}} +\ii \eta_2 \\ \zeta_2 + \omega_2 + \ii
\left(\psi_2 + \omega_{\text{CP}} \right)  
\end{pmatrix}   \;, \qquad  \Phi_S = \zeta_3 + \omega_S\,,
\end{equation}
where we have expanded the Higgs fields in terms of the charged field
components $\rho_i$ and $\eta_i$ ($i=1,2$), and the neutral CP-even
and CP-odd fields $\zeta_i$ and $\psi_i$, respectively. As mentioned
above, already the tree-level zero-temperature vacuum can exhibit
a quite general structure as compared to the 2HDM. For simplicity we
choose for our present vacuum at zero temperature a vacuum structure
with the same properties as in the C2HDM, given by
\begin{align}
\langle\Phi_1\rangle\big\vert_{T=0}=\frac{1}{\sqrt{2}}\left(
\begin{array}{c}
0\\
v_1
\end{array}\right),\qquad
\langle\Phi_2\rangle\big\vert_{T=0}=\frac{1}{\sqrt{2}}\left(
\begin{array}{c}
0\\
v_2
\end{array}\right)
,\qquad \langle S\rangle \big\vert_{T=0} = v_S\,,
\label{N2HDM::TzeroVeVconfig}
\end{align}
with the zero temperature VEVs
\begin{eqnarray}
   v_i \equiv \omega_i\big\vert_{T=0}\,, \quad i=1,2,S,\mbox{CP},\mbox{CB}\,.
\end{eqnarray}
The CP-violating and CB VEVs are hence chosen to be vanishing at zero temperature
\begin{equation}
  v_{\text{CP}} = v_{\text{CB}} = 0 \,.
\end{equation}
And the electroweak VEVs $v_{1,2}$ are related to the SM VEV $v\approx 246\gev$ by
\begin{equation}
  v_1^2+ v_2^2 \equiv v^2\,.
\end{equation}
As in the (C)2HDM, the ratio of the electroweak VEVs $v_1$ and $v_2$
is defined by the angle $\beta$, 
\begin{equation}
  t_\beta = \tan\beta\equiv\frac{v_2}{v_1}\,.
\end{equation}
Requiring the tree-level potential in \eqref{N2HDM_Pot} to be
minimized at the electroweak vacuum Eq.~(\ref{N2HDM::TzeroVeVconfig}) yields 
\begin{subequations}
\label{N2HDM::TadpolCondition}
\begin{align}
v_2 m_{12}^2- v_1 m_{11}^2 & = \frac{v_1}{2} \left(v_1^2\lambda_1+v_2^2\lambda_{345} +v_S^2\lambda_7\right)\\
v_1 m_{12}^2- v_2 m_{22}^2 & = \frac{v_2}{2}\left(v_1^2\lambda_{345}+v_2^2\lambda_2+v_S^2\lambda_8\right)\\
-m_S^2 v_s  & = \frac{v_s}{2} \left(v_1^2\lambda_7+v_2^2\lambda_8+v_S^2\lambda_6\right)\,,
\end{align}
\end{subequations}
with
\begin{equation}
  \lambda_{345} \equiv \lambda_3+\lambda_4+ \lambda_5\,.
\end{equation}
Equation~(\ref{N2HDM::TadpolCondition}) allows to trade the parameters
$m_{11}^2$, $m_{22}^2$ and $m_S^2$ for the VEVs $v_{1,2,S}$ resulting
in the mass matrix for the CP-even neutral fields $\zeta_{1,2,3}$ at
zero temperature 
\begin{equation}
M^2_{\text{Scalar}}=
\left(
\begin{array}{ccc}
 v^2 \lambda_1 c^2_{\beta}+m_{12}^2 t_{\beta} & v^2 \lambda_{345} c_{\beta} s_{\beta}-m_{12}^2 & v v_S
   \lambda_7 c_{\beta} \\
 v^2 \lambda_{345} c_{\beta} s_{\beta}
   -m_{12}^2 & v^2 \lambda_2 s^2_{\beta}+m_{12}^2 /t_{\beta} & v v_S
   \lambda_8 s_{\beta} \\
 v v_S \lambda_7 c_{\beta} & v v_S \lambda_8 s_{\beta} &
   v_S^2 \lambda_6 \\
\end{array}
\right)\,.
\end{equation}
The matrix $R$ rotates the mass matrix $M^2_{\text{Scalar}}$ in the
mass eigenstates $H_{1,2,3}$ 
\begin{equation}
\left(
\begin{array}{c}
H_1\\
H_2\\
H_3
\end{array}
\right)
=
R \left(
\begin{array}{c}
\rho_1\\
\rho_2\\
\rho_S
\end{array}
\right)\;,
\label{N2HDM::rotation}
\end{equation}
with the mass eigenvalues
\begin{equation}
R M^2_{\text{Scalar}} R^T = \operatorname{diag}(m_{H_1}^2,m_{H_2}^2,m_{H_3}^2)\,.
\end{equation}
We use the same convention as in the C2HDM where the mass eigenstates
are ordered by ascending masses as 
\begin{equation}
  m_{H_1}\leq m_{H_2}\leq m_{H_3}\,.
\end{equation}
The N2HDM has 12 real independent parameters where we chose as many
parameters as possible with a physical meaning. We take
\eqref{N2HDM::TadpolCondition} to trade the potential parameters
$m_{11}^2$, $m_{22}^2$ and $m_S^2$ for the SM and singlet VEVs $v$ and $v_S$, and
$\tan\beta$. Moreover, we express the quartic couplings $\lambda_i$ in
terms of the physical masses and mixing angles. The $\mathbb{Z}_2$
breaking mass term $m_{12}^2$ is kept as independent parameter. This
yields the following set of input parameters 
\begin{equation}
\alpha_1 \,,\quad \alpha_2\,,\quad \alpha_3 \,,\quad t_{\beta}\,,\quad v \,,\quad v_S \,,\quad m_{H_{1,2,3}} \,,\quad m_A \,,\quad m_{H^{\pm}} \,,\quad m_{12}^2\,.
\end{equation}
To avoid FCNC in the N2HDM the same types as in the C2HDM, given in
\cref{tab::TypeTable}, can be used since the singlet field $\Phi_S$ does
not couple to fermions. For more details and a phenomenological
discussion of the N2HDM we refer to \cite{Muhlleitner:2016mzt}. 

\subsection{One-loop Effective Potential at Finite Temperature}
In the following we briefly repeat the main ingredients for the
one-loop effective potential at non-zero temperature. For details, we
refer to \cite{C2HDM,Basler:2016obg,Basler:2018cwe} which can be generalized to the
N2HDM. \s 

The loop-corrected effective potential splits into the tree-level
potential $V_{\text{tree}}$, the one-loop correction at zero
temperature given by the Coleman-Weinberg (CW) potential
$V_{\text{CW}}$, and the temperature-dependent
part $ V_T$, so that it reads 
\begin{equation} 
  V\equiv V_{\text{tree}} + V_{\text{CW}} + V_{T}\,,
\end{equation}
where $V_{\text{tree}}$ is given by Eq.~(\ref{2HDM:Pot}) for the C2HDM
and Eq.~(\ref{N2HDM_Pot}) for the N2HDM, after replacing the
doublet fields of the C2HDM with their classical constant field
configuration given in
Eqs.~(\ref{eq:2hdmdoublets1}), (\ref{eq:2hdmdoublets2}), and the doublet 
and singlet fields of the N2HDM with \eqref{N2HDM::TzeroVeVconfig}. 
The CW potential \cite{Coleman:1973jx} in the $\MSb$ scheme is given by
\begin{equation}
V_{\text{CW}}=\sum_j
\frac{n_j}{64\pi^2}(-1)^{2s_i}m_j^4\left[\ln\left(\frac{m_j^2}{\mu^2}\right)-c_j\right]\,, 
\label{VCW_Potential}
\end{equation}
where $n_j$ are the degrees of freedom, $s_j$ the spin and $m_j$ the
mass of the specific particle $j$. 
The sum extends over the Higgs and Goldstone bosons, the massive gauge
bosons, the photon and the fermions, so that both for the
C2HDM and the N2HDM we have
\begin{equation}
	j\in\lbrace H_1, H_2, H_3,A, H^{\pm}, G^0, G^{\pm}, W^{\pm}, Z, \gamma, f\rbrace\,.
\end{equation}
The mass $m_j$ is the mass eigenvalue of particle $j$
obtained from the tree-level mass matrix expressed in terms of the
general VEV configuration $\omega_i$, with
$i=1,2,\mathrm{CP},\mathrm{CB}$ for the C2HDM and
$i=1,2,S,\mathrm{CP},\mathrm{CB}$ for the N2HDM.  
Applying the Landau gauge in the calculation of the CW potential
allows us to drop the ghost contributions in the analysis, but we need to
account for the possibility that the Goldstone bosons get massive. The
Goldstone bosons as well the photon are massless at $T=0$, but for
field configurations different from the tree-level VEVs at $T=0$,
which is needed for the minimization procedure, they can acquire an
effective mass term. Furthermore, we allow for unphysical vacuum
structures with non-zero $\omega_{\text{CB}}$ inducing additionally
unphysical masses. We also account for the possibility
  of the generation of a CP-violating VEV $\omega_{\text{CP}}$.
For the neutral scalars $\Phi^0=H_1,H_2,H_3,A,G^0$, the charged
scalars $\Phi^{\pm}=H^{+},H^-,G^{+},G^-$, the leptons $l^+$,$l^-$,
quarks $q,\overline{q}$ and longitudinal and transversal gauge bosons
$V_L=Z_L,W_L^+,W_L^-,\gamma_L$ and $V_T=Z_T,W_T^+,W_T^-,\gamma_T$ the
degrees of freedom in \eqref{VCW_Potential} read 
\begin{table}[h]
\center
\begin{tabular}{c c c c }
$n_{\Phi^0}=1$,	&	$n_{\Phi^{\pm}}=2$,	&	$n_{V_T}=2$,	&	$n_{V_L}=1$, \\
$n_{l^+}=2$,	&	$n_{l^-}=2$,	&$n_q=6$, & $n_{\bar q}=6$ \,.
\end{tabular}
\end{table}\\
The renormalisation scale is chosen as $\mu=v\approx 246.22\gev$ and the
$\overline{\mbox{MS}}$ renormalisation constants read
\begin{equation}
c_j=
\begin{cases}
\frac{5}{6}\,,\quad j=W^{\pm},Z,\gamma\\
\frac{3}{2}\,,\quad \text{otherwise}
\end{cases}
\,.
\end{equation}
The thermal corrections $V_T$ comprise the daisy resummation
\cite{Carrington:1991hz} of the $n=0$ Matsubara modes of the
longitudinal components of the gauge bosons $W^{\pm}_L, Z_L, \gamma_L$
and the bosons $\Phi^0$, $\Phi^\pm$. This 
Debye correction adds to their masses at zero
temperature. The thermal corrections $V_T$ can be cast in the form
\cite{Dolan:1973qd,Quiros:1999jp} 
\begin{equation}
V_T= \sum_k n_k\frac{T^4}{2\pi^2}J_{\pm}^{(k)}\,.
\end{equation}
The sum extends over $k = W^{\pm}_{T,L}, Z_{L,T}, \gamma_{L,T},
\Phi^0, \Phi^\pm,l^{\pm}, q , \overline{q}$. Denoting
the mass eigenvalues including 
the thermal corrections for the particles $k$ by $\overline{m_k}$, we
have for $J_{\pm}$ 
(\textit{cf.~e.g.}\cite{MSSM2})\footnote{We use the
    'Arnold-Espinosa' \cite{Arnold:1992rz,Parwani:1991gq} 
  approach for the inclusion of the Debye corrected masses. For
  further remarks on the approach and how it compares to the 'Parwani'
approach, see \cite{}. Further discussions and comparisons can also
be found in \cite{Cline:1996mga,Cline:2011mm}.}
\begin{equation}
J_{\pm}^{(k)} =
\begin{cases}
J_-(\frac{m_k^2}{T^2})-\frac{\pi}{6}\left(\overline{m}_k^3/T^3 - m_k^3/T^3\right), \quad & k=W_L,Z_L,\gamma_L,\Phi^0,\Phi^{\pm}\\
J_-(\frac{m_k^2}{T^2}) \quad & k=W_T, Z_T, \gamma_T\\
J_+(\frac{m_k^2}{T^2}) \quad & k=\text{fermion}\,\\
\end{cases}
\label{thermalIntegral}
\end{equation}
with the thermal integrals
\begin{equation}
J_{\pm}(x) = \mp\int_0^{\infty}dx
\,x^2\log\left[1\pm\exp(-\sqrt{x^2+m_k^2/T^2})\right]\,,
\end{equation}
where $J_+$ ($J_-$) applies for $k$ representing a fermion (boson).
The general formulae for the thermal masses can be
  found in \cite{Basler:2018cwe}.
For the numerical evaluation of the effective potential at finite
temperatures and the further minimization procedure we use the code
{\tt BSMPT v1.1.2}~\cite{Basler:2018cwe} and we refer to
\cite{Basler:2018cwe} for a more detailed discussion of the used
numerical approximations in the thermal integrals in
\eqref{thermalIntegral}. 

\section{Renormalisation \label{sec:renormalisation}}
At one-loop level the masses and mixing angles differ from those
extracted from the tree-level potential. In order to take into account
the one-loop effects and at the same time enable an efficient scan
in the parameter space of the model, we renormalize the loop-corrected
masses and mixing angles such that they are equal to their tree-level
values. This allows us to use them directly as input values for our
scan. The scheme has been introduced in Ref.~\cite{Basler:2016obg} where it
was applied to the 2HDM. In \cite{C2HDM} it was extended to the
C2HDM. We therefore show here only the procedure
for the N2HDM. The renormalised loop-corrected potential $\hat{V}$ is 
obtained by adding the counterterm potential $V_{\text{CT}}$
\begin{equation}
  \hat{V}=V+V_{\text{CT}}=V_{\text{tree}}+V_{\text{CW}}+V_{T}+V_{\text{CT}}\,
  \label{N2HDM::FULLPOTENTIAL}
\end{equation}
with the counterterm potential given by
\begin{equation}
  V_{\text{CT}} = \sum\limits_{i=1} \frac{\partial V_{\text{tree}}}{\partial
    p_i} \delta p_i + \sum\limits_{k=1} \left(\phi_k +
    \omega_k\right)\delta T_k \;,
\end{equation}
where $p_i$ stands for the parameters of the tree-level
potential. Furthermore, for each field direction $\phi_k$ which is
allowed to develop a VEV, a tadpole counterterm $\delta T_k$ is
introduced. For the N2HDM these are $~\delta T_1\,,~ \delta
T_2\,,~\delta T_S\,,~\delta T_{\text{CP}}$ and $\delta T_{\text{CB}}$. 
 The renormalisation conditions for the scheme described above read
\begin{equation}
\partial_{\phi_i}V_{\text{CT}}(\phi)\big\vert_{\langle\phi\rangle_{T=0}}
=-\partial_{\phi_i}V_{\text{CW}}(\phi)\big\vert_{\langle\phi\rangle_{T=0}}  
\label{CT::mini_condition}
\end{equation}
and
\begin{equation}
\partial_{\phi_i}\partial_{\phi_j}V_{\text{CT}}(\phi)\big\vert_{\langle\phi\rangle_{T=0}}
=
- \partial_{\phi_i}\partial_{\phi_j}V_{\text{CW}}(\phi)\big\vert_{\langle\phi\rangle_{T=0}}\,, 
\label{CT::mass_condition}
\end{equation}
where $\langle\phi\rangle_{T=0}$ denotes the tree-level vacuum state
at zero temperature. The conditions \eqref{CT::mini_condition} and
\cref{CT::mass_condition} ensure that at zero temperature the
tree-level minimum remains a local minimum at one-loop level. We check
numerically if it also the global one. Additionally, the second set of
conditions in \eqref{CT::mass_condition} ensures that the masses
and mixing angles derived from the loop-corrected effective potential
remain at their tree-level values. In general, the 
renormalisation conditions result in an overconstrained system of
equations which can be solved by imposing additional assumptions. In
the case of the N2HDM Eqs.~(\ref{CT::mini_condition}) and
(\ref{CT::mass_condition}) lead to a two-dimensional solution space
which can be fixed by imposing 
\begin{equation}
  \delta\lambda_4 =0 \qquad \text{and} \qquad \delta T_S = 0\,.
\end{equation}
Solving then for the counterterm parameters yields
\begin{subequations}\begin{align}
\delta m_{11}^2 &= \frac{1}{2} \left[ \frac{v_s}{v_1} H^{\text{CW}}_{\zeta_1,\zeta_S} + \frac{v_2}{v_1} \left( H^{\text{CW}}_{\zeta_1,\zeta_2} - H^{\text{CW}}_{\eta_1,\eta_2} \right) + 2H^{\text{CW}}_{\eta_1,\eta_1} - 5H^{\text{CW}}_{\eta_1,\eta_1} + H^{\text{CW}}_{\zeta_1,\zeta_1} \right] \\
		\delta m_{22}^2 &= \frac{1}{2} \left[ \frac{v_s}{v_2} H^{\text{CW}}_{\zeta_2,\zeta_S} + H^{\text{CW}}_{\zeta_2,\zeta_2} - 3H^{\text{CW}}_{\eta_2,\eta_2}  + \frac{v_1}{v_2} \left( H^{\text{CW}}_{\zeta_1,\zeta_2}  - H^{\text{CW}}_{\eta_1,\eta_2} \right) + 5\frac{v_1^2}{v_2^2} \left(H^{\text{CW}}_{\eta_1,\eta_1}-H^{\text{CW}}_{\eta_1,\eta_1}\right) \right]\\
		\delta m_{12}^2 &= H^{\text{CW}}_{\eta_1,\eta_2} + \frac{v_1}{v_2} \left(H^{\text{CW}}_{\eta_1,\eta_1} - H^{\text{CW}}_{\eta_1,\eta_1} \right)\\
		\delta \lambda_1 &= \frac{1}{v_1^2} \left( 2H^{\text{CW}}_{\eta_1,\eta_1} - H^{\text{CW}}_{\eta_1,\eta_1} - H^{\text{CW}}_{\zeta_1,\zeta_1}\right)  \\
		\delta \lambda_2 &= \frac{1}{v_2^2} \left(H^{\text{CW}}_{\eta_2,\eta_2}-H^{\text{CW}}_{\zeta_2,\zeta_2}\right) + 2\frac{v_1^2}{v_2^4} \left( H^{\text{CW}}_{\eta_1,\eta_1}-H^{\text{CW}}_{\eta_1,\eta_1}\right) \\
		\delta \lambda_3 &= \frac{1}{v_2^2} \left( H^{\text{CW}}_{\eta_1,\eta_1} - H^{\text{CW}}_{\eta_1,\eta_1} \right) + \frac{1}{v_1v_2} \left( H^{\text{CW}}_{\eta_1,\eta_2} - H^{\text{CW}}_{\zeta_1,\zeta_2} \right)\\
		\delta \lambda_4 &= 0 \\
		\delta \lambda_5 &= \frac{2}{v_2^2} \left(H^{\text{CW}}_{\psi_1,\psi_1} - 2H^{\text{CW}}_{\eta_1,\eta_1}\right)  \\
		\delta m_S^2 &= \frac{1}{2} \left( H^{\text{CW}}_{\zeta_S,\zeta_S} + \frac{v_2}{v_s} H^{\text{CW}}_{\zeta_2,\zeta_S} + \frac{v_1}{v_s} H^{\text{CW}}_{\zeta_1,\zeta_S} - \frac{3}{v_S} N^{\text{CW}}_{\zeta_S} \right)  \\
		\delta \lambda_6 &= \frac{1}{v_s^3} \left(N^{\text{CW}}_{\zeta_S} - v_s H^{\text{CW}}_{\zeta_S,\zeta_S} \right)\\
		\delta \lambda_7 &=  - \frac{1}{v_sv_1} H^{\text{CW}}_{\zeta_1,\zeta_S} \\
		\delta \lambda_8 &= - \frac{1}{v_sv_2} H^{\text{CW}}_{\zeta_2,\zeta_S} \\
		\delta T_1 &= H^{\text{CW}}_{\eta_1,\eta_1}v_1 + H^{\text{CW}}_{\eta_1,\eta_2} v_2 - N^{\text{CW}}_{\zeta_1} \\
		\delta T_2 &= \frac{v_1^2}{v_2} \left(H^{\text{CW}}_{\eta_1,\eta_1} - H^{\text{CW}}_{\psi_1,\psi_1}\right) + H^{\text{CW}}_{\eta_1,\eta_2} v_1 + H_{\eta_2,\eta_2} v_2 - N^{\text{CW}}_{\rho_2} \\
		\delta T_S &= 0\\
		\delta T_3 &= \frac{v_1^2}{v_2} H^{\text{CW}}_{\zeta_1,\psi_1} + H^{\text{CW}}_{\zeta_1,\psi_2} v_1 - N^{\text{CW}}_{\psi_2} \\
	\delta T_{CB} &= -N^{\text{CW}}_{\rho_2}  \,,
\end{align}
\label{eq:countertermsn2hdm}
	\end{subequations}
with the shorthand notations
\begin{equation}
N_{\phi_i}^{\text{CW}}\equiv \partial_{\phi_i}V_{\text{CW}}
\end{equation}
and
\begin{equation}
H_{\phi_i\phi_j}^{\text{CW}}=\partial_{\phi_i}\partial_{\phi_j}V_{\text{CW}}\,.
\end{equation}
In Ref.~\cite{MarcoBible} formulae for both the first and the second
derivatives of the CW potential have been derived in the Landau gauge
basis. 

\section{Numerical Analysis \label{sec:numerical}}
\subsection{Minimisation of the Effective Potential}
The EWPT is of strong first order if the baryon-washout condition is
met which requires that the ratio $\xi_c$ of the critical VEV $v_c$
and the critical temperature $T_c$ is larger than one
\cite{Quiros:1994dr,Quiros:1999jp,Moore:1998swa}, 
\begin{equation}
  \xic \equiv \frac{v_c}{T_c} \geq 1\,.
\end{equation}
The VEV $v$ at the temperature $T$ is given by
\begin{equation}
  v(T) = \sqrt{\omega_1^2(T)+\omega_2^2(T)+\omega_{\text{CP}}^2(T)+\omega_{\text{CB}}^2(T)}\,,
  \label{NUM::EW_VEV}
\end{equation}
where the $\omega_i$ are the field configurations that minimise the
loop-corrected effective potential at finite temperature $T$. Note that
we do not include the singlet VEV $\omega_S$ (present in the N2HDM) in
\eqref{NUM::EW_VEV}, 
but we take $\omega_S$ into account for the minimisation
procedure. Since the electroweak sphaleron couples only to particles
charged under $\mathrm{SU(2)}_L$, the singlet VEV can be dropped in
the calculation of the critical VEV in \cref{NUM::EW_VEV}. 
The critical temperature $T_c$ is defined as the
temperature where the potential has developed two degenerate
minima. In order to compute the global electroweak minimum of
the one-loop corrected effective C2HDM and N2HDM potentials of
\eqref{N2HDM::FULLPOTENTIAL} we implemented both models in 
{\tt BSMPT v1.1.2} \cite{Basler:2018cwe} which also calculates 
the strength $\xic$ of the phase transition.

\subsection{Constraints and Parameter Scan} \label{Sec:ScanConstraints}
For simplicity, in the numerical analysis we discuss only Type I and II of the models. The parameter samples used for the numerical
investigation in this paper have to satisfy theoretical and
experimental constraints. We obtained them by performing scans in the
parameter spaces of the C2HDM and N2HDM, respectively. For the scans
we required that one of the neutral Higgs bosons, denoted by $h$ in
the following, behaves SM-like and has a mass of $m_h=125.09\gev$
\cite{Aad:2015zhl}. The scan ranges for all the input parameters are given in
\cref{N2HDM::ParamT1} for Type I (T1) and in \cref{N2HDM::ParamT2} for
Type II (T2) of the N2HDM, for the C2HDM they are given in
\cref{C2HDM::ParamT1} for Type I and in \cref{C2HDM::ParamT2} for Type
II. We introduce here the notation $m_{H_\downarrow/H_\uparrow}$ for the masses
of the lighter/heavier of the two non-SM-like neutral Higgs bosons. \s
\begin{table}[t]
    \centering
    \begin{tabular}{c c c c c c c c }
    \toprule
    $m_{h}$ 	&	$\mdown$ 	& 	$\mup$	& $m_A$	&	$\mHc$	&	$m_{12}^2$ \\
    && in $\gev$&&& in $\gev^2$&  \\\midrule
    $125.09$ & $\left[30,1500\right]$ & $\left[30,1500\right]$ & $\left[30,1500\right]$ & $\left[30,1500\right]$ & $\left[10^{-3},10^5\right]$ & \\\bottomrule
    $\alpha_1$  & $\alpha_2$ & $\alpha_3$ & $\tan\beta$ & $v_S \left[\gev\right]$\\\midrule
    $\left[-\frac{\pi}{2} , \frac{\pi}{2}\right)$ &  $\left[-\frac{\pi}{2} , \frac{\pi}{2}\right)$ &  $\left[-\frac{\pi}{2} , \frac{\pi}{2}\right)$ & $\left[0.8,20\right]$ & $\left[1,3000\right]$\\\bottomrule
    \end{tabular}
    \caption{Parameter ranges for the N2HDM T1 input parameters used
      in {\tt ScannerS}.}
    \label{N2HDM::ParamT1}
\vspace*{0.5cm}
    \end{table}

\begin{table}[t]
    \centering
    \begin{tabular}{c c c c c c c c }
        \toprule
        $m_{h}$ 	&	$\mdown$ 	& 	$\mup$	& $m_A$	&	$\mHc$	&	$m_{12}^2$ \\
        && in $\gev$&&& in $\gev^2$&  \\\midrule
        $125.09$ & $\left[30,1500\right]$ & $\left[30,1500\right]$ & $\left[30,1500\right]$ & $\left[580,1500\right]$ & $\left[10^{-3},10^5\right]$ & \\\bottomrule
        $\alpha_1$  & $\alpha_2$ & $\alpha_3$ & $\tan\beta$&$v_S \left[\gev\right]$\\\midrule
        $\left[-\frac{\pi}{2} , \frac{\pi}{2}\right)$ &  $\left[-\frac{\pi}{2} , \frac{\pi}{2}\right)$ &  $\left[-\frac{\pi}{2} , \frac{\pi}{2}\right)$ & $\left[0.8,20\right]$& $\left[1,3000\right]$\\\bottomrule
        \end{tabular}
    \caption{Parameter ranges for the N2HDM T2 input parameters used
      in {\tt ScannerS}.}
    \label{N2HDM::ParamT2}
    \end{table}
%
\begin{table}[t]
    \centering
    \begin{tabular}{c c c c c   }
        \toprule
        $m_{h}$ 	&	$\mdown$ 	& 	$\mup$	&	$\mHc$	&	$\Re m_{12}^2$ \\
        && in $\gev$&& in $\gev^2$  \\\midrule
        $125.09$ &  $\left[30,1500\right]$ & $\left[30,1500\right]$ &  $\left[30,1500\right]$ & $\left[10^{-3},10^5\right]$  \\\bottomrule
        $\alpha_1$  & $\alpha_2$ & $\alpha_3$ & $\tan\beta$\\\midrule
        $\left[-\frac{\pi}{2} , \frac{\pi}{2}\right)$ &  $\left[-\frac{\pi}{2} , \frac{\pi}{2}\right)$ &  $\left[-\frac{\pi}{2} , \frac{\pi}{2}\right)$ & $\left[0.8,20\right]$\\\bottomrule
        \end{tabular}
    \caption{Parameter ranges for the C2HDM T1 input parameters used
      in {\tt ScannerS}.}
    \label{C2HDM::ParamT1}
\vspace*{0.5cm}
    \end{table}

    \begin{table}[t]
        \centering
        \begin{tabular}{c c c c c   }
            \toprule
            $m_{h}$ 	&	$\mdown$ 	& 	$\mup$	&	$\mHc$	&	$\Re m_{12}^2$ \\
            && in $\gev$&& in $\gev^2$  \\\midrule
            $125.09$ &  $\left[30,1500\right]$ & $\left[30,1500\right]$ &  $\left[580,1500\right]$ & $\left[10^{-3},10^5\right]$  \\\bottomrule
            $\alpha_1$  & $\alpha_2$ & $\alpha_3$ & $\tan\beta$\\\midrule
            $\left[-\frac{\pi}{2} , \frac{\pi}{2}\right)$ &  $\left[-\frac{\pi}{2} , \frac{\pi}{2}\right)$ &  $\left[-\frac{\pi}{2} , \frac{\pi}{2}\right)$ & $\left[0.8,20\right]$\\\bottomrule
            \end{tabular}
        \caption{Parameter ranges for the C2HDM T2 input parameters
          used in {\tt ScannerS}.}
        \label{C2HDM::ParamT2}
        \end{table}

As for the SM parameters of our analysis, 
we use the fine structure constant taken at the $Z$ boson mass scale
\cite{Agashe:2014kda,LHCHXSWG}, 
\begin{equation}
\alpha_{\text{EM}}^{-1}(M_Z^2) = 128.962 \;,
\end{equation}
and the masses for the massive gauge bosons are chosen as
\begin{equation}
m_W = 80.385\gev \quad \text{and }\quad m_Z=91.1876\gev\,.
\end{equation}
The lepton masses are set to
\begin{equation}
m_e=0.511~\mathrm{MeV},\quad m_{\mu}=105.658~\mathrm{MeV},\quad m_{\tau} = 1.777\gev\,,
\end{equation}
and the light quark masses to
\begin{equation}
m_u = m_d = m_s = 100~\mathrm{MeV}\,.
\end{equation}
To be consistent with the CMS and ATLAS analyses, we take the on-shell
top quark mass as \cite{Dittmaier:2011ti} 
\begin{equation}
m_t = 172.5\gev\,
\end{equation}
and the recommended charm and bottom quark on-shell masses
\begin{equation}
m_c=1.51\gev \quad \text{and} \quad m_b = 4.92\gev\,.
\end{equation}
The CKM matrix is taken to be real, with the CKM matrix elements given by \cite{Agashe:2014kda}
\begin{equation}
  V_{\text{CKM}} =
  \begin{pmatrix}
    &V_{ud} & V_{us} & V_{ub}&\\
    &V_{cd} & V_{cs} & V_{cb}&\\
    &V_{td} & V_{ts} & V_{tb}&
  \end{pmatrix}
  =
  \begin{pmatrix}
    &0.97427 & 0.22536 & 0.00355 &\\
    &-0.22522 & 0.97343 & 0.0414 &\\
    &0.00886 & -0.0405 & 0.99914 &
  \end{pmatrix}
\end{equation}
Finally, the electroweak VEV is set to
\begin{equation}
v=1/\sqrt{\sqrt{2}G_F}=246.22\gev\,.
\end{equation}
To be consistent with recent flavour constraints, we test for the compatibility
with $\mathcal{R}_b$ \cite{Haber:1999zh,Deschamps:2009rh} and
$B\rightarrow X_s \gamma $
\cite{Deschamps:2009rh,flavor1,flavor2,flavor3,charged580} in the
$m_{H^{\pm}}-\tan\beta$ plane. For T2, this implies that the charged
Higgs mass has to be above 580~GeV \cite{charged580} whereas in T1
this bound is much weaker and is strongly correlated with 
$\tan\beta$.\footnote{Many of the experimental constraints applied on
    the 2HDM also hold for the N2HDM, since these constraints are only
    sensitive to the charged Higgs boson so that the calculation of the
    2HDM can be taken over to the N2HDM \cite{Muhlleitner:2016mzt}. In
    this way we are able to use the same constraints for the C2HDM and
    N2HDM.} 
Compatibility with the electroweak precision data is checked
through the oblique parameters $S,T$ and $U$ \cite{Peskin:1991sw}
where we apply the general procedure for extended Higgs sectors given
in \cite{STU1,STU2}. Including the full correlations, we demand
$2\sigma$ compatibility with the SM fit \cite{STU3}. \s

In order to avoid degenerate Higgs signals we impose a mass window between
the non-SM- and SM-like Higgs bosons so that masses $\vert m_{H_i\neq
  h } - m_h\vert < 5\gev$ are excluded from the analysis. 
We use the program {\tt ScannerS}
\cite{Coimbra:2013qq,Ferreira:2014dya,Costa:2015llh,Muhlleitner:2016mzt} to search
for valid parameter points. The program allows to check for 
boundedness from below of the tree-level potential and,
for the C2HDM, uses the tree-level
discriminant of \cite{Ivanov:2015nea} to ensure 
the electroweak vacuum to be the global minimum at tree
level.  For the N2HDM, all tree-level minima have been
implemented in {\tt ScannerS} and are compared numerically to find the
global minimum.
To check for consistency with the Higgs exclusion limits from LEP,
Tevatron and LHC {\tt HiggsBounds5.5.0} \cite{HiggsBound1,
  HiggsBound2, HiggsBound3} is used, and the SM-like Higgs rates are required to
be within the 2$\sigma$ range of the SM which is checked by {\tt
  HiggsSignals2.3.0} \cite{Bechtle:2013xfa}.  
The required decay widths and branching ratios are obtained from {\tt
  C2HDM\_HDECAY} \cite{Fontes:2017zfn} for the C2HDM and from {\tt
  N2HDECAY} \cite{Muhlleitner:2016mzt,Engeln:2018mbg} for the N2HDM. Both
codes are based on the implementation of the C2HDM and the N2HDM each
in the existing code {\tt HDECAY} \cite{Djouadi:1997yw,Djouadi:2018xqq}. 
Due to the CP violation, in the C2HDM also compatibility with the
electric dipole moments 
\cite{Inoue:2014nva} has to be checked.The most stringent limit is
provided by the ACME collaboration \cite{Andreev:2018ayy}. 
Moreover, we take into account for both models the impact of the
recent di-Higgs searches in the final states $4b$ \cite{DIHIGS1,
  DIHIGS2}, $(2b)(2\tau)$ \cite{DIHIGS3,DIHIGS4} and
$(2b)(2\gamma)$ \cite{CMS:2017ihs} on the viable parameter
space. These searches are also 
implemented in {\tt HiggsBounds} and {\tt HiggsSignals}. \s

In addition to these checks, we impose the requirement that the
tree-level minimum of the potential is still the global electroweak
minimum at one-loop level. The one-loop minimum is determined by
numerical minimisation of the one-loop potential at zero temperature
and checked against the tree-level value. In the following, all
parameter points providing an SFOEWPT also have an NLO stable global
minimum at zero temperature. Furthermore, we demand an approximated
NLO unitarity. The tree-level perturbative unitarity relations for the N2HDM read
\cite{Muhlleitner:2016mzt,Basler:2018cwe}  
\begin{align}
    \left\vert \frac{1}{2} a_{1,2,3} , b_\pm, c_\pm, e_{1,2}, f_\pm, f_1, p_1 , s_1, s_2 \right\vert & < 8\pi \label{N2HDM:Unitarity} \,,
\end{align}
with 
\begin{subequations}
\begin{align}
   b_\pm ={}& \frac{1}{2} \left(\lambda_1 + \lambda_2 \right) \pm \frac{1}{2} \sqrt{\left(\lambda_1-\lambda_2\right)^2 + 4\lambda_4^2} \,,\\
   c_\pm ={}& \frac{1}{2} \left(\lambda_1 + \lambda_2\right) \pm \frac{1}{2} \sqrt{\left(\lambda_1-\lambda_2\right)^2 + 4 \lambda_5^2} \,,\\
   e_1 ={}& \lambda_3+2\lambda_4-3\lambda_5 \,,\\
   e_2 ={}& \lambda_3-\lambda_5 \,,\\
   f_+ ={}& \lambda_3 + 2\lambda_4 + 3\lambda_5 \,,\\
   f_- ={}& \lambda_3+\lambda_5 \,, \\
   f_1 ={}& \lambda_3 + \lambda_4 \,,\\
   p_1 ={}& \lambda_3 - \lambda_4  \,,\\
   s_1 ={}& \lambda_7 \,,\\
   s_2 ={}& \lambda_8 
\end{align}
\label{UNIT}
\end{subequations}
and the eigenvalues $a_{1,2,3}$ that are the real roots of the cubic polynomial 
\begin{align}
   f(x)  ={}& x^3 + x^2 \left[ - 6\left(\lambda_1+\lambda_2\right) - 3\lambda_6 \right] \notag \\ &+ x \left[ 36\lambda_1\lambda_2-16\lambda_3^2-16\lambda_3\lambda_4-4\lambda_4^2+18\lambda_1\lambda_6+18\lambda_2\lambda_6-4\lambda_7^2-4\lambda_8^2 \right] \notag \\ &+ 4\left( -27\lambda_1\lambda_2\lambda_6+12\lambda_3^2\lambda_6+12\lambda_3\lambda_4\lambda_6+3\lambda_4^2\lambda_6+6\lambda_2\lambda_7^2
   -8\lambda_3\lambda_7\lambda_8 \right. \notag \\  &\left.- 4\lambda_4\lambda_7\lambda_8 + 6\lambda_1\lambda_8^2 \right) \,.
\end{align}
By replacing $\lambda_i\rightarrow \lambda_i +\delta \lambda_i$  in
\cref{UNIT}, the NLO effects on the unitarity of the $S$-matrix can be
approximated and checked in the parameter scan. 
The counterterms $\delta \lambda_i$ in the N2HDM are given in
Eq.~(\ref{eq:countertermsn2hdm})
and are calculated numerically during the minimisation procedure. The 
corresponding relations for the C2HDM counterterms and perturbative 
unitarity relations are given in 
\cite{Branco:2011iw,Basler:2018cwe,Fontes:2017zfn}.
By imposing these additional constraints at NLO, it can happen that
the parameter sample is significantly reduced. In \cref{tab:EWPTPheno:Reduction} a list
of the remaining parameter sample is given after imposing the NLO
constraints and the requirement for an SFOEWPT. These numbers have to
be taken with a grain of salt, 
however, as dedicated scans adapted to specific requirements like {\it
e.g.}~not loosing points due to NLO unitarity, would change this
picture. The numbers are simply meant to show that NLO constraints and/or
the requirement of an SFOEWPT have an effect on the parameter sample. 
\s
\begin{table}
    \centering
    \begin{tabular}{p{0.33\linewidth}ccccc}
    \toprule {\small Applied constraint} & {\small C2HDM~(T1)} & {\small C2HDM~(T2)} & {\small N2HDM~(T1)} & {\small N2HDM~(T2)} \\ \midrule 
    {\small Total number of parameter points} & 233163& 1029538 & 271743 & 302653\\\midrule
    {\small NLO vacuum stability} & $97.32\%$ & $90.36\%$ & $83.64\%$& $87.38\%$\\\midrule
    {\small NLO vacuum stability} + NLO  perturbative unitarity & $91.00\%$ & $79.451\%$ & $80.32\%$&$85.96\%$ \\\midrule
    {\small SFOEWPT + NLO vacuum stability + NLO  perturbative unitarity} & $0.012\%$ & $0.003\%$& $0.379\%$  & $0.022\%$\\
    \bottomrule
    \end{tabular}
    \caption{Reduction of the number of parameter points before and
      after applying NLO vacuum stability, NLO perturbative unitarity
      and an SFOEWPT.} 
    \label{tab:EWPTPheno:Reduction}
    \end{table}

The Higgs spectra of the C2HDM and N2HDM consist of at least three
neutral Higgs bosons (in the N2HDM, we additionally have the
pseudoscalar). Denoting the lighter and heavier non-SM-like Higgs
bosons with masses $m_{H_\downarrow}$ and $m_{H_\uparrow}$ by $H_\downarrow$
and $H_\uparrow$, we can have three different mass configurations that
we will refer to in the following as \textit{heavy mass hierarchy} with
the mass hierarchy
\begin{equation}
\quad m_h < \mdown < \mup\,,
\end{equation}
\textit{semi-inverted mass hierarchy} with
\begin{equation}
 \quad \mdown < m_h < \mup\,,
\end{equation}
and \textit{inverted mass hierarchy} with the hierarchy
\begin{equation}
  \quad \mdown<\mup < m_h\,.
\end{equation}

\section{Results \label{sec:results}}
In the following analysis we investigate to which extent the viable
parameter spaces of the C2HDM and N2HDM are constrained by the
additional requirement of a strong first order EWPT. This also allows
us to investigate the differences that arise due to CP violation on
the one hand and singlet admixture on the other hand. Since we
discussed the implications for the phenomenology of the C2HDM already
in detail in Ref.~\cite{C2HDM}, we start by providing a rather short
update of our analysis of the C2HDM by taking into account the
newest results for the Higgs data. Subsequently, we discuss in detail
the interplay between a strong first order phase transition and the
collider phenomenology of the N2HDM. In particular, we study the
impact of $\xic \ge 1$ on the size of the trilinear Higgs
self-couplings and the overall mass spectrum. We will provide
benchmark scenarios that highlight the connection between a successful
SFOEWPT and collider phenomenology with emphasis on Higgs pair production
at the LHC. In the end a comparison between the two models will be
drawn and the characteristic differences in their
phenomenology with respect to a successful SFOEWPT will be discussed.
For our analysis we produced for each model of each type about half a
million parameter points that respect the theoretical and
experimental constraints listed in \cref{Sec:ScanConstraints}. These
parameter points are checked for a successful SFOEWPT by using the
program {\tt BSMPT v1.1.2} \cite{Basler:2018cwe}. \s

\subsection{C2HDM -- Update \label{subsec:c2hdm}}
We start with an update of our analysis in Ref.~\cite{C2HDM} by taking into
account the new collider constraints that have been implemented in
{\tt HiggsSignals 2.3.0} and {\tt HiggsBounds 5.5.0}. In addition, we
increased the scan ranges of all scalar masses from $1\tev$ up to
$1.5\tev$ with the aim to find new valid parameter points featuring a
heavy scalar spectrum that provides an SFOEWPT. The subsequent
discussion will show that no additional heavy parameter points were
found fulfilling the requirement of an SFOEWPT and compatibility with
recent collider constraints. \s

\begin{figure}[t]
  \centering
  \subfigure[\label{C2HDM::mcharged_a}]{\includegraphics[width=0.5\textwidth]{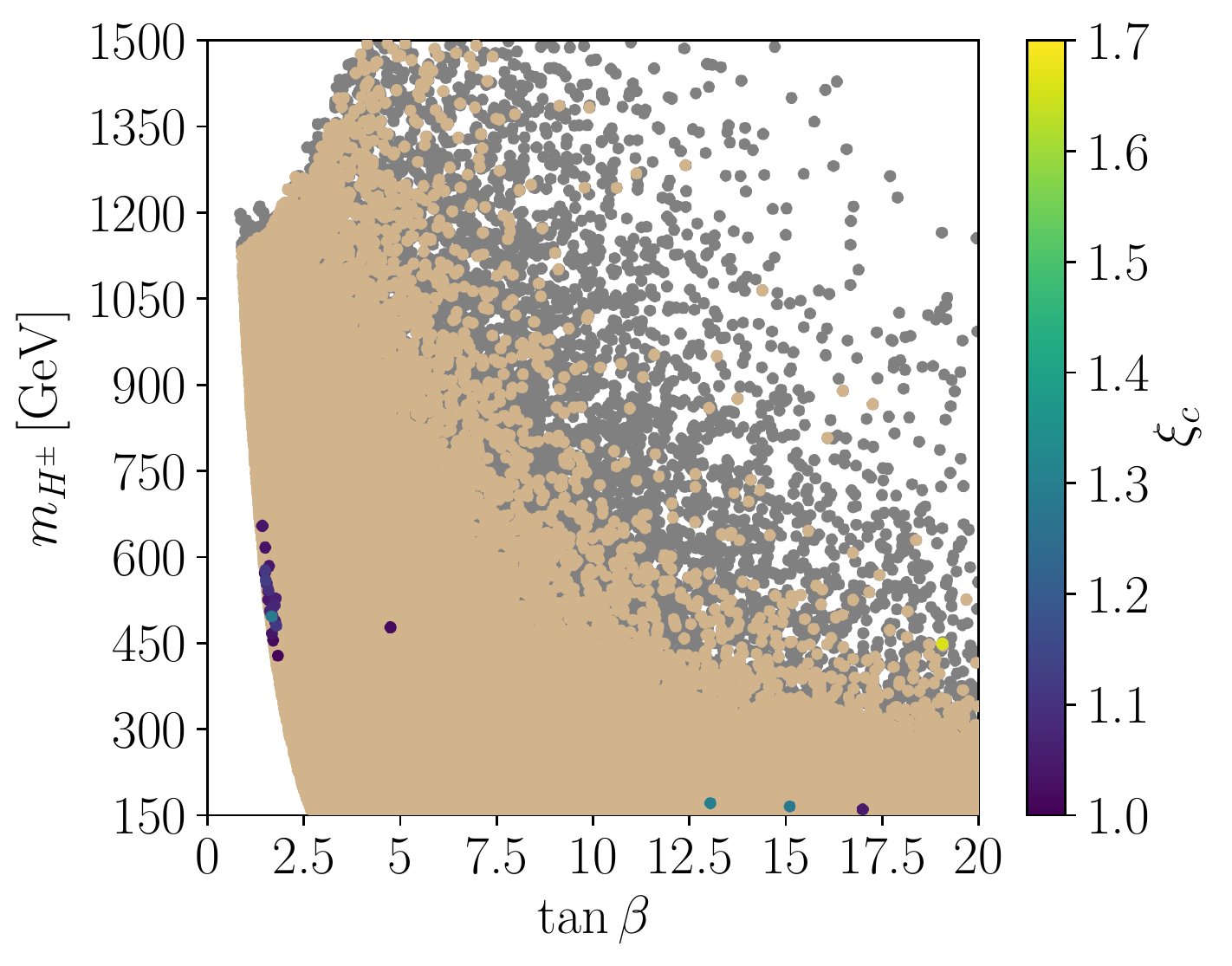}}%
  \subfigure[\label{C2HDM::mcharged_b}]{\includegraphics[width=0.5\textwidth]{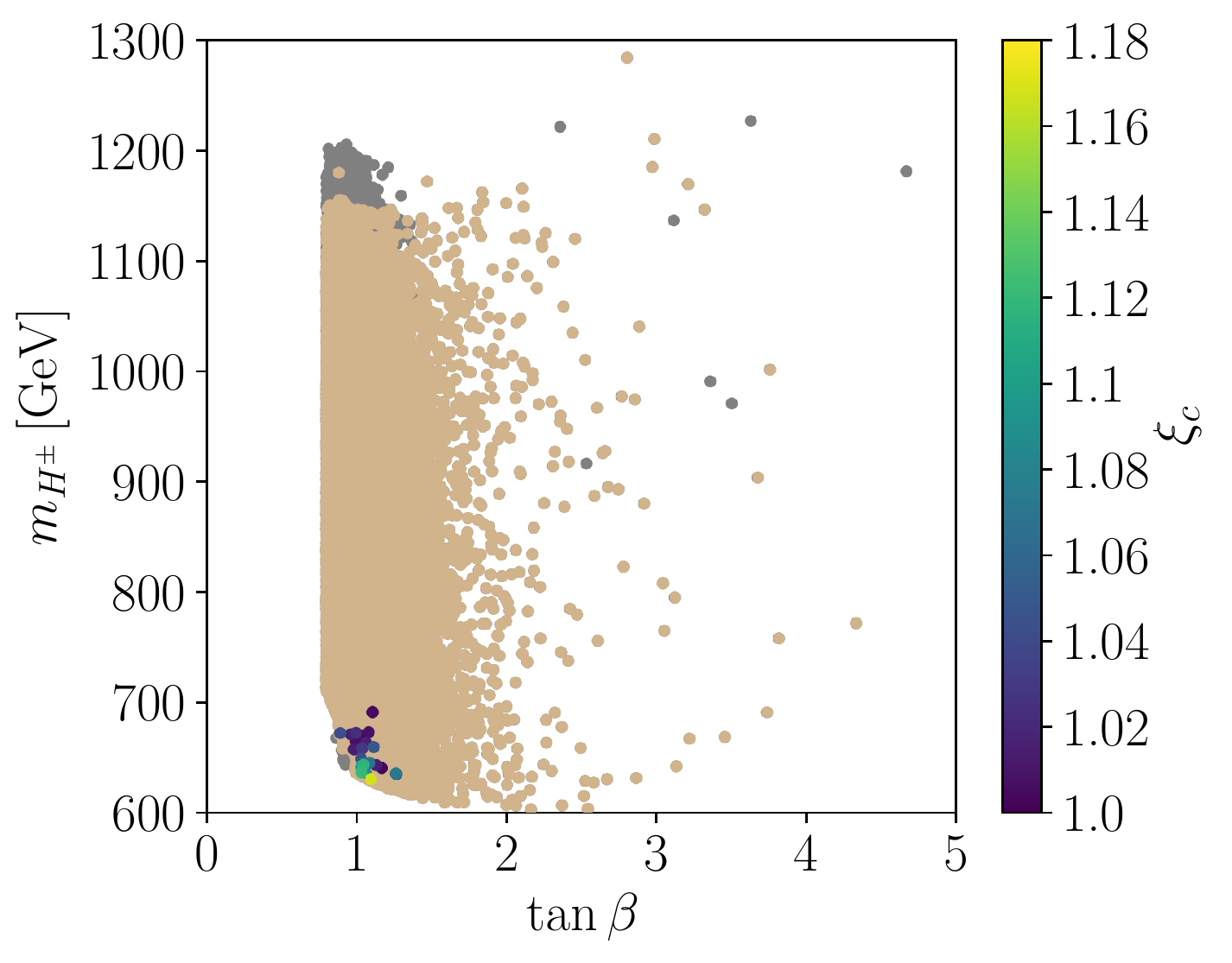}} %
  \caption{C2HDM: The mass $\mHc$ versus $\tan\beta$ for T1 (left) and
  T2 (right). The grey points are all parameter points passing the
  experimental and theoretical checks of {\tt ScannerS}. The brown
  points additionally fulfill NLO vacuum stability and NLO perturbative unitarity
  constraints. The color code
  denotes the strength of the phase transition $\xic$ for $\xic \ge
  1$. The di-Higgs search constraints are included.}
  \label{C2HDM::mcharged}
\end{figure}

We start the discussion with \cref{C2HDM::mcharged}, where the charged
Higgs boson mass $\mHc$ is shown as a function of $\tan\beta$ for the
C2HDM T1 (left) and T2 (right). The grey points are all parameter
points compatible with the theoretical and recent experimental
constraints as described in Sec.~\ref{Sec:ScanConstraints}.
The brown points additionally provide an NLO stable vacuum and
  fulfill NLO perturbative unitarity (see also Sec.~\ref{Sec:ScanConstraints}).
The color code indicates parameter points with values of $\xi_c \ge 1$
and thereby all points with an SFOEWPT. 
In the C2HDM T1 two distinct possible scenarios for 
parameter points providing an SFOEWPT can be observed. The first
region has charged Higgs boson masses of $\sim 450\gev$ up to $\sim
\unit[690]{GeV}$ and quite small $\tan\beta$ values around 1. Only one
point provides an SFOEWPT with a medium charged Higgs boson mass and a
$\tan\beta$ value around 5. All the other points of the second region
with larger values of $\tan\beta$ have a charged Higgs boson mass
below $200\gev$. Compared to our previous analysis \cite{C2HDM},
parameter points with medium charged Higgs boson masses and large
$\tan\beta$ values could not be found any more, so that we have this
strict separation of small masses in combination with large
$\tan\beta$ values and medium masses in combination with small
$\tan\beta$. The maximum strength of the phase transition that we found for the
C2HDM T1 is $\xi_{c} \approx 1.7$, and for the C2HDM T2 it
is $\xi_{c} \approx 1.18$, which is compatible with our
findings in \cite{C2HDM}.\s

In the C2HDM T2, the flavor constraints $B\rightarrow X_s \gamma$
\cite{Deschamps:2009rh,flavor1,flavor2,flavor3,charged580} require the
charged Higgs boson mass to 
be above $580\gev$, which is reflected in Fig.~\ref{C2HDM::mcharged}
(right). The figure shows that most parameter points compatible with
theoretical and experimental constraints have a rather small
$\tan\beta$ of the order of $\mathcal{O}(\tan\beta)\approx 1-4$ and
charged Higgs boson masses up to $\sim 1.1 \tev$. The requirement an
SFOEWPT like in the C2HDM T1 sets an upper bound on the charged Higgs
boson mass which is $\sim 700\gev$. All valid parameter points that
have an SFOEWPT gather in the lower left corner of the plot, with
small $\tan\beta$ values and as light as possible charged Higgs boson masses. Future
updates in the flavor sector that constrain this specific corner of
the parameter space might rule out the C2HDM T2 in combination with an
SFOEWPT. \s

\begin{figure}[t]
  \centering
    \subfigure[\label{C2HDM::massspectrum_a}]{\includegraphics[width=0.5\textwidth]{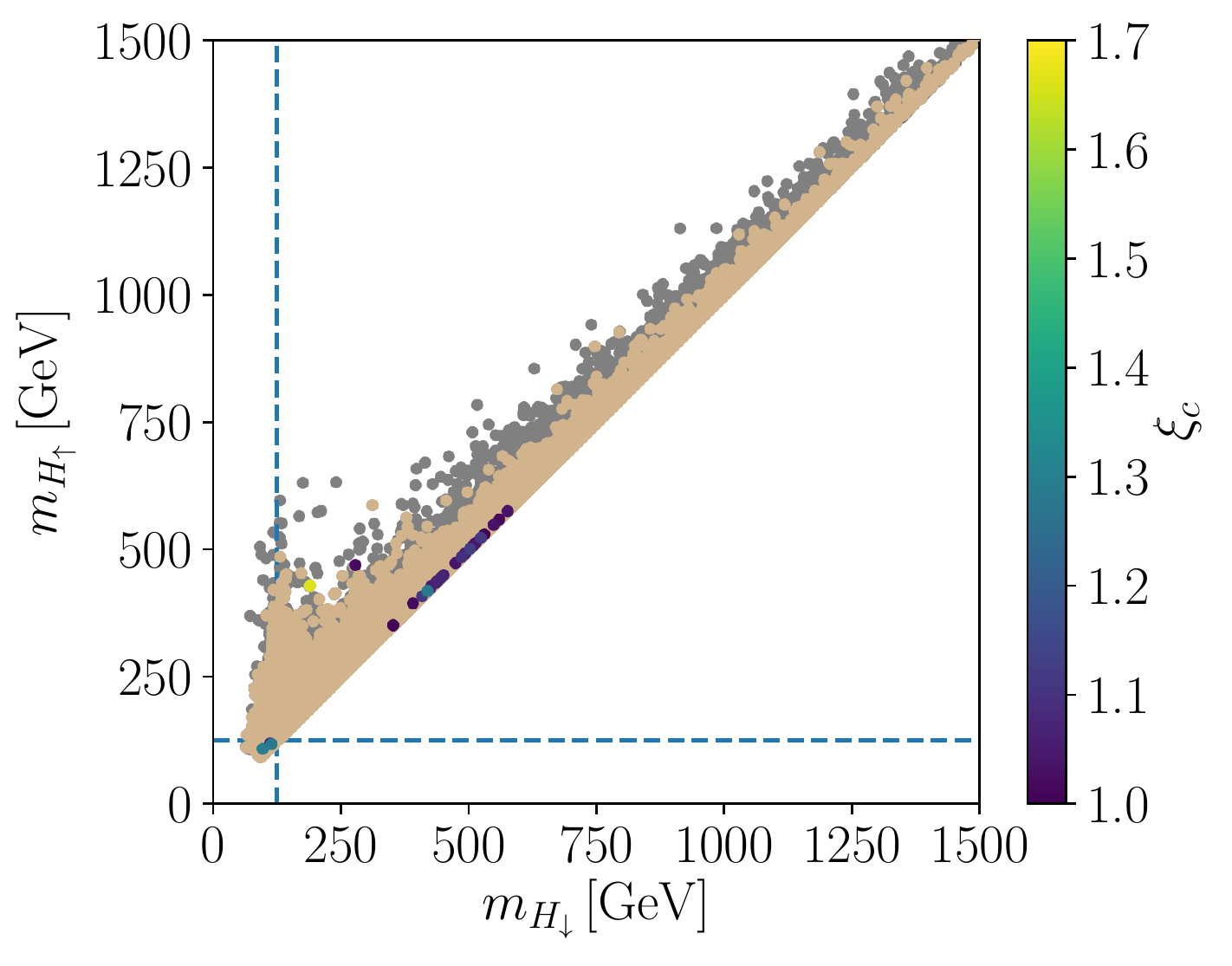}}%
    \subfigure[\label{C2HDM::massspectrum_b}]{\includegraphics[width=0.5\textwidth]{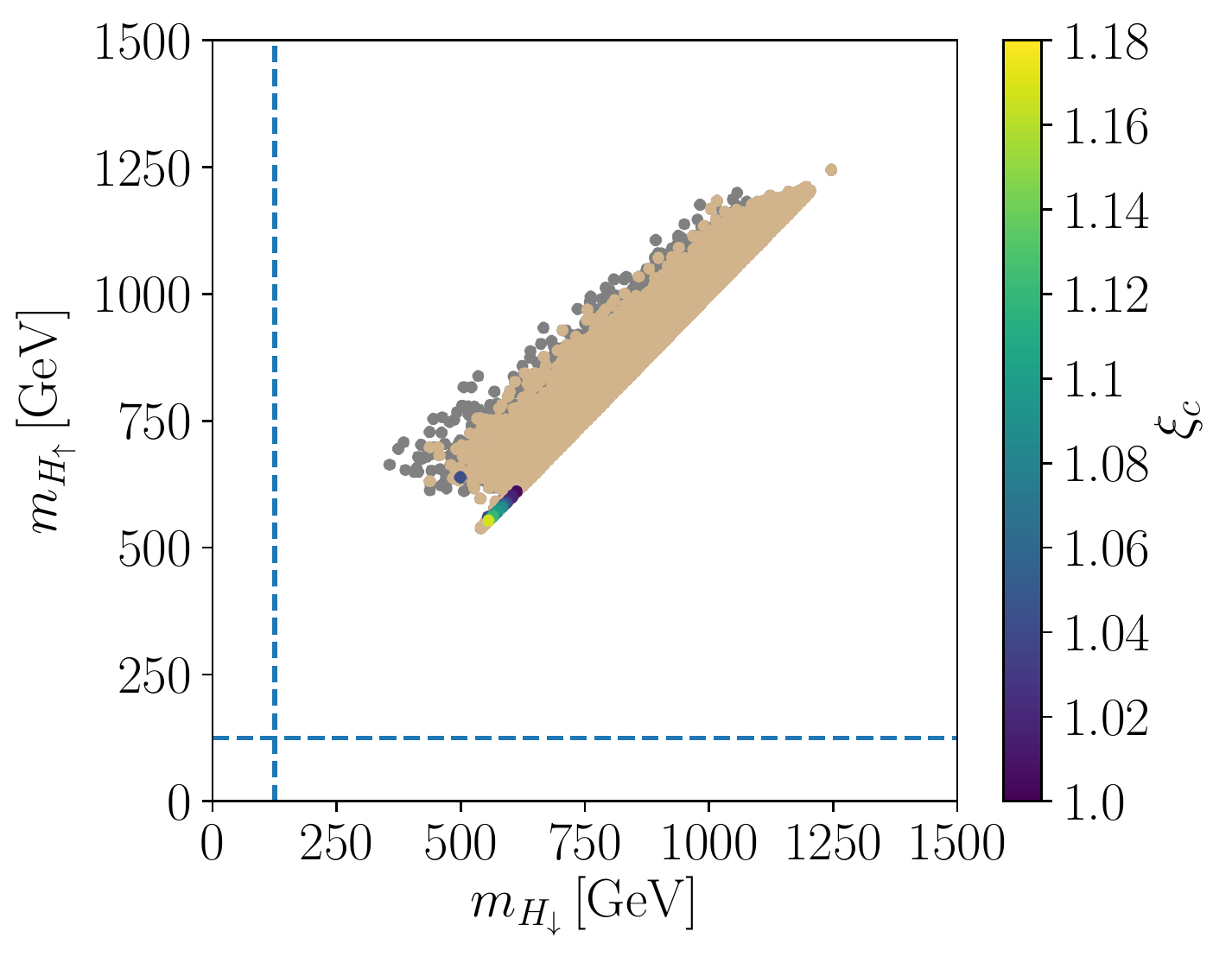}} %
  \caption{C2HDM: The mass $\mdown$ versus $\mup$ for T1 (left) and T2
    (right). The grey points are all parameter points passing the
  experimental and theoretical checks of {\tt ScannerS}. The color code
  denotes the strength of the phase transition $\xic$ for $\xic \ge
  1$. The di-Higgs search constraints are included. The blue dashed
  lines indicate the SM Higgs mass value of $m_h 
  \approx 125$~GeV.}
  \label{C2HDM::massspectrum}
\end{figure}

Compatibility with the electroweak precision observables enforces the
degeneracy of two Higgs boson masses, so either $\mHc \approx
m_{\hdown,\hup,h}$ or a pair of the neutral Higgs bosons are mass
degenerate. In order to quantify this effect, we first of all look for
that pair of Higgs masses $(m_i,m_j)$ that has the minimum mass
difference out of all possible Higgs mass pairings, {\it i.e.}~we
define the mass gap of the (almost) degenerate pair $(m_{H_i},m_{H_j})$ as
\begin{align}
\Delta_{m_{H_i},m_{H_j}} =& \min\limits_{m_{H_i} \neq
m_{H_j}} \left\vert m_{H_i} -  m_{H_j} \right\vert \;, \qquad
                            \mbox{with }
H_{i,j} \in\lbrace h, H_\downarrow, H_\uparrow, H^\pm \rbrace \,. 
\label{eq:massgap}
\end{align}
The requirement of an SFOEWPT tightens this mass gap to even smaller
values 
\begin{align}
  \text{Type I}: &\max\limits_{\mathrm{sample}} \Delta_{m_{H_i},m_{H_j}} \approx 61\gev \xrightarrow{\text{EWPT}} \,\approx 21\gev\,, \\
  \text{Type II}: &\max\limits_{\mathrm{sample}} \Delta_{m_{H_i},m_{H_j}} \approx 62\gev \xrightarrow{\text{EWPT}} \,\approx 33\gev\,.
\end{align}
In both types of the C2HDM the mass degeneracy that is realized in
most of the points is the one of $H_\downarrow$ and $H_\uparrow$, as
shown in \cref{C2HDM::massspectrum}. It shows the non-SM-like Higgs
boson mass plane $m_{H_\uparrow}-m_{H_\downarrow}$ with the color
code denoting the values of $\xi_c$ above 1, hence an SFOEWPT. In the
other cases the mass degeneracy occurs with the charged Higgs boson,
{\it i.e.} $\mHc\approx m_{H_\downarrow}$ or $\mHc\approx
m_{H_\uparrow}$. In the C2HDM T1 in \cref{C2HDM::massspectrum_a} two
mass hierarchies are possible, the heavy mass hierarchy
$m_h<m_{H_\downarrow}<\mup$ and the fully inverted mass hierarchy
$\mdown<\mup<m_h$. As we will see later, in the N2HDM
the semi-inverted and the heavy mass hierarchies are possible in T1 and T2. \s

As noted in \cite{Basler:2016obg} the SFOEWPT favors a light scalar
spectrum, where additional heavy degrees of freedom, that do not
obtain an $\mathrm{SU}(2)_L$ VEV, can help to strengthen the phase
transition. At the same time, the mass degeneracy of $\hup$ and
$\hdown$ enforces the overall scalar spectrum to be in a medium range
of $\mathcal{O}(\mdown,\mup)\sim 500\gev$. The
parameter region with $m_h<\mdown<\mup$ with a heavy $\hup$ (
$\mdown\ll\mup$) could also produce an SFOEWPT as remarked in
\cite{Basler:2016obg}, since this region has a light $\hdown$ enabling
the SFOEWPT and in the meantime the heavy degree of freedom $\hup$
could build up a deep potential barrier between the symmetric and
broken minimum. Since this parameter region is restricted by the
di-Higgs searches through Higgs decays $\hup\rightarrow
\hdown\hdown/hh$, the parameter space in this region is already sparse
due to the collider constraints. Consequently, with the updated {\tt
  HiggsSignals 2.3.0} and {\tt HiggsBound 5.5.0} versions taking into
account the recent di-Higgs searches, it is more involved to find
parameter points compatible with the collider constraints and an
SFOEWPT. As we will see later, this restriction can be circumvented in
the N2HDM due to the singlet admixture. \s 

In the C2HDM T2 the overall scalar spectrum in
\cref{C2HDM::massspectrum_b} is heavier compared to T1. This is due to
the already required heavy charged Higgs boson mass and the small mass
gaps. As the SFOEWPT still favors a light scalar spectrum, the only
parameter points providing an SFOEWPT are found in the edge of
smallest masses. The overall order of neutral non-SM like Higgs masses
providing an SFOEWPT is also $\mathcal{O}(m_{h_i})\sim 500\gev$ as in the 
C2HDM T1. \s

To conclude the C2HDM update, we find that the mass spectrum
compatible with the recent collider and theoretical constraints, for
T1 is mainly constrained by the recent di-Higgs measurements whereas
for T2 the flavor constraints are the most restrictive ones. Future analyses
with increasing constraining power in these mass regions might exclude
significant regions of the C2HDM parameter space providing an SFOEWPT.  

\subsection{N2HDM - Phenomenology of the SFOEWPT \label{subsec:n2hdmpheno}}
In the following we discuss the implications of an SFOEWPT on the
phenomenology of the N2HDM. The N2HDM  has one more degree of freedom
compared to the C2HDM due to the additional singlet in the Higgs
sector. The larger number of free parameters in this model reduces the 
influence of the constraints on the parameter space so that heavier
Higgs spectra compatible with an SFOEWPT are still possible. In this
kind of scenarios, however, one of the non-SM like Higgs bosons is almost
completely singlet-like. The maximum values of $\xi_c$ that we find
are $\xi_c^{\text{N2HDM T1}} = 2.04$ and $\xi_c^{\text{N2HDM,T2}}=1.43$
for the N2HDM T1 and T2, respectively. We start with the investigation of the mass
spectrum followed by the discussion of the trilinear Higgs
self-couplings. Afterwards, we will present several benchmark points
providing interesting scenarios with different features.

\subsubsection{Mass Spectrum of the N2HDM T1}
\begin{figure}
  \centering
    \subfigure[\label{N2HDM::massspectrum_a}]{\includegraphics[width=0.5\textwidth]{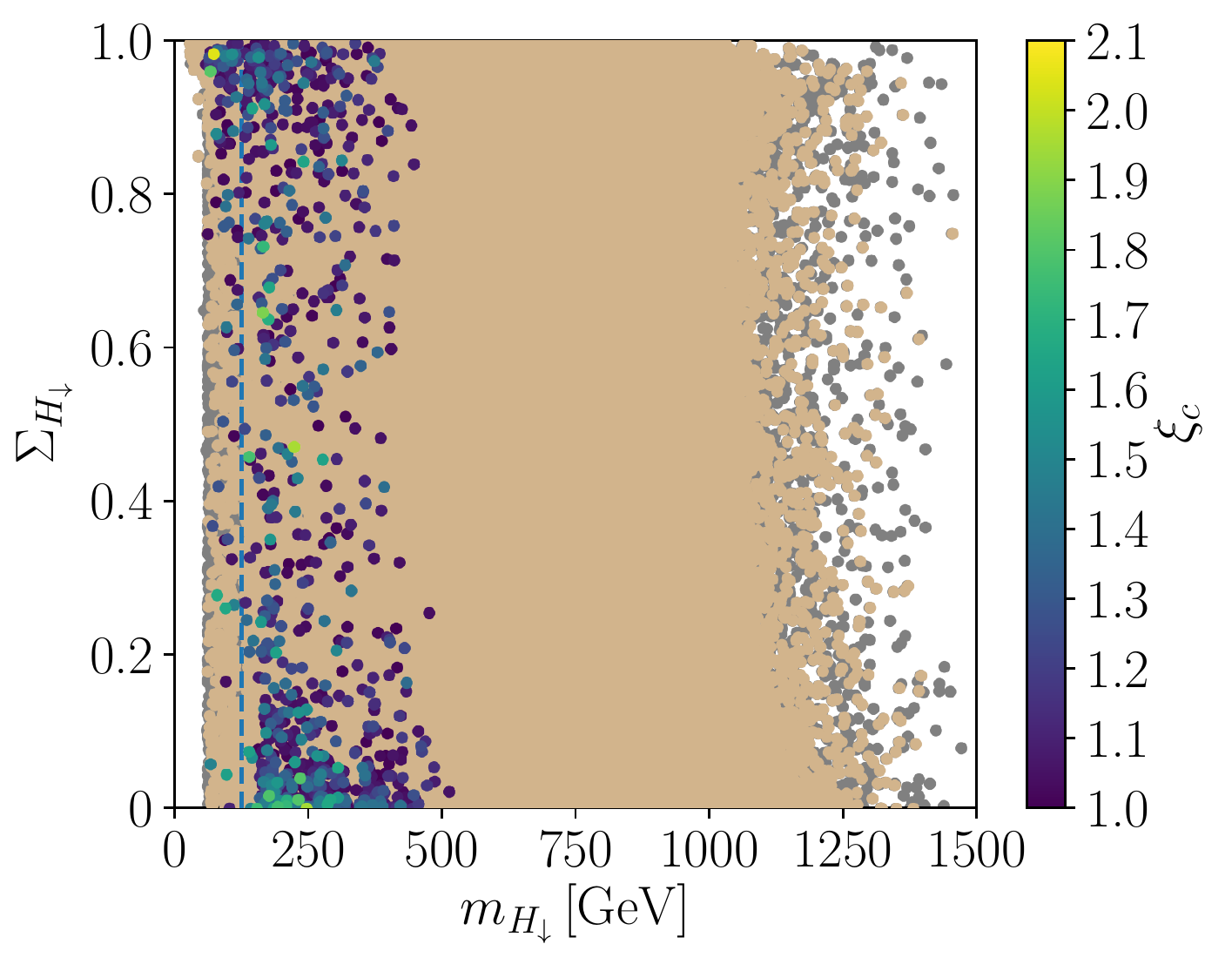}}%
    \subfigure[\label{N2HDM::massspectrum_b}]{\includegraphics[width=0.5\textwidth]{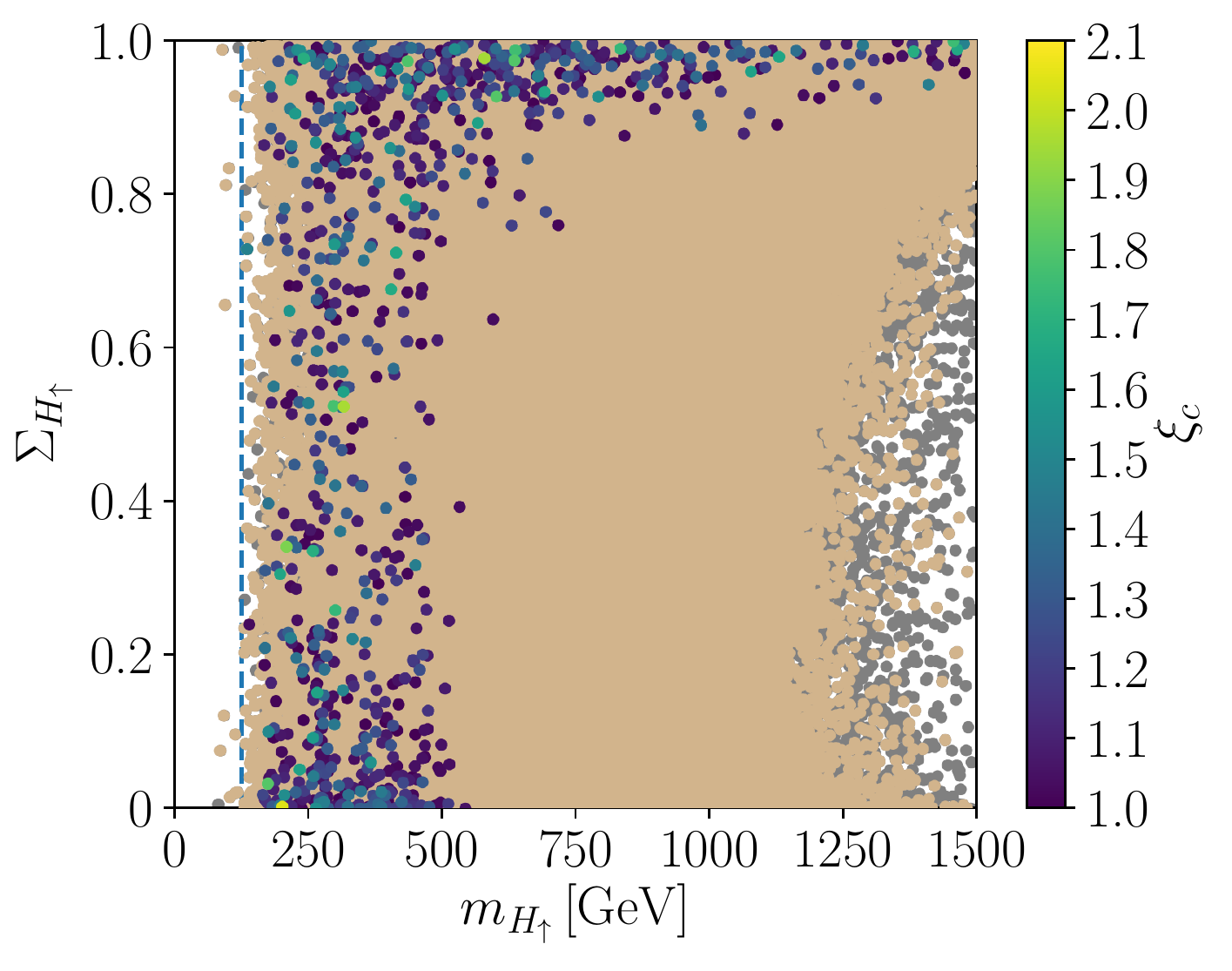}} %
  \caption{N2HDM T1: The singlet admixture $\Sigma$  versus the mass
    for $H_\downarrow$ (left) and $H_\uparrow$ (right). The grey
    points are all parameter points passing the experimental and
    theoretical checks of {\tt ScannerS}. The brown points
    additionally fulfill NLO vacuum stability and NLO perturbative
    unitarity constraints. The color code shows the strength of the
    phase transition $\xi_c$ for $\xi_c \ge 1$. The di-Higgs search constraints are
    included. The blue dashed line indicates the SM Higgs mass value of $m_h 
  \approx 125$~GeV.}
  \label{N2HDM::massspectrum}
\end{figure}

For the following discussion of the N2HDM we introduce the
singlet admixture $\Sigma_{H_i} $ of the respective
CP-even Higgs boson $H_i$ as 
\begin{equation}
	\Sigma_{H_i} = R_{i3}^2 \;.
	\label{SingletAdmixture}
\end{equation}
It describes the amount of admixture of the singlet field $\rho_S$
to the corresponding mass eigenstate $H_i$. 
In \cref{N2HDM::massspectrum} left (right) we plot the singlet admixture
$\Sigma_{H_\downarrow}$ ($\Sigma_{H_\uparrow}$) of $H_\downarrow$
($H_\uparrow$) versus its mass. The grey points denote parameters
points compatible with the theoretical and experimental constraints,
the brown points additionally provide an NLO stable vacuum and 
NLO perturbative unitarity. The color code indicates the strength of
the phase transition for $\xi_c \ge 1$. The masses that provide an SFOEWPT are
\begin{align}
  \mdown \in \left[53\,, 513\right]\gev\,, \qquad &
  \mup\in\left[136\,,1479\right]\gev\,. 
\end{align}
As these mass windows show, in the N2HDM T1 the heavy mass hierarchy ($m_h <
  \mdown <\mup$) and the semi-inverted mass hierarchy
  ($\mdown<m_h<\mup$) are possible, whereas the inverted hierarchy
  ($\mdown<\mup<m_h$) is not realised. We will provide benchmark 
scenarios for all possible cases. Like in the C2HDM, an SFOEWPT favors
light Higgs mass spectra below $\sim 500\gev$ except for spectra with
\textit{singlet-like} $\hup$. If the heaviest CP-even non-SM like
Higgs boson $\hup$ has a singlet admixture above
$\Sigma_{H_\uparrow}\gtrsim 80\%$, the SFOEWPT opens the window for
larger masses $\mup$. In case of singlet admixtures below about 80\%,
$\Sigma_{H_\uparrow} \lesssim 80\%$, on the other
hand the same preference for 
intermediate Higgs mass spectra as in the C2HDM can be observed for
$\xi_c \ge 1$. The possibility of large neutral Higgs boson masses
allows for heavy Higgs decays into pairs of lighter Higgs bosons,
$\hup\rightarrow H_i H_j$. Since almost all of these heavy states are
singlet-like, their couplings to SM particles are suppressed and the
Higgs-to-Higgs decay channel becomes an important discovery channel. 
We will provide benchmark scenarios where these channels may become
accessible at the LHC. \s
\begin{figure}
  \centering
    \subfigure[\label{N2HDM::massgap}]{\includegraphics[width=0.47\textwidth]{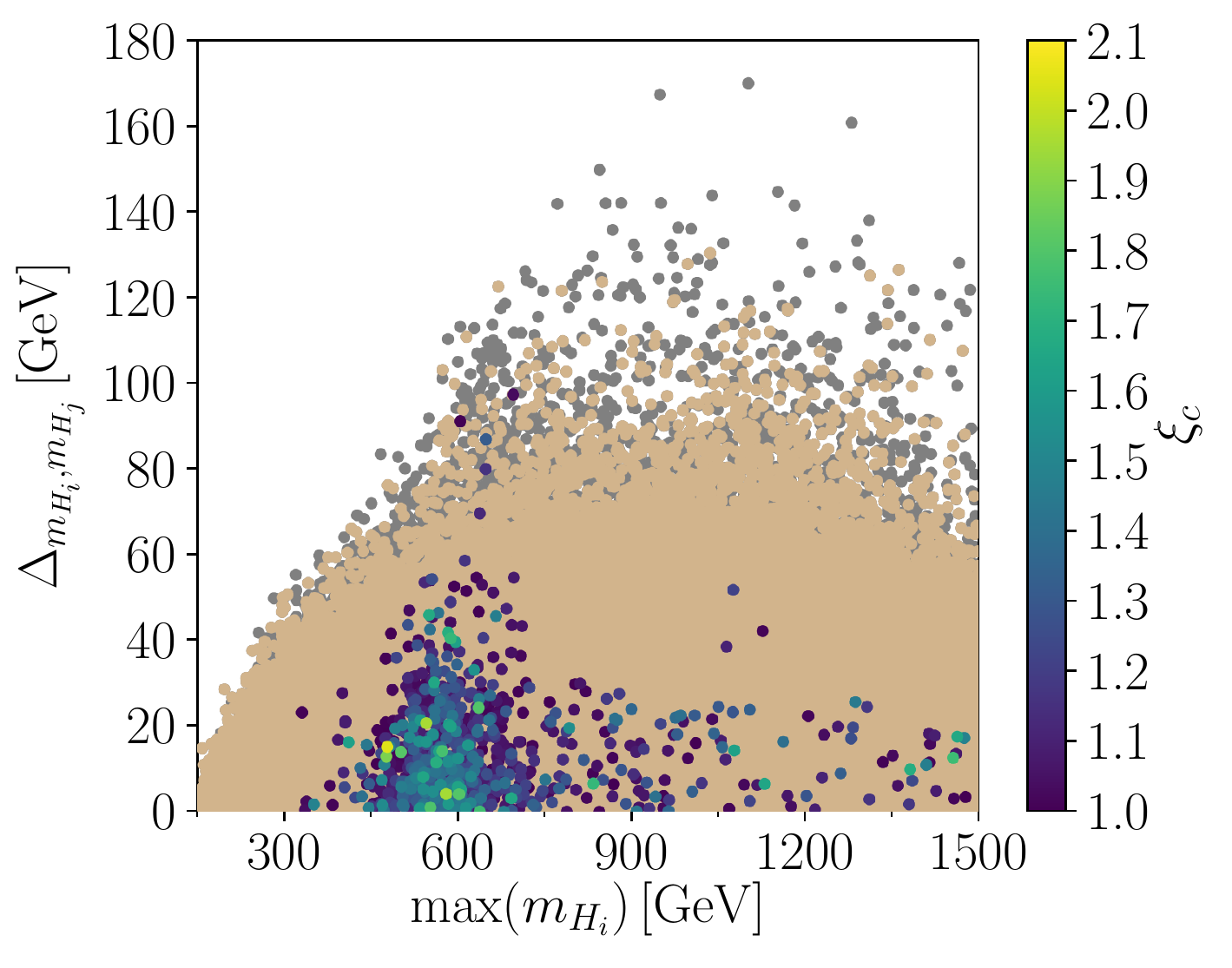}} %
    \subfigure[\label{N2HDM::MHC}]{\includegraphics[width=0.47\textwidth]{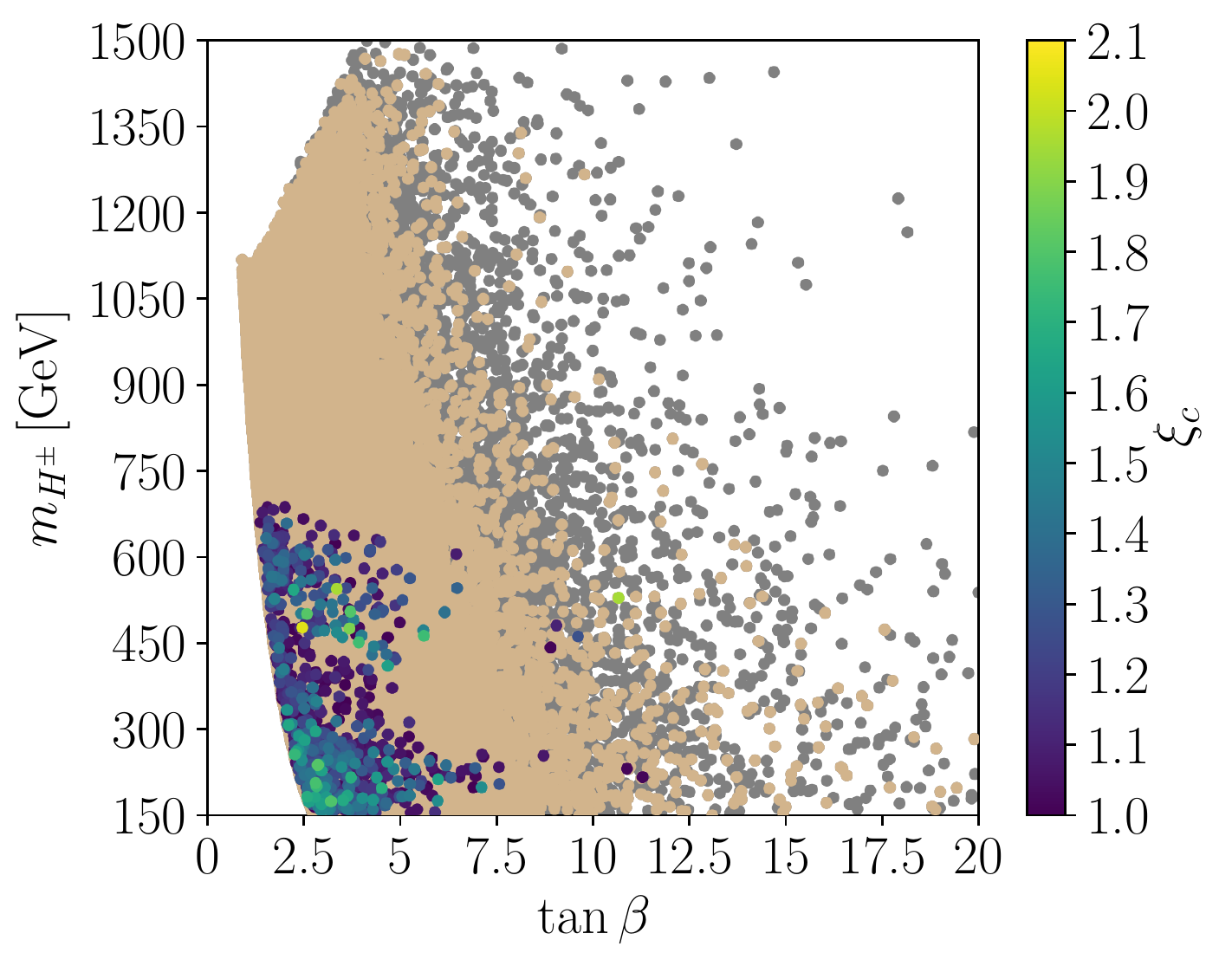}}%
  \caption{N2HDM T1: Left: The minimum mass gap
    $\Delta_{m_{H_i},m_{H_j}}$ versus the maximum of the scalar
    masses. Right: Charged Higgs boson mass $m_{H^\pm}$ versus
    $\tan\beta$. The color code is the same as in
    Fig.~\ref{N2HDM::massspectrum}. The di-Higgs search constraints are included.}
\end{figure}

As already mentioned, compatibility with the EW precision data checked
through the $S,T$ and $U$ parameters requires a small mass gap between
the charged Higgs boson $H^{\pm}$ and one of the neutral Higgs bosons
or between two neutral Higgs bosons, 
and the SFOEWPT enforces even more the mass degeneracy between at
least one non-SM like Higgs boson pair. In \cref{N2HDM::massgap} we
show the minimum mass gap out of all possible neutral Higgs
pairings, $\Delta_{m_{H_i},m_{H_j}}$, defined analogously to Eq.~(\ref{eq:massgap}),
\begin{equation}
\Delta_{m_{H_i},m_{H_j}} = \min\nolimits_{i\neq j} \vert m_{H_i} - m_{H_j} \vert\, \quad
  \mbox{with } H_{i,j}\in\lbrace
  h,\hdown,\hup,H^{\pm},A\rbrace\,, 
\end{equation}
versus the maximum mass in the spectrum, max($m_{H_i}$). The color
code is the same as in 
Fig.~\ref{N2HDM::massspectrum}. The experimental and theoretical
constraints allow for mass gaps even above $150\gev$ and the NLO
stable vacuum and NLO perturbative unitarity are compatible with mass gaps up
to about 130~GeV, while the SFOEWPT reduces this
upper bound down to $\mathcal{O}(50\gev)$ with a few
  exceptions of up to 100~GeV. Scenarios with degenerate neutral Higgs
  boson masses are rather rare so that in
  general the charged Higgs boson mass lies in the same region as the
neutral Higgs boson masses. Consequently, the mass spectrum of the
charged Higgs boson is also reduced in case of an SFOEWPT. This is
reflected in \cref{N2HDM::MHC} where the charged mass $\mHc$ is
depicted versus $\tan\beta$, with the same color coding as in
\cref{N2HDM::massgap}. \s 

\begin{figure}
  \centering
  \includegraphics[width=0.7\textwidth]{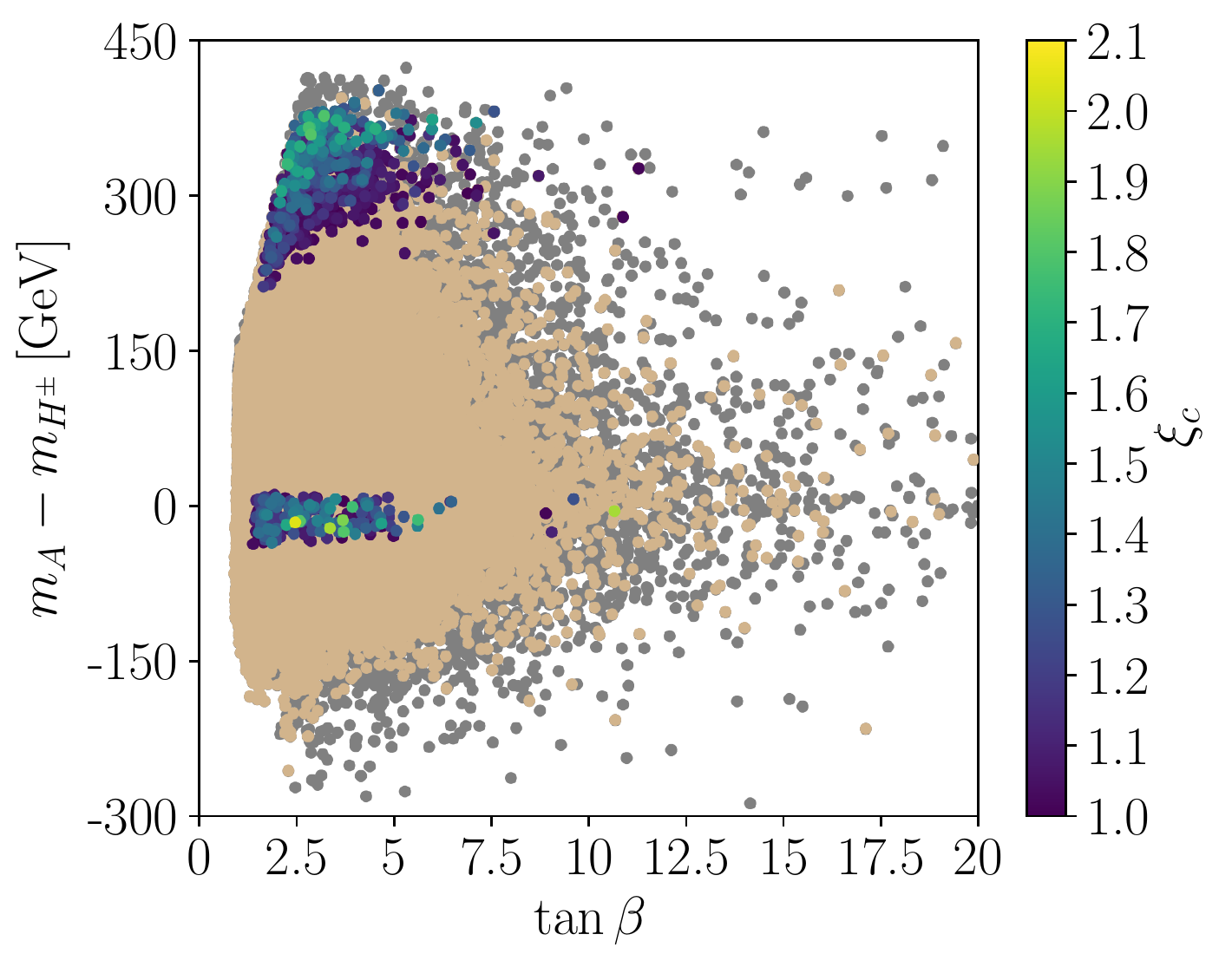}
  \caption{N2HDM T1: The mass difference $m_A-\mHc$ versus
    $\tan\beta$. The color code is the same as in
    Fig.~\ref{N2HDM::massspectrum}. The di-Higgs search constraints are included.}
  \label{N2HDM::T1::Masssplit}
\end{figure}

The last phenomenological effect on the N2HDM mass spectrum induced by
the SFOEWPT we want to discuss, is shown in
\cref{N2HDM::T1::Masssplit}. It displays the mass difference $m_A-\mHc$
versus $\tan\beta$. As can be inferred from the figure, the SFOEWPT
cuts out two distinct regions in the viable parameter space, $\mdeg$
and $\msep$, with 
\begin{equation}
  \mdeg: m_A \approx \mHc\quad  \text{and } \quad \msep: m_A
  \ge \mHc + 220\mbox{ GeV}\,.
\end{equation}
This plots shows that an SFOEWPT does not only allow
  for scenarios with $m_A\approx \mHc$, but also mass spectra with
  large gaps between $m_A$ and $\mHc$ are allowed. This possibility
  should also be taken into account when the EWPT in investigated.
For parameter points in $\mdeg$,
stronger EWPTs with $\xi_c$ up to $\xic\lesssim 2.1$ can be observed
compared to the points in $\msep$ with $\xic\lesssim 1.93$. To discuss
the slight tendency of a stronger EWPT in ${\cal M}_{\text{sep}}$ for
increasing mass gaps $m_A - \mHc$, we first note the observations made
in Refs.~\cite{C2HDM,MSSM2}: The 
strength of the phase transition $\xic$ increases with
  the size of the couplings of the light bosonic particles to the
  SM-like Higgs boson and decreases with the Higgs boson mass.
Additionally, particles that contribute to the EWPT necessarily have
a non-vanishing electroweak VEV. All non-SM-like neutral Higgs bosons
$\hdown, \hup$ receive an electroweak VEV through mixing and
therefore, for an SFOEWPT, their masses $\mdown,\mup$ have to be
either light or their VEV has to be small. We note here again
explicitly that we take the singlet VEV $v_S$ into account for the
minimisation of the effective potential, but we do not include $v_S$
in the calculation of the electroweak VEV in \cref{NUM::EW_VEV}. So
the EWPT is not directly affected by the singlet VEV, just indirectly
through the minimisation. Thus particles that do not obtain an
electroweak VEV ($A$, $H^{\pm}$) and \textit{singlet-like} CP-even
neutral Higgs bosons ($\Sigma_i \approx 1$) are still allowed to be
heavy without decreasing the strength of the EWPT. The heavy degrees
of freedom even help to strengthen the EWPT by enabling a deeper
potential barrier between the broken and symmetric phase
\cite{Dorsch:2017nza}. For increasing mass gaps 
$m_A-\mHc$ and therefore increasing $m_A$ masses, we have an
additional heavy degree of freedom in the spectrum, whereas the
bosonic degrees of freedom that obtain an electroweak VEV remain
light. Consequently, the strength of the EWPT will increase for the
parameter points in $\msep$ with increasing mass gap. Since there are
additional interplays  
between $\xic$ and the mass spectrum or other effects, the effect on
the size of $\xic$ is not significantly enhanced, however.  

\subsubsection{N2HDM T1 - Trilinear Higgs Self-Couplings}
\label{N2HDM::T1::Tripple}
\begin{figure}
  \centering
  \includegraphics[width=0.7\textwidth]{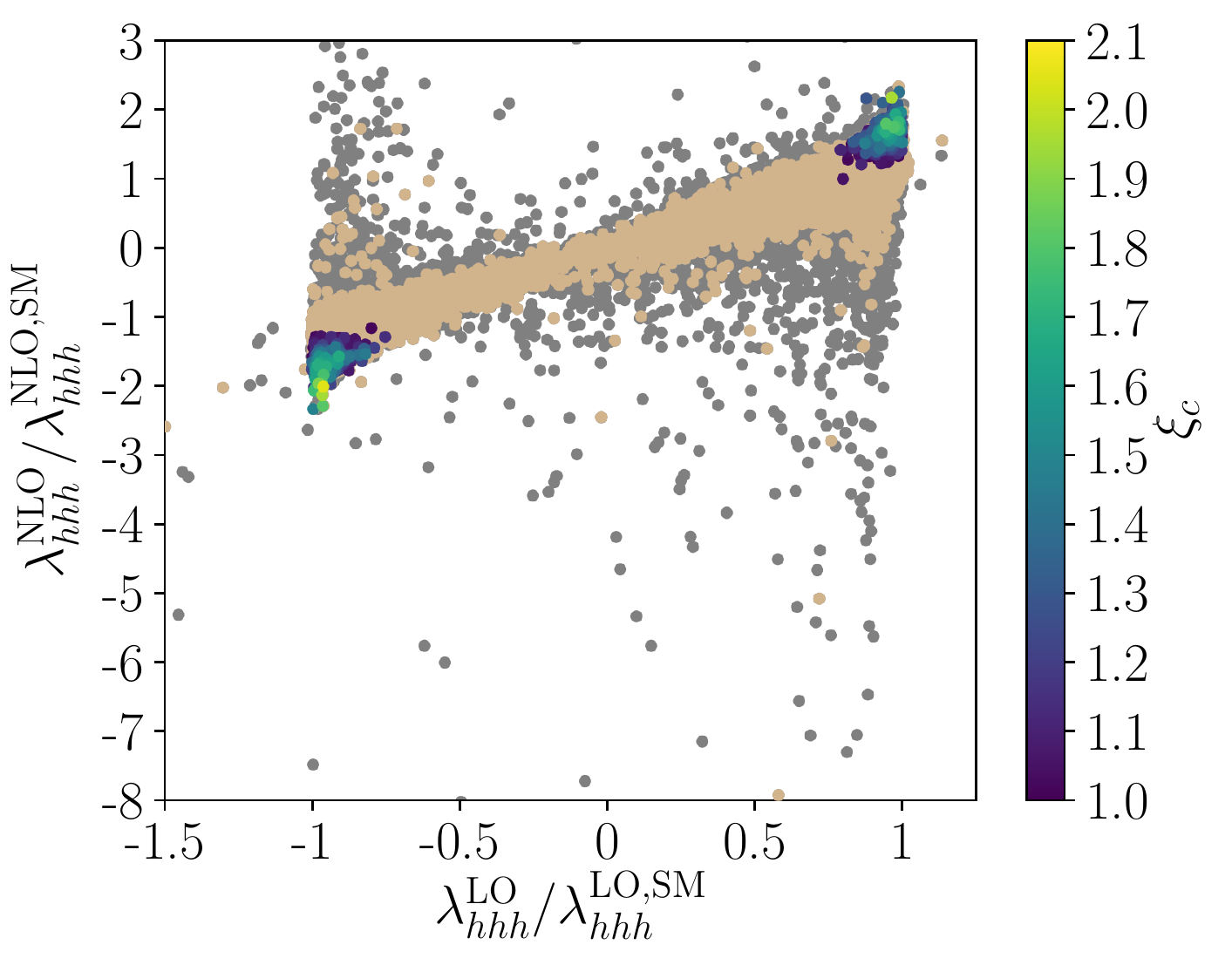}
  \caption{N2HDM T1: The leading-order trilinear self-coupling between
    three SM-like Higgs bosons normalised to the SM reference value
    versus the NLO trilinear self-coupling between three SM-like Higgs
    bosons normalised to the SM reference value. The color code is the same as in
    Fig.~\ref{N2HDM::massspectrum}. The di-Higgs search constraints are
    included.}
  \label{N2HDM::trilinear_full}
\end{figure}
The one-loop-corrected trilinear Higgs self-couplings are obtained from the one-loop
effective potential by performing the third derivative with respect to 
the Higgs fields. The problem of infrared divergences related to the
Goldstone bosons in the Landau gauge is treated analogously to the
extraction of the masses from the second derivative of the potential, {\it
  cf}~Ref.~\cite{MarcoBible} for details. In
\cref{N2HDM::trilinear_full} the next-to-leading order (NLO) trilinear
self-coupling between three SM-like Higgs bosons normalised to the SM
reference is plotted versus the leading-order (LO) coupling. The color
code is the same as in the previous plots. For the SM reference we
take the value of \cite{Kanemura:2002vm} which takes into account the
dominant NLO top-quark contribution. The NLO corrections can both
suppress and enhance the LO values quite significantly. The
experimental and theoretical constraints allow for a largely enhanced
NLO Higgs self-coupling compared to the SM NLO value, between a factor
of -7.9 and 2.4. By requiring an SFOEWPT this upper bound is 
reduced down to a factor $\pm 2.4$. At the same time the SFOEWPT
disfavors trilinear self-couplings below the SM value, hence
\begin{equation}
  \lambda^{\text{NLO}}_{hhh}/\lambda^{\text{NLO,SM}}_{hhh} \in
  \left[-2.4 , -1.2 \right] \cup \left[1.0 , 2.4 \right]\,, \quad
  \mbox{for } \xi_c \ge 1.
\end{equation}

As observed in \cite{C2HDM} the SFOEWPT favors large trilinear Higgs
self-couplings which is also observed here, since the strongest EWPTs
are located at the maximum values of the trilinear Higgs
self-couplings. On the other hand, the upper bound of the trilinear
Higgs self-coupling is significantly reduced by the SFOEWPT which can
be explained by the interplay between the quartic coupling and the
masses of the Higgs bosons participating in the EWPT. Note, that besides the
dominant top-loop contributions to the NLO coupling, also the Higgs-loop contributions
present in the C2HDM  and N2HDM  can be large. The masses of the heavy
N2HDM Higgs bosons $\Phi$ can be cast in the following schematic form (see
\cite{Kanemura:2002vm} for the 2HDM) 
\begin{equation}
  m_{\Phi}^2 = M^2 + f_v (\lambda_i) v^2 + f_m(\lambda_i) v v_s +  f_s
  (\lambda_i) v_s^2 + \mathcal{O}(v^4/M^2,v_s^4/M^2) 
  \label{N2HDM::Mass}
\end{equation}
with $M^2$ denoting the mass scale independent of the VEVs and
$f_{v,m,s} (\lambda_i)$ a linear combination of the quartic couplings
of the Higgs potential. For an SFOEWPT large couplings
$\lambda_i$ are required. On the other hand, the masses should not
become too heavy, which we observed in the
previous discussion of the general mass spectrum, thus limiting the
maximum values for the quartic coupling due to
\eqref{N2HDM::Mass}. This explains why we observe the strongest EWPT
for the largest trilinear coupling, but the maximum enhancement of the
trilinear Higgs self-coupling for a
strong EWPT remains below the value compatible with the applied
constraints. \s

\begin{figure}[t]
  \centering
    \subfigure[\label{N2HDM::couplingadmixture1}]{\includegraphics[width=0.5\textwidth]{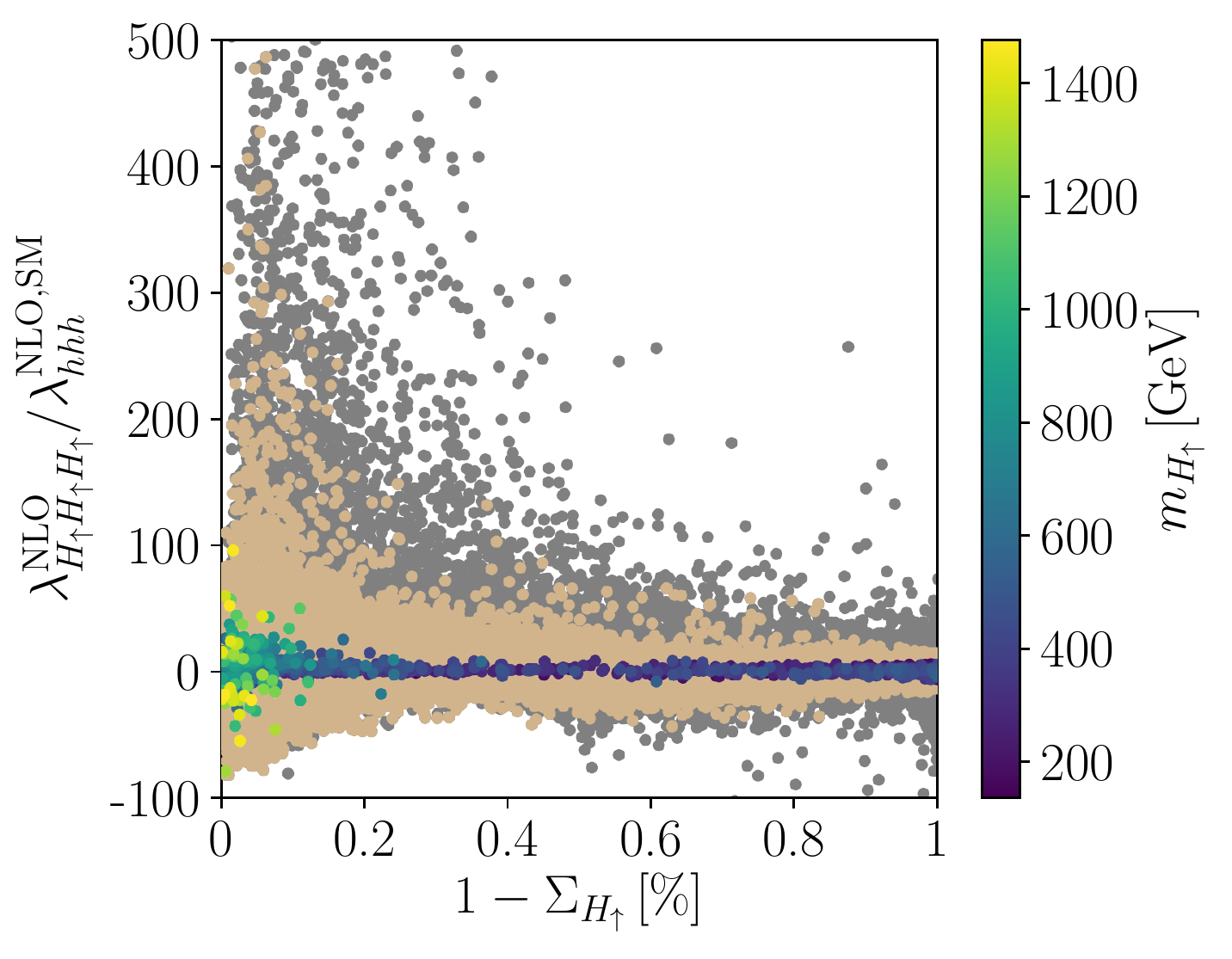}}%
    \subfigure[\label{N2HDM::couplingadmixture2}]{\includegraphics[width=0.5\textwidth]{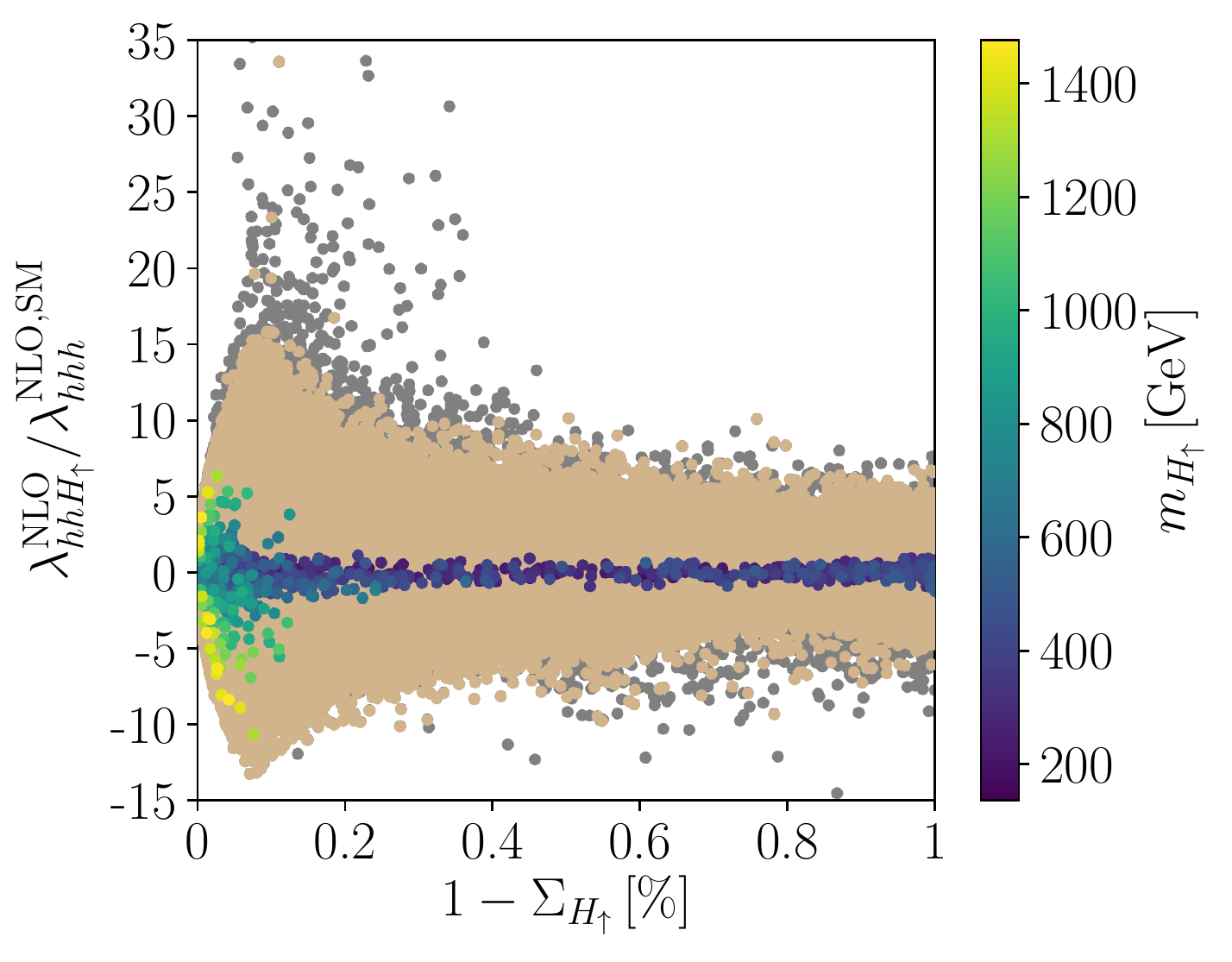}} %
  \caption{N2HDM T1: Left: The NLO trilinear self-coupling between
    three heavy neutral Higgs bosons $\hup$ normalised to the SM
    reference value versus the doublet admixture of the heavy neutral
    Higgs boson $\Sigma_{\hup}$. Right: The NLO trilinear
    self-coupling between two SM-like Higgs bosons and one heavy
    neutral Higgs boson $\hup$ normalised to the NLO SM reference value.
    The color scale show the mass value
    $\mup$ for those points that are compatible with an SFOEWPT.}
  \label{N2HDM::couplingadmixture}
\end{figure} 
Since the N2HDM, in contrast to the C2HDM, has a singlet VEV which
does not contribute to the EWPT, heavy degrees of freedom are
compatible with an SFOEWPT if they are \textit{singlet-like}. In
\cref{N2HDM::couplingadmixture1} we display the NLO trilinear Higgs
self-coupling between three heavy Higgs bosons $H_\uparrow$ normalised
to the NLO trilinear Higgs self-coupling of the SM versus one minus
the singlet admixture $\Sigma_{H_\uparrow}$ of $H_\uparrow$ (which
corresponds to the doublet admixture of $H_\uparrow$). The meaning for the grey
and brown points is the same as in the previous plots. The colored
points now, however, denote the mass value $m_{H_\uparrow}$
for those parameter points that are compatible with an
SFOEWPT. We observe that all singlet admixtures from $0\%$ up to $100\%$ are
possible for intermediate mass ranges, but only for the
\textit{singlet-like} heavy states we observe large masses $\mup$ of
up to $1490\gev$. For these heavy masses the trilinear self-coupling
$\lambda_{H_\uparrow H_\uparrow H_\uparrow}^{\text{NLO}}$ is enhanced
compared to the parameter 
points with intermediate masses. Simultaneously these large masses
enhance the trilinear coupling $\lambda_{hh\hup}$ opening a window for
heavy Higgs decay chains like $\hup\rightarrow h h \rightarrow X X
$. The enhancement can be observed in \cref{N2HDM::couplingadmixture2}
where the NLO trilinear self-coupling $\lambda_{hh\hup}$ normalised to
the NLO trilinear Higgs self-coupling of the SM is plotted against one
minus the singlet admixture $\Sigma_{H_\uparrow}$, with the same color code as
in \cref{N2HDM::couplingadmixture1}. Like for the self-coupling
$\lambda_{\hup\hup\hup}$ the largest enhancements of $\lambda_{hh
  \hup}$ compatible with an SFOEWPT are observed for large singlet
admixtures. These heavy
Higgs decays allow to distinguish between the C2HDM and the N2HDM as
we will discuss later.

\subsubsection{N2HDM T1 - Benchmark Points}
In the following we will present benchmark points that provide an
SFOEWPT and have interesting phenomenological features. In total we
generated 271743 parameter points with {\tt ScannerS}, fulfilling theoretical and
experimental constraints. After applying the NLO constraints, NLO vacuum stability and
NLO unitarity, and demanding an SFOEWPT 920 parameter points are left in the sample.
836 of these points feature the {\it heavy mass hierarchy} while 84 have the
{\it semi-inverted}.
\subsubsection*{Semi-inverted Mass Hierarchy}
\begin{table}[t]
\centering
\begin{tabular}{c c c c c c c c c c }
\toprule
 & $\mdown $    & $\mup$ & $\mHc$& $m_A$ & $\tan\beta$ & $\alpha_1$ & $\alpha_2$ & $\alpha_3$ & $v_S$  \\\midrule
BPSep   &101.22  &270.97& 230.89 & 558.03 & 2.462 & -0.721  & 1.062  & 0.213 &705.78\\
BPDeg   &67.00  & 178.76 & 348.95 &  350.68  & 2.826  & -0.388 & -0.243& -0.052 & 1723\\
\bottomrule
\end{tabular}
\caption{N2HDM T1 benchmark points BPSep and BPDeg with semi-inverted
  mass hierarchy. The masses and the singlet VEV $v_s$ are given in [GeV].}
\label{N2HDM::T1::M2}
\end{table}
In \cref{N2HDM::T1::M2} two benchmark points are listed with a
semi-inverted mass hierarchy where 
 \begin{equation}
  \mdown \lesssim m_h \lesssim \mup\,.
\end{equation}
The first parameter point BPSep is in the region of the parameter
space where the pseudoscalar mass and the charged mass have a large
mass gap, denoted by $\msep$ in the previous discussion. The neutral
CP-even and the charged Higgs boson masses are all light and below
about 271~GeV whereas the pseudoscalar mass is significantly heavier
with a mass of $558\gev$. BPsep provides a rather strong SFOEWPT with $\xi_c = 1.31$ ($T_c=133.21\gev\,,~v_c=174.59\gev$). This is in agreement with
the already discussed observation that a light mass spectrum in
combination with a heavy scalar degree of freedom strengthens the phase
transition. 
The singlet admixtures of the neutral non-SM-like CP-even Higgs bosons
of BPSep are 
\begin{equation}
  \Sigma_{\hdown}= 76.3\%\,,\quad \Sigma_{\hup}=22.7 \%\,. 
\end{equation}
The mass hierarchy allows for the Higgs-to-Higgs decays
$\hup\rightarrow\hdown h$, $\hup\rightarrow\hdown \hdown$ and
$\hup\rightarrow h h$ with branching ratios of
\begin{equation}
  \br\cbrak{\hup\rightarrow\hdown h}=74.6\%\,,\,\br\cbrak{\hup\rightarrow\hdown \hdown}=13.2\%\,,\,\br\cbrak{\hup\rightarrow hh}= 2.5\%\,.
\end{equation}
With $H_\uparrow$ being doublet-like and rather
light also the production cross section is reasonably large so that
for a c.m.~energy of $\sqrt{s}= 13$~TeV we have a signal rate of 
\begin{equation}
  \sigma\cbrak{pp\rightarrow \hup \rightarrow\cbrak{\hdown \rightarrow \overline b b }\cbrak{h \rightarrow \overline b b }}= 740.3 \fb\,.
\end{equation}
This is phenomenologically very interesting, as we have two different
Higgs boson masses in the final state and a rather large cross
section. To put this into context, we remind the reader that the
production cross section for a pair of SM-like Higgs bosons including
NLO QCD corrections taking into account the full top-quark mass
dependence is 32.91~fb at $\sqrt{s}=14$~TeV
\cite{Borowka:2016ehy,Borowka:2016ypz,Baglio:2018lrj}. \s 

The second parameter point BPDeg belongs to the phase space region
$\mdeg$ where $m_A\approx \mHc$ and also provides a rather high $\xi_c
= 1.38$ ($T_c=135.92\gev, v_c= 188.17\gev$). While the non-SM-like
CP-even Higgs bosons are somewhat lighter than in BPSep, the pseudoscalar and
charged Higgs boson masses are around
350~GeV, so that the decay of $A$ into a gauge plus Higgs boson pair
is kinematically possible. With the singlet admixtures of $H_\uparrow$
and $H_\downarrow$ given by
\begin{equation}
 \Sigma_{\hdown}= 5.8 \%\,,\quad \Sigma_{\hup}= 94.0\%\,,
\end{equation}
the dominant decay is into the doublet-like Higgs, hence $A\rightarrow
Z \hdown$. The parameter point has been chosen as it provides the
largest signal rate for $A\rightarrow Z \hdown$ among our parameter sample, with 
\begin{equation}
  \sigma\cbrak{pp\rightarrow A \rightarrow Z\hdown}= 6.91\pb
\end{equation}
at $\sqrt{s}= 13$~TeV. This signature would be a clear sign of
beyond-the-SM physics and should be 
accessible at the LHC in view of the large cross
section. We add that the dominant branching ratios
  of $\hdown$ are given by  
  \begin{align}
    \br(\hdown\rightarrow\bar b b) = 84.6\%\,,\quad& 
    \br(\hdown\rightarrow \tau\tau)= 7\% \;.
  \end{align}
\subsubsection*{Heavy Mass Hierarchy}
In \cref{N2HDM::T1::M1}, we provide two benchmark points for the heavy
mass hierarchy.
\begin{table}[b]
\centering
\begin{tabular}{c c c c c c c c c c }
\toprule
 & $\mdown$    & $\mup$ & $\mHc$& $m_A $ & $\tan\beta$ & $\alpha_1$ & $\alpha_2$ & $\alpha_3$ & $v_S$  \\\midrule
BPii1  &285.26 & 1461.94  &  543.24 & 525.72  & 2.226 & 1.189 & 0.081  & 0.072 & 757.08\\
BPii2  &221.71 & 269.93 &217.75 & 570.91  & 6.522 & 1.319 & -0.227  & -0.387 & 945.55\\
\bottomrule
\end{tabular}
\caption{N2HDM T1 benchmark points BPii1 and BPii2 with heavy 
  mass hierarchy. The masses and the singlet VEV $v_s$ are given in [GeV].}
\label{N2HDM::T1::M1}
\end{table}
The parameter point BPii1 features a quite heavy $\hup$, while the
other Higgs-boson masses are in the  intermediate mass range so that
the mass gap between the neutral non-SM like Higgs bosons $\hup$ and 
$\hdown$ is very large. In contrast, BPii2 has an overall light
Higgs spectrum apart from the pseudoscalar Higgs boson with a mass of
$571\gev$. For both benchmark points the singlet admixture of the heavy
CP-even Higgs boson is quite high so that $\hup$ is singlet-like with 
\begin{equation}
  \Sigma^{\text{BPii1}}_{\hup}= 98.8\%\,\quad\text{and}\quad
  \Sigma^{\text{BPii2}}_{\hup}= 81.1\%\,.
\end{equation}
The enhanced mass of $H_\uparrow$ with $\mup = 1.46$~TeV of BPii1 is only
possible for an almost completely singlet-like state, so that this
heavy degree of freedom does not contribute to the EWPT. This allows
us to have a strong EWPT with $\xi_c = 1.66$ ($T_c=130.1\gev\,,
v_c=216.3\gev$).  The benchmark point BPii2 with an intermediate Higgs mass spectrum
only has a $\xi_c = 1.04$ ($T_c = 136.54\gev\,,~v_c = 
141.96\gev$). Among the parameter points with this mass 
hierarchy, BPii2 has the largest signal rate for the production of a
pair of SM-like Higgs bosons through $\hup\rightarrow h h
$. It has a larger branching ratio
for this final state than BPii1, with $\br\cbrak{\hup\rightarrow h h }_{\text{BPii2}}
= 31.3\%$ compared to 
$\br\cbrak{\hup\rightarrow h h }_{\text{BPii1}} = 10.9\%$ in BPii1, and
the gluon fusion cross section is larger because of the lighter mass
$m_{H_\uparrow}$ compared to BPii1,
\begin{equation}
  \sigma\cbrak{gg\rightarrow \hup}_{\text{BPii1}} = 0.13\fb\,\quad \text{and}\quad \sigma\cbrak{gg\rightarrow \hup}_{\text{BPii2}} = 1.01\pb
\end{equation}
for a c.m.~energy of $\sqrt{s}= 13$~TeV.
We then have for BPii2 the following signal rates for SM-like di-Higgs
production in the $4b$, $(\bar{b}b)(\tau\tau)$, $(\bar{b}b)(WW)$, $4W$ and
$(\bar{b}b)(\gamma\gamma)$ final states, 
\begin{align}
  \sigma\cbrak{pp\rightarrow \hup \rightarrow \cbrak{h\rightarrow \overline b b}\cbrak{h \rightarrow \overline b b}} &= 109.3 \fb\\
  \sigma\cbrak{pp\rightarrow \hup \rightarrow \cbrak{h\rightarrow
  \overline b b}\cbrak{h \rightarrow \overline \tau \tau}} &= 11.8 \fb\\
  \sigma\cbrak{pp\rightarrow \hup \rightarrow \cbrak{h\rightarrow
  \overline b b }\cbrak{h \rightarrow WW }} &=39.9 \fb \\
  \sigma\cbrak{pp\rightarrow \hup \rightarrow \cbrak{h\rightarrow  W W}\cbrak{h \rightarrow WW }} &= 14.6 \fb\\
  \sigma\cbrak{pp\rightarrow \hup \rightarrow \cbrak{h\rightarrow
  \overline b b}\cbrak{h \rightarrow \gamma \gamma}} &= 0.4 \fb \;.
\end{align}
In BPii1, due to the smaller $H_\uparrow$ production cross section we
have for the $4b$ final state the much smaller rate
\begin{equation}
  \sigma\cbrak{pp\rightarrow \hup \rightarrow \cbrak{h\rightarrow \overline b b}\cbrak{h \rightarrow \overline b b}} = 0.005\fb\,.
\end{equation}
The benchmark scenario BPii1 features enhanced trilinear Higgs
self-couplings
$\lambda_{hhh}^{\text{NLO,N2HDM}}/\lambda_{hhh}^{\text{NLO,SM}}=1.93$
between three SM-like Higgs bosons, and the largest absolute value for
the coupling between $\hup$ and two SM-like Higgs bosons $h$ of the
sample, with
$\lambda_{hh\hup}^{\text{NLO,N2HDM}}/\lambda_{hhh}^{\text{NLO,SM}} =
-3.97$. Despite the significantly enhanced trilinear Higgs
self-couplings the expected signals in the di-Higgs rates are small
due to the small production cross section of the heavy Higgs state. 
On the other hand, BPii2 with its trilinear Higgs self-couplings of 
\begin{equation}
    \lambda_{hhh}^{\text{NLO,N2HDM}}/\lambda_{hhh}^{\text{NLO,SM}}= -1.47 \,,\quad  \lambda_{hh\hup}^{\text{NLO,N2HDM}}/\lambda_{hhh}^{\text{NLO,SM}}= 0.67\,,
\end{equation}
allows for the largest expected di-Higgs signals in $pp\rightarrow
\hup \rightarrow \cbrak{h\rightarrow X}\cbrak{h\rightarrow Y}$\,. 
To conclude, singlet-like\footnote{Singlet-like Higgs states do not
  directly contribute to the EWPT since the singlet VEV $v_S$ is not
  included in the determination of $\omega_c$.} heavy Higgs bosons are
interesting in the sense that they can strengthen the EWPT by
providing a heavy scalar degree of freedom whereby the remaining
neutral Higgs bosons can have light or intermediate mass values. Those
heavy degrees of freedom can have significantly enhanced couplings to
the SM-like Higgs boson. Simultaneously the production cross-section
of the heavy state is reduced, however, so that the expected di-Higgs
signals are suppressed. After all there exist also points having a
compromise with an intermediate mass spectrum and decent up to large expected
signals like BPii2. 

\subsubsection{Mass Spectrum of the N2HDM T2}
\begin{figure}
  \centering
    \subfigure[\label{N2HDM::T2::massdiff}]{\includegraphics[width=0.5\textwidth]{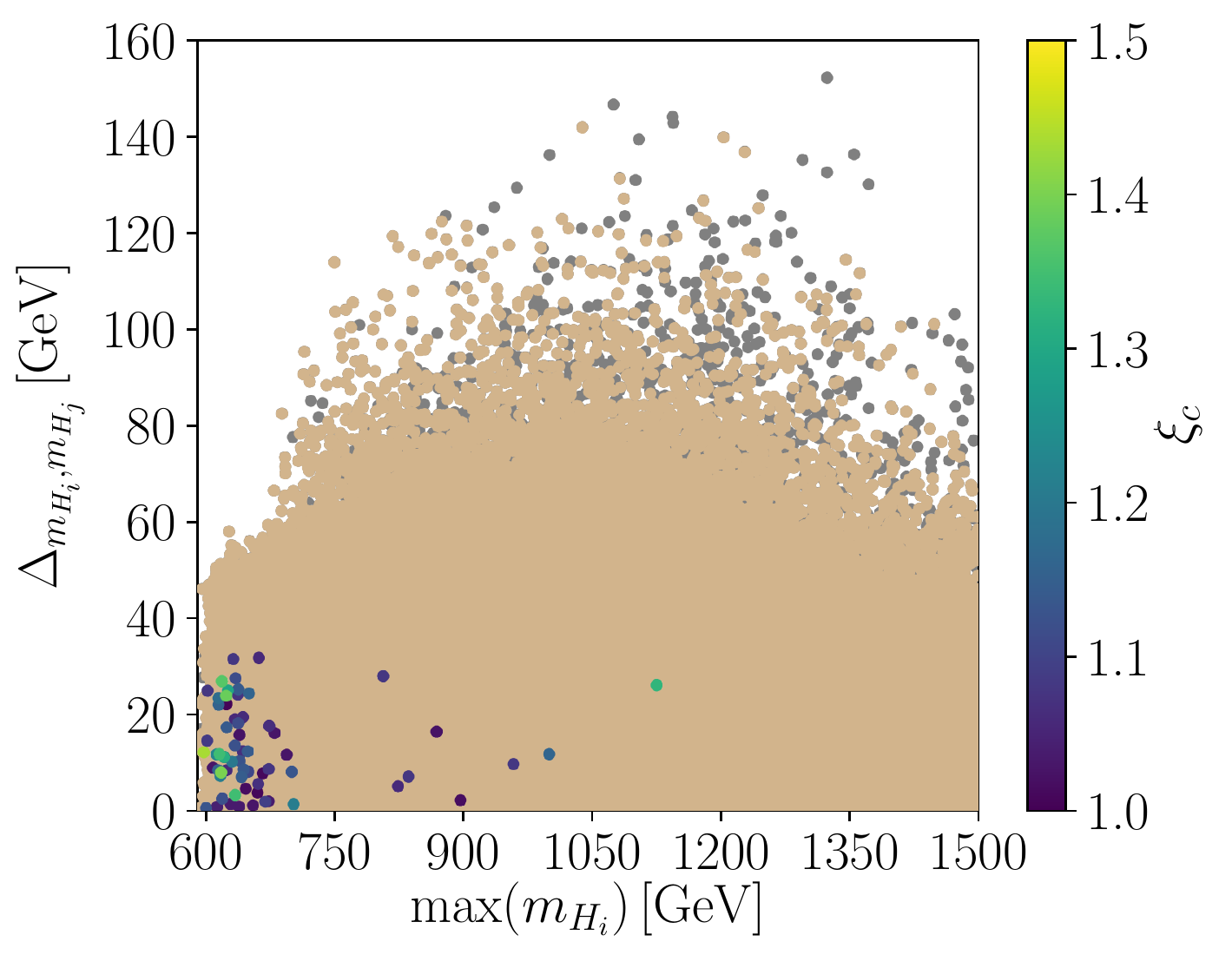}}%
    \subfigure[\label{N2HDM::T2::mupmdown}]{\includegraphics[width=0.5\textwidth]{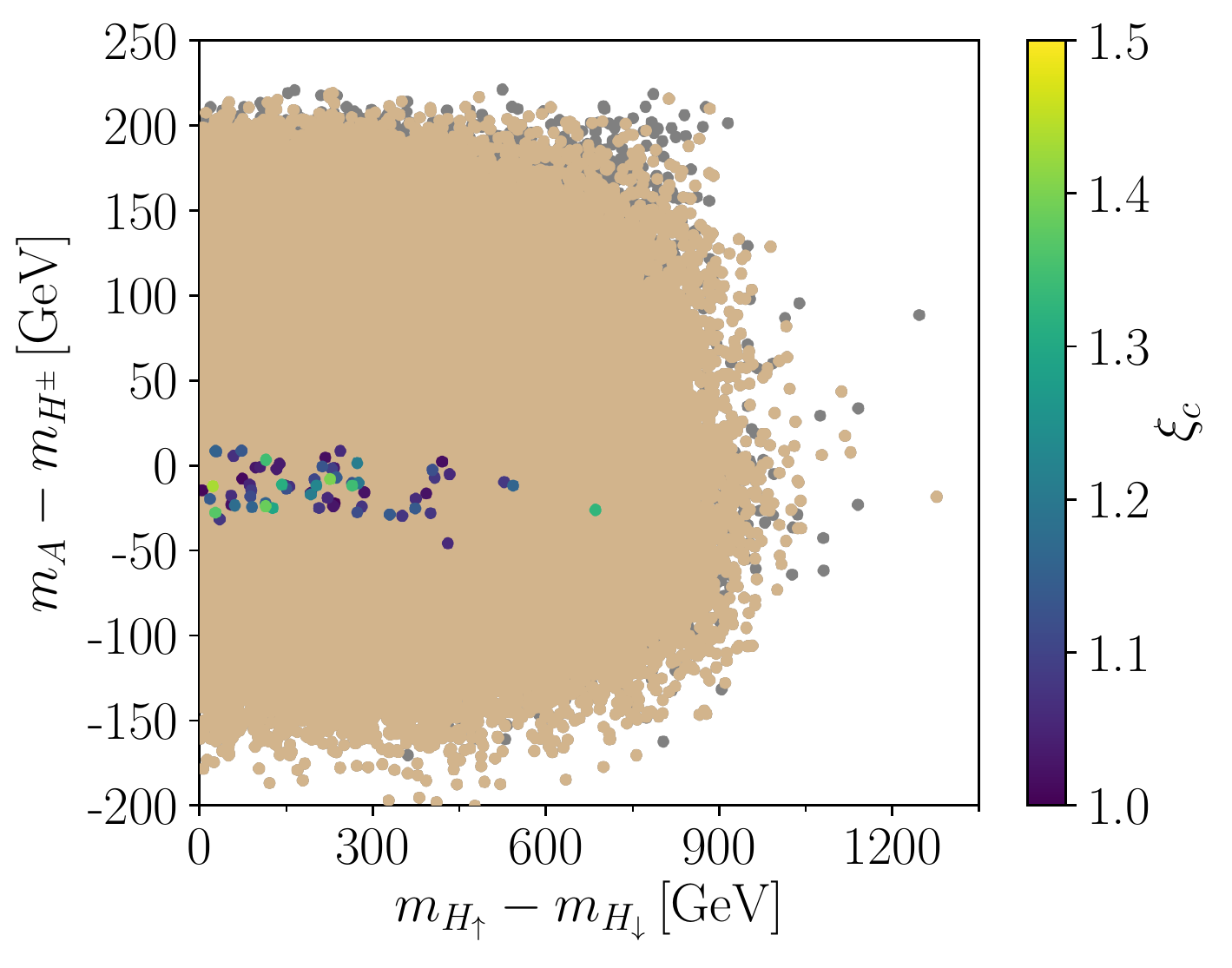}} %
  \caption{N2HDM T2: Left: The minimum mass gap $\Delta_{m_{H_i},m_{H_j}}$
    versus the maximum of the scalar masses. Right: The mass gap
    between the CP-odd and charged Higgs bosons, $A$ and $\hHc$,
    versus the mass gap between the CP-even
    neutral Higgs bosons $\hup$ and $\hdown$. The color code is the same as in
    Fig.~\ref{N2HDM::massspectrum}. The color bar indicates
    the strength of the phase transition for $\xi_c \ge 1$. The
    di-Higgs search constraints are included.} 
  \label{N2HDM::T2::massspectrum}
\end{figure}
We start our discussion of the N2HDM T2 mass spectrum with the minimum
mass gap $\Delta_{m_{H_i},m_{H_j}} $ among all possible pairings of the
N2HDM Higgs bosons. Figure~\ref{N2HDM::T2::massdiff} shows
$\Delta_{m_{H_i},m_{H_j}} $ versus the maximum of the masses,
max$(m_{H_i})$. The color bar indicates the strength of
the phase transition for points providing an 
SFOEWPT. As observed in the discussion of the N2HDM T1 the minimal
mass gap is already reduced through the requirement of 
compatibility with the EW precision data and even more by an SFOEWPT. In the N2HDM
T2 an even more reduced mass gap with $\Delta_{m_{H_i},m_{H_j}}$
below $\nullel(40\gev)$ is favored by an SFOEWPT, so that at least one
pair of the Higgs bosons has to be almost mass
degenerate. Additionally, except for some parameter points, the mass
spectrum is not too heavy with max$(m_{H_i})$
below $\nullel(750\gev)$.  
The heavy Higgs bosons with masses above $\sim 800\gev$ are again
singlet-like states. In \cref{N2HDM::T2::mupmdown} the mass gap
between $A$ and $\hHc$ versus the mass gap between $\hdown$ and $\hup$
is shown. Unlike the results in the N2HDM T1, we do not find the
separation into the two distinct parameter regions $\mathcal M_{\text{deg}}$
and $\mathcal{M}_{\text{sep}}$. In the N2HDM T2, only parameter points with nearly mass
degenerate $A$ and $H^\pm$ fulfill the experimental and
theoretical constraints and simultaneously provide an SFOEWPT. \s

\begin{figure}[t!]
  \centering
    \subfigure[\label{N2HDM::T2::mdown}]{\includegraphics[width=0.5\textwidth]{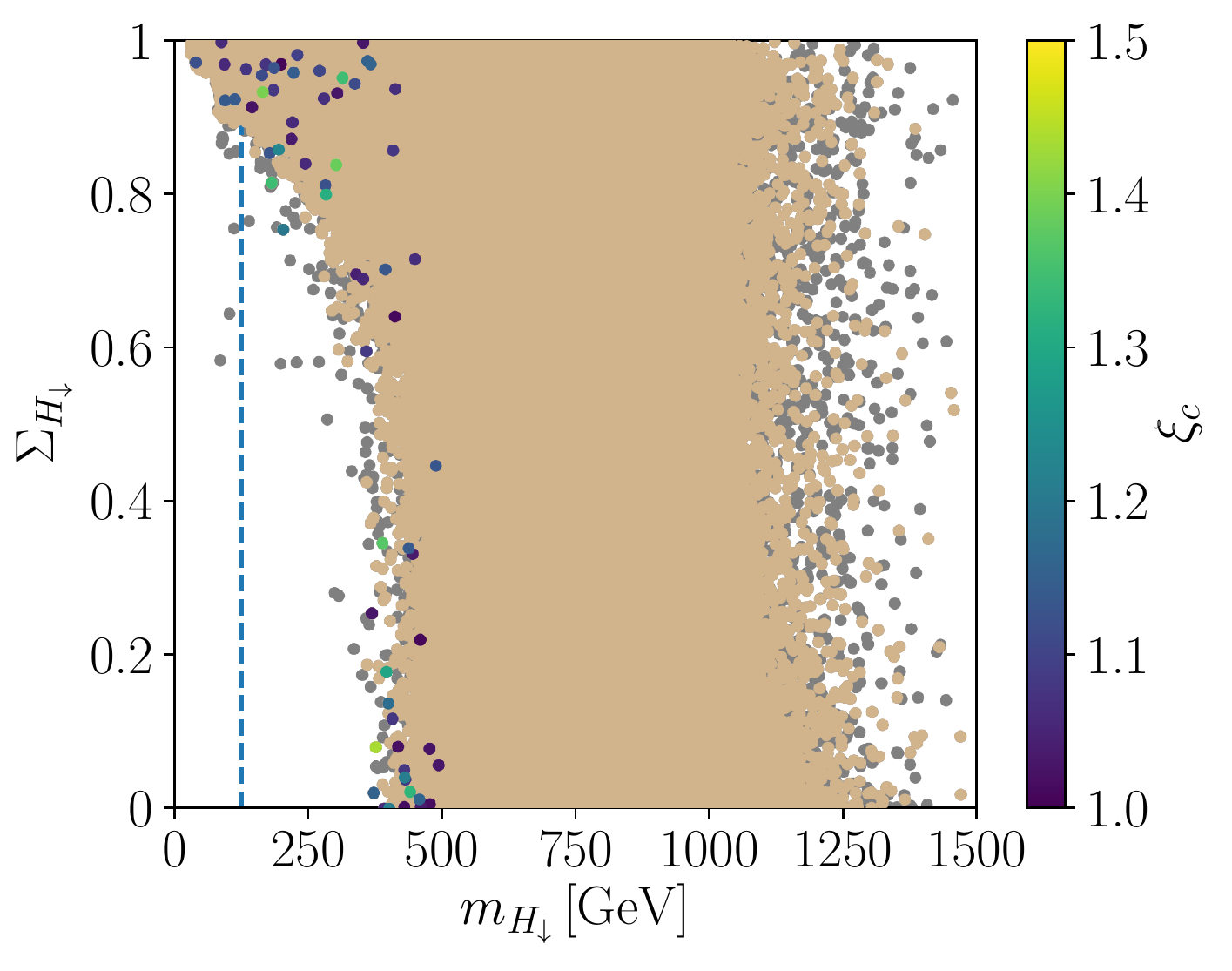}}%
    \subfigure[\label{N2HDM::T2::mup}]{\includegraphics[width=0.5\textwidth]{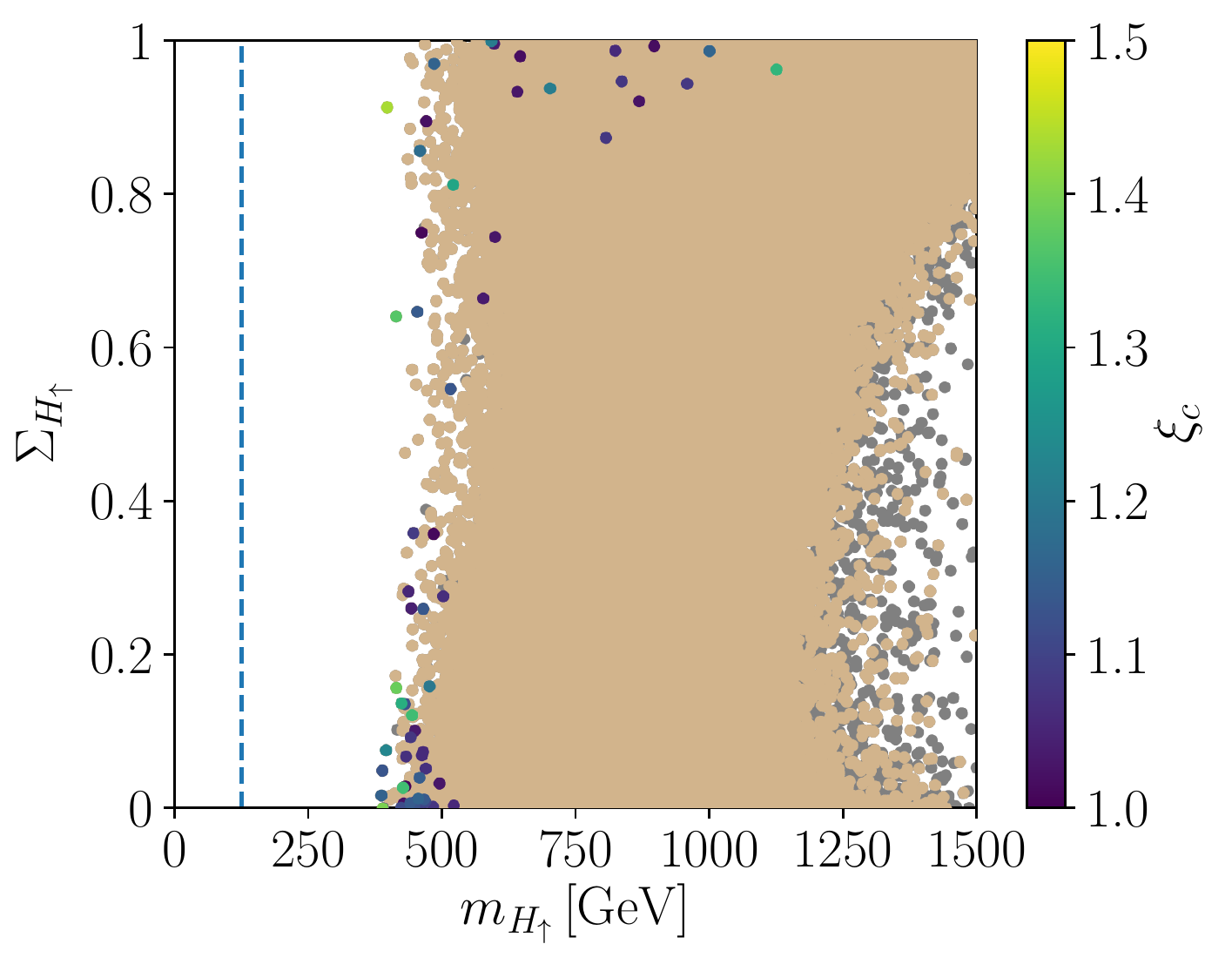}} %
  \caption{N2HDM T2: Left: singlet admixture $\Sigma_{H_\downarrow}$
    versus $m_{H_\downarrow}$; Right: $\Sigma_{H_\uparrow}$ versus
    $m_{H_\uparrow}$. The color code is the same as in
    Fig.~\ref{N2HDM::massspectrum}. The color bar indicates
    the strength of the phase transition for $\xi_c \ge 1$. The
    di-Higgs search constraints are included. The blue dashed
  line indicates the SM Higgs mass value of $m_h \approx 125$~GeV.}
  \label{N2HDM::T2::CPevenMasses}
\end{figure}
\begin{figure}[b!]
\centering
\includegraphics[width=0.6\textwidth]{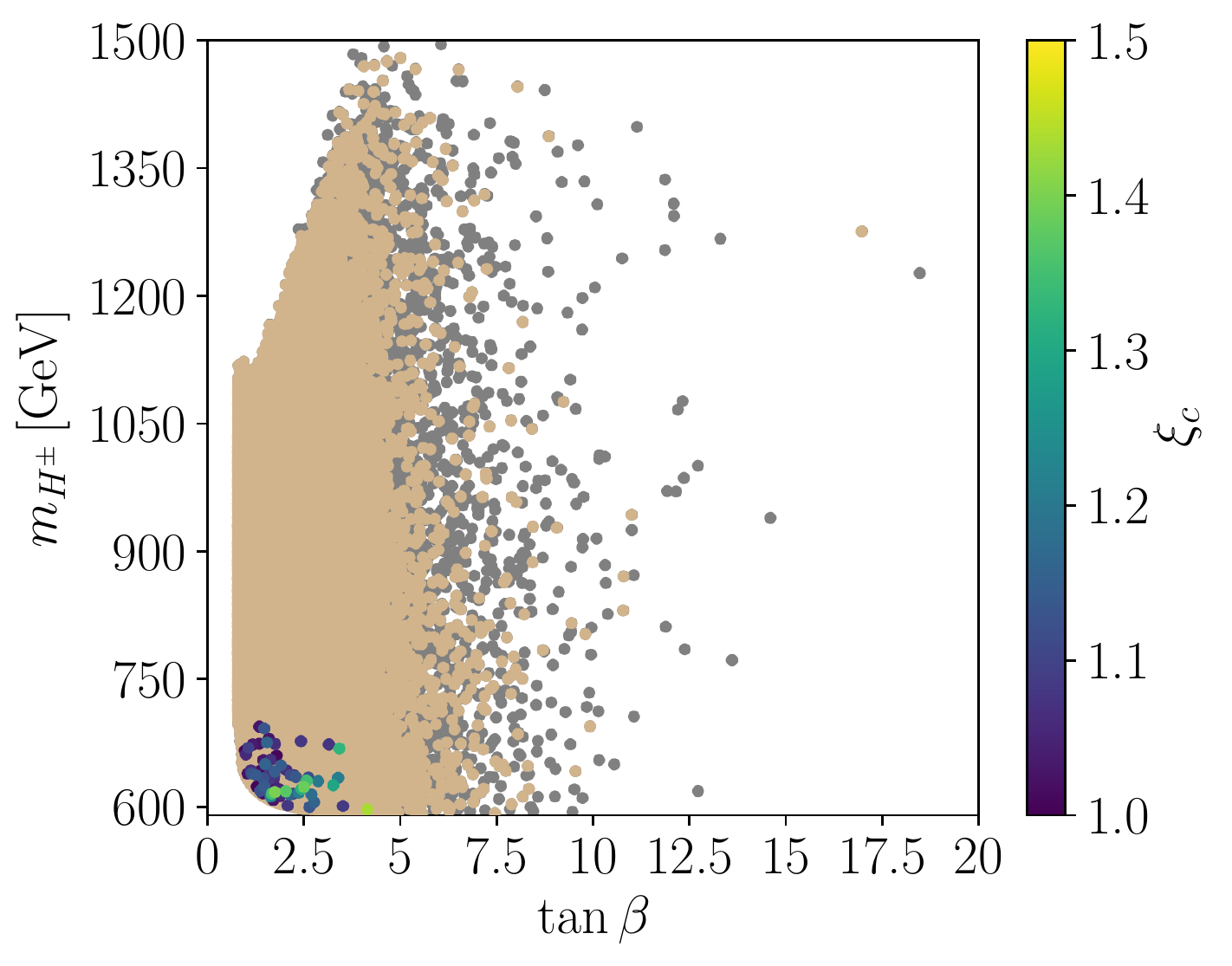}
\caption{N2HDM T2: The charged mass $\mHc$ versus $\tan\beta$. The
  color code is the same as in 
    Fig.~\ref{N2HDM::massspectrum}. The color bar indicates
    the strength of the phase transition for $\xi_c \ge 1$. The
    di-Higgs search constraints are included.}
\label{N2HDM::T2::mHctbeta}
\end{figure}
In \cref{N2HDM::T2::CPevenMasses} we show the singlet admixture $\Sigma$ of
the CP-even neutral Higgs bosons $H_\downarrow$ (left) and $H_\uparrow$ (right) 
versus their respective mass values. The blue dashed
line indicates the SM mass of $m_h = 125.09\gev$. For both
$H_\downarrow$ and $H_\uparrow$ the requirement of an SFOEWPT reduces
their mass values as follows
\begin{align}
  &\mdown\in\sbrak{31 , 1469}\gev\, \xrightarrow{\text{SFOEWPT}}
    \sbrak{40,493}\gev\\ 
  &\mup\in\sbrak{387,1500}\gev \, \xrightarrow{\text{SFOEWPT}}
    \sbrak{387,1124}\gev\,. 
\end{align}
As observed in the N2HDM T1 masses $\mup$ of the order
$\nullel\cbrak{1\tev}$ can only be realised for a singlet-like $\hup$
with a singlet-admixture of at least $\sim 80\%$. The
semi-inverted mass hierarchy with
$\mdown\lesssim m_h$ is possible in the N2HDM T2, but only for
singlet-like $\hdown$ with $\Sigma^S_{\hdown} \gtrsim 92\% $ whereas
the inverted mass hierarchy is not realized. 
In \cref{N2HDM::T2::mHctbeta} $\mHc$ is plotted versus $\tan\beta$. As
can be inferred from the figure, the requirement of an SFOEWPT reduces
the upper bound of the charged mass quite significantly, 
\begin{equation}
  \mHc\in\sbrak{592 , 1480}\gev\, \xrightarrow{\text{SFOEWPT}}
  \sbrak{597,694}\gev\,. 
\end{equation}
Additionally, the SFOEWPT favors small values of $\tan\beta$ in the range of
\begin{equation}
  \tan\beta\big\vert_{\text{SFOEWPT}}\in\sbrak{0.98,4.11}\,.
\end{equation}
Due to the favored intermediate mass regions for the charged mass,
upcoming flavor constraints updates pushing the constraint on the
charged mass higher might have an important constraining power on the
N2HDM  T2 with respect to an SFOEWPT as already observed in the C2HDM T2.

\subsubsection{N2HDM T2 - Trilinear Higgs Self-Couplings}
\begin{figure}
\centering
\includegraphics[width=0.6\textwidth]{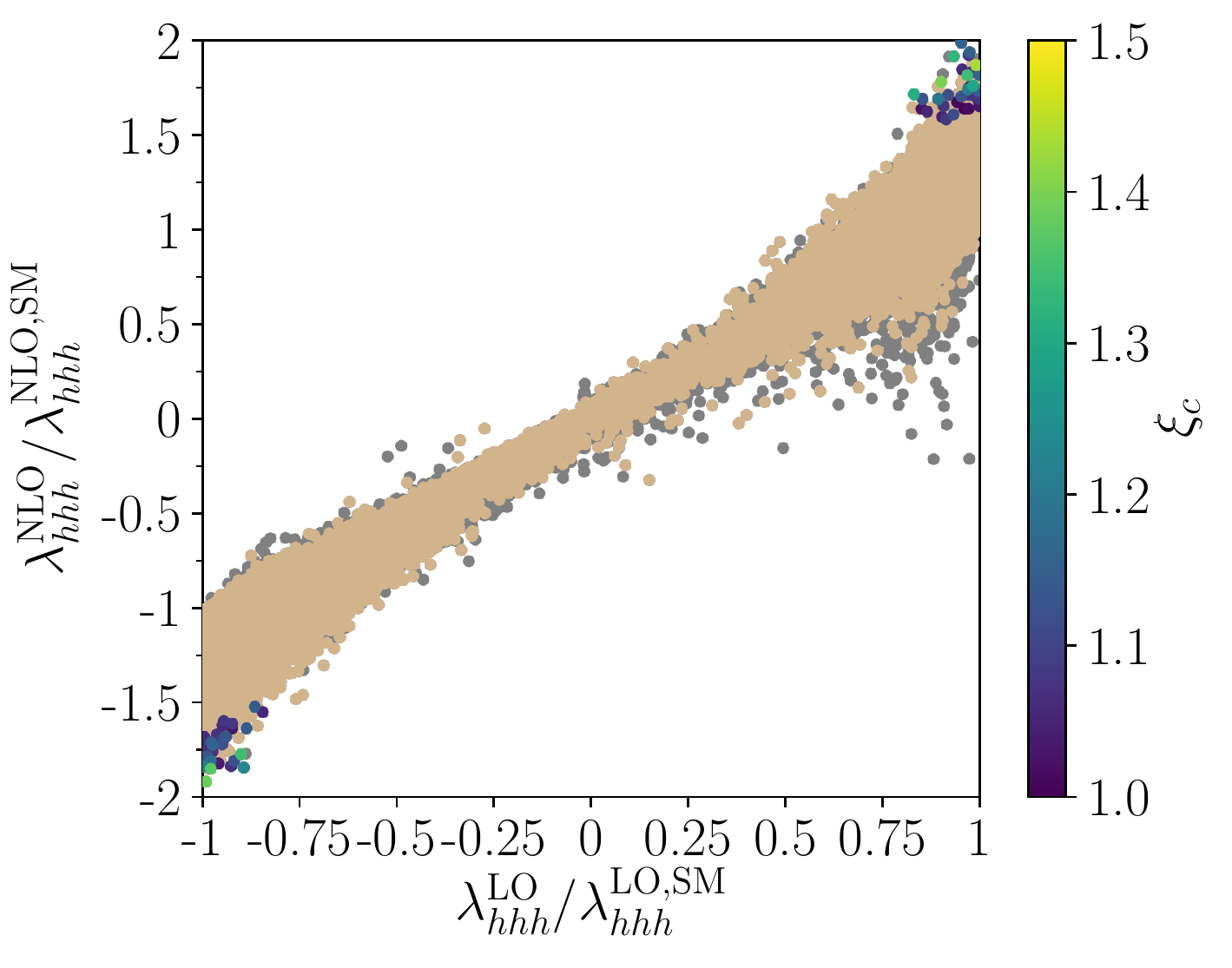}
\caption{N2HDM T2: NLO trilinear self-coupling between
  three SM-like Higgs bosons normalised to the SM reference value
  versus the corresponding LO ratio. The
  color code is the same as in 
    Fig.~\ref{N2HDM::massspectrum}. The color bar indicates
    the strength of the phase transition for $\xi_c \ge 1$. The
    di-Higgs search constraints are included.}
\label{N2HDM::T2::triplehhh}
\end{figure}
In \cref{N2HDM::T2::triplehhh} the NLO trilinear coupling of the
SM-like Higgs boson normalised to the NLO SM value is displayed versus
the corresponding LO ratio. The requirement of an SFOEWPT reduces the
range allowed by the theoretical and experimental
constraints as follows, 
\begin{equation}
  \lambda^{\text{NLO}}_{hhh}/\lambda_{hhh}^{\text{NLO,SM}} \in \sbrak{-1.92, 1.99} \xrightarrow{\text{SFOEWPT}} \sbrak{-1.92, -1.52}\cup \sbrak{1.58,1.98}\,.
\end{equation}
As already mentioned in the discussion of the N2HDM T1, the interplay
of the requirement of large quartic couplings and small to medium
Higgs boson masses, pushes the trilinear couplings to enhanced values,
but their upper limit remains under the allowed possibilities
concerning theoretical and experimental constraints. 
In contrast to the N2HDM T1, in the N2HDM T2 trilinear couplings of at
least $1.52$ times the SM value are required for an SFOEWPT, the SM size is not sufficient. 

\subsection{Comparison of the C2HDM and the N2HDM Di-Higgs Rates \label{subsec:comparison}}
So far we discussed both models independently with respect to their
phenomenology of the mass spectra and the trilinear Higgs
self-couplings. In the following we will compare the two models with
respect to their expected signal rates in the di-Higgs final
states. Both models feature three neutral Higgs bosons
$\cbrak{h,\hdown,\hup}$ and one charged Higgs boson. Recent di-Higgs
searches in the final states $4b$ \cite{DIHIGS1,DIHIGS2},
$(2b)(2\tau)$ \cite{DIHIGS3,DIHIGS4} and $(2b)(2\gamma)$ \cite{CMS:2017ihs} put strict
constraints on the viable parameter space of both models so that we want to
investigate to which extend future studies of this kind might help to
tighten the viable parameter space of the N2HDM and C2HDM. All rates
given in the following have been computed for $\sqrt{s}=13$~TeV.

\subsubsection{C2HDM and N2HDM T1}
\label{bbbb}
\begin{figure}[b!]
  \centering
    \subfigure[C2HDM T1]{\includegraphics[width=0.5\textwidth]{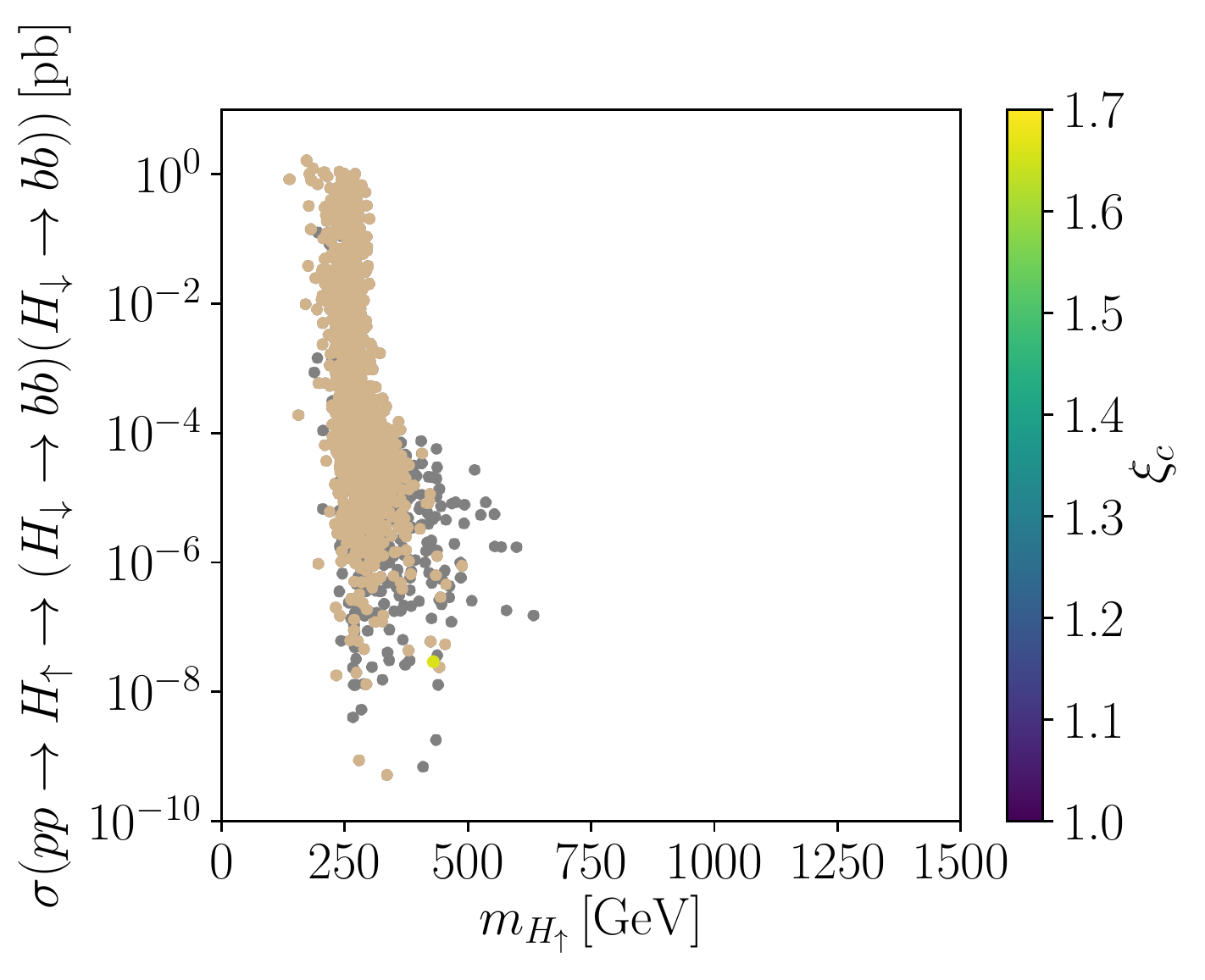}}%
    \subfigure[N2HDM T1]{\includegraphics[width=0.5\textwidth]{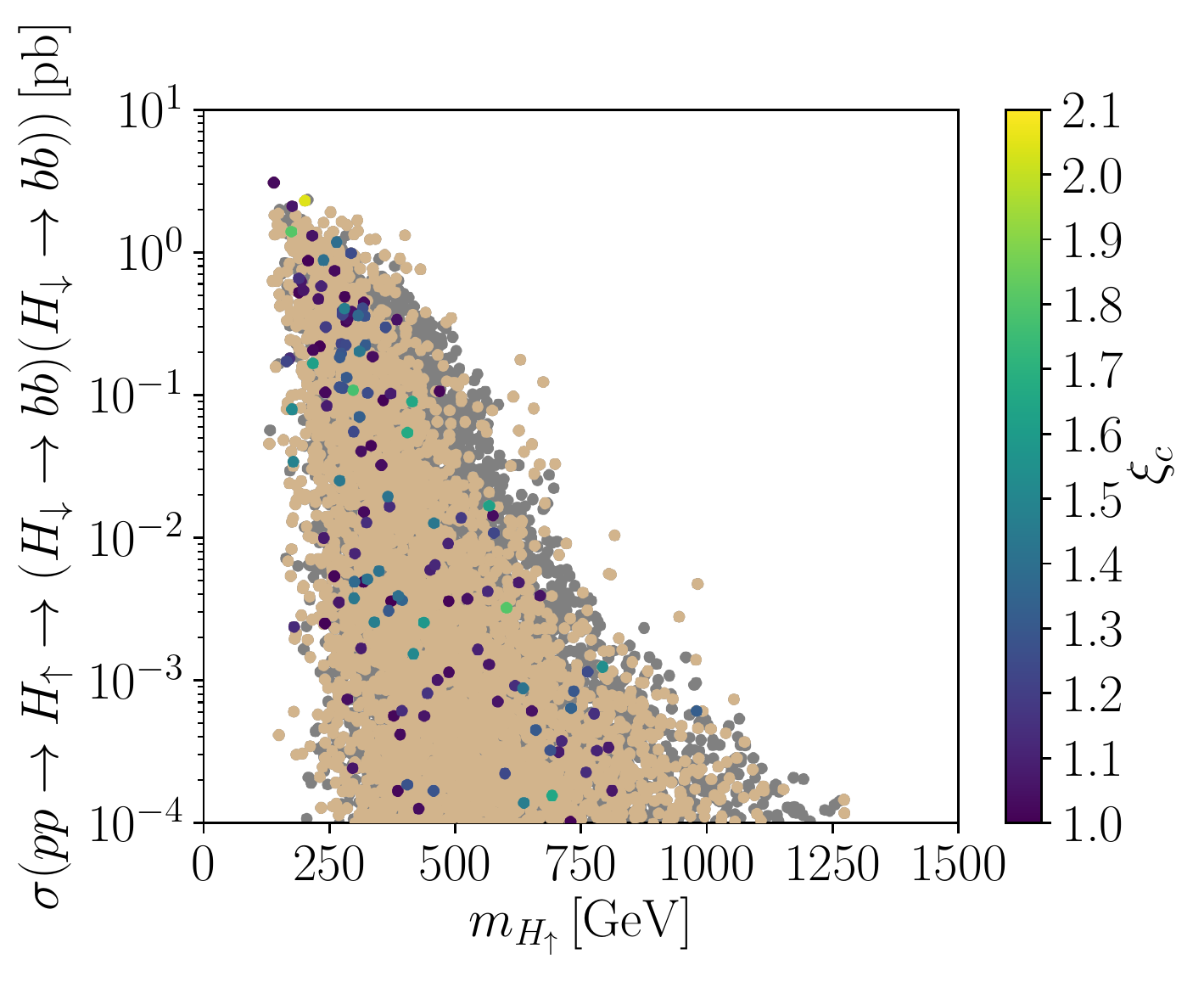}} %
  \caption{T1: Production rate of $H_\uparrow$ with subsequent decay
    into $H_\downarrow H_\downarrow$ in the $4b$ final state versus
    $m_{H_\uparrow}$ for the C2HDM (left) and the N2HDM 
    (right). The color code is the same as in
    Fig.~\ref{N2HDM::massspectrum}. The color bar indicates 
    the strength of the phase transition for $\xi_c \ge 1$. The
    di-Higgs search constraints are included.}
  \label{COMP::HUP::hdownhdown::BBBB}
\end{figure}
\begin{figure}[t!]
  \centering
    \subfigure[C2HDM T1]{\includegraphics[width=0.5\textwidth]{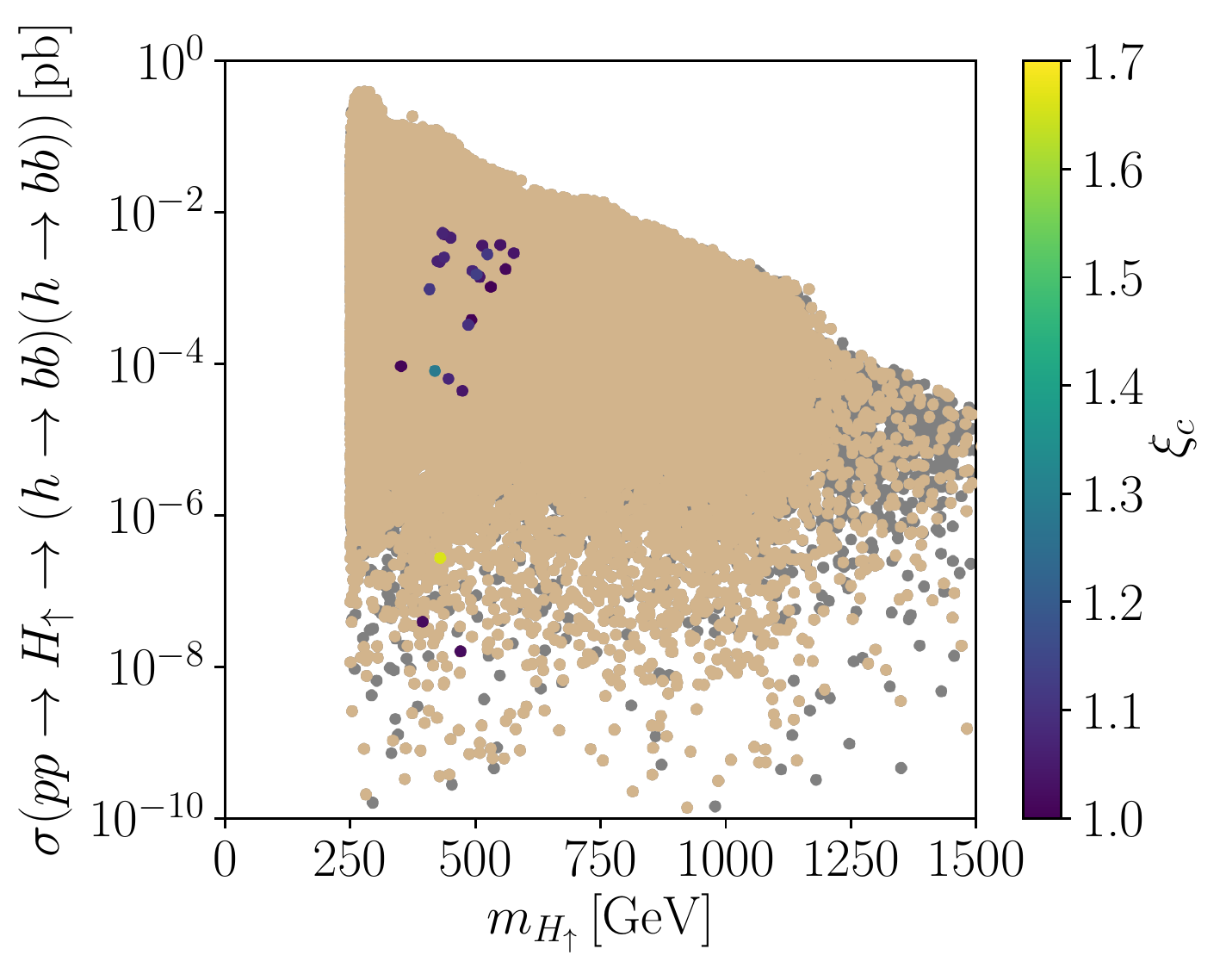}}%
    \subfigure[N2HDM T1]{\includegraphics[width=0.5\textwidth]{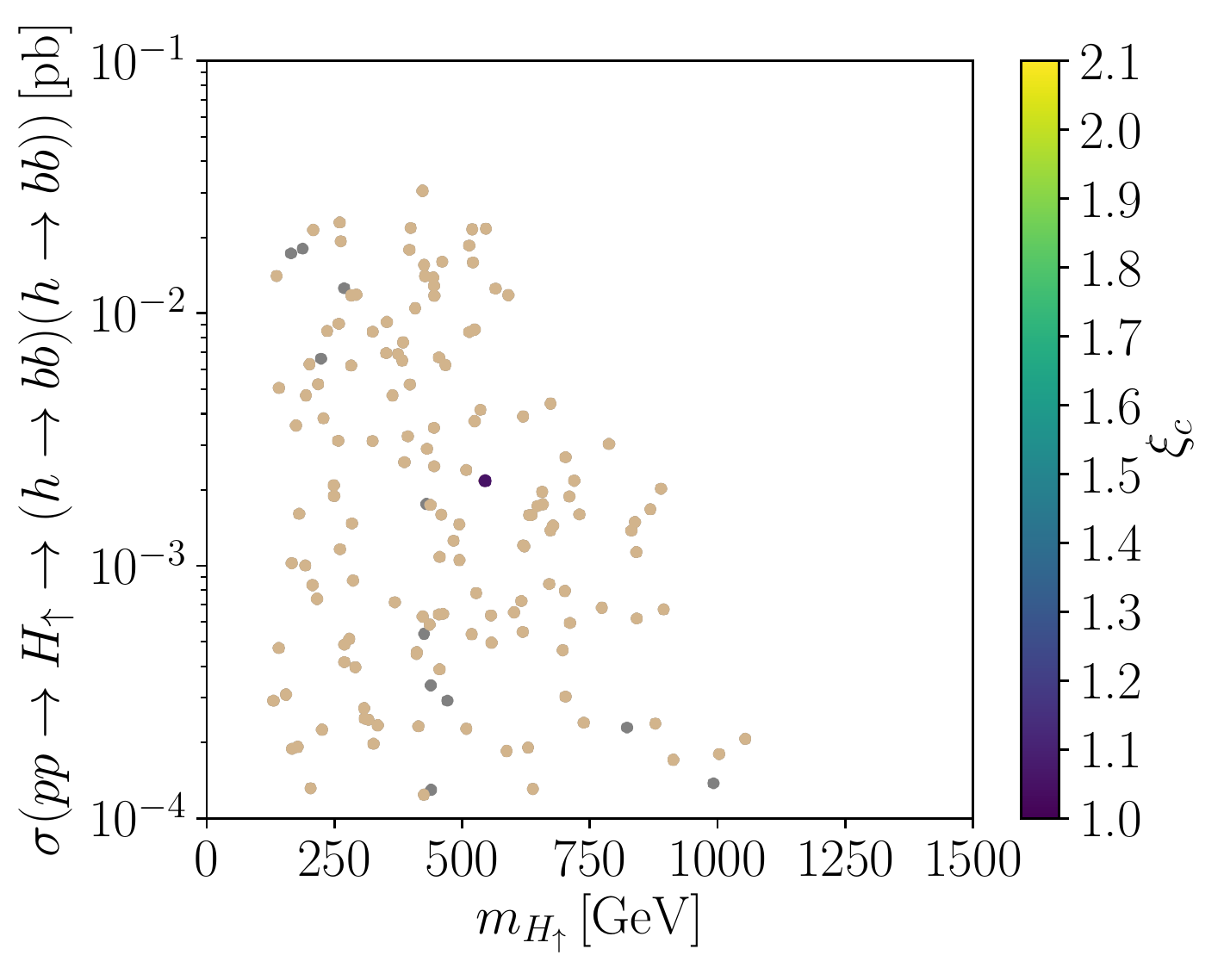}} %
  \caption{T1: Production rate of $H_\uparrow$ with subsequent decay
    into $hh$ in the $4b$ final state versus
    $m_{H_\uparrow}$ for the C2HDM (left) and the N2HDM 
    (right). The color code is the same as in
    Fig.~\ref{N2HDM::massspectrum}. The color bar indicates 
    the strength of the phase transition for $\xi_c \ge 1$. The
    di-Higgs search constraints are included}
  \label{COMP::HUP::hh::BBBB}
\end{figure}
\subsubsection*{$\cbrak{\overline b b}\cbrak{\overline b b}$ Final State}
 We start the comparison for the $(4b)$ final state of the T1 models. In
\cref{COMP::HUP::hdownhdown::BBBB} we show the signal rates for the process
$pp\rightarrow\hup\rightarrow\cbrak{\hdown
  \rightarrow\overline b b}\cbrak{\hdown \rightarrow\overline b b}$
versus the mass of the heaviest neutral Higgs boson $\hup$ for the
C2HDM (left) and the N2HDM (right). Figure \ref{COMP::HUP::hh::BBBB}
displays the signal rates $pp\rightarrow\hup\rightarrow\cbrak{h
  \rightarrow\overline b b}\cbrak{h \rightarrow\overline b b}$ versus
$m_{H_\uparrow}$. The rates decrease with increasing mass $m_{H_\uparrow}$. 
In Fig.~\ref{COMP::HUP::hdownhdown::BBBB} (left) a strong reduction of the
C2HDM rates can be observed at the threshold for the decay into a
top-quark pair, $m_{H_\uparrow}= 350$~GeV, down from ${\cal
  O}(1~\mbox{pb})$ by more than three orders of
magnitude.\footnote{Where necessary to show certain effect, we display rates down to
  $10^{-10}$ in the plots. Otherwise, the plots are cut at $10^{-4}$.} In the
N2HDM, no such reduction appears as the addition of the singlet to the
Higgs sector allows for
singlet-like $\hup$ granting Higgs-to-Higgs decays being of the order
$\nullel\cbrak{1\pb}$. Reaching $\mup\approx 700\gev $ also 
$\hdown$ can be heavier than $350\gev$ so that the decays $H_\downarrow
\to t\bar{t}$ can become possible. In the C2HDM, this possibility reduces
the overall cross section so effectively that no Higgs-to-Higgs decays
above $\nullel\cbrak{10^{-10} \mbox{pb}}$ are realized in the C2HDM for masses
above $700\gev$. With respect to a successful SFOEWPT the differences
of both models are significant. In the C2HDM the requirement of an
SFOEWPT reduces the maximal signal rate 
\begin{equation}
  \max\sigma\cbrak{pp\rightarrow\hup\rightarrow\cbrak{\hdown \rightarrow\overline b b}\cbrak{\hdown \rightarrow\overline b b}}_{\text{C2HDM}} = 1611\fb
\end{equation}
to a signal rate of the oder $\mathcal{O}(10^{-8}\pb)$, while no such
reduction is observed in the N2HDM. The SFOEWPT parameter sample covers
almost the full sample and we have the following maximum rates
both with and without an SFOEWPT
\begin{equation}
  \max\sigma\cbrak{pp\rightarrow\hup\rightarrow\cbrak{\hdown \rightarrow\overline b b}\cbrak{\hdown \rightarrow\overline b b}}_{\text{N2HDM}} = 3070\fb \,.
\end{equation}
An analogous behaviour of the maximum rates is observed in
\cref{COMP::HUP::hh::BBBB} in the decay channel
$\hup\rightarrow\cbrak{h\rightarrow\overline b b
}\cbrak{h\rightarrow\overline b b }$ 
\begin{subequations}
\begin{align}
\label{maxCX}
&\max\sigma\cbrak{pp\rightarrow\hup\rightarrow\cbrak{h \rightarrow\overline b b}\cbrak{h \rightarrow\overline b b}}_{\text{C2HDM}} = 385.92\fb \xrightarrow{\text{SFOEWPT}}  5.24 \fb\\
&\max\sigma\cbrak{pp\rightarrow\hup\rightarrow\cbrak{h \rightarrow\overline b b}\cbrak{h \rightarrow\overline b b}}_{\text{N2HDM}} = 413.20\fb \xrightarrow{\text{SFOEWPT}}  285.70 \fb\,.
\end{align}
\end{subequations}
In the SM-like decay chain $\hup\rightarrow h h \rightarrow XX $ (and
also in the decay chain $\hup\rightarrow \hdown\hdown\rightarrow XX$)
increasing sensitivity in the di-Higgs searches for intermediate Higgs
boson masses (up to $500\gev$) could allow for probing valid N2HDM
candidates with a successful strong first order EWPT, in contrast to the
C2HDM. In the C2HDM the cross section for the SFOEWPT parameter points is orders of
magnitudes below the LHC sensitivity.
Since in C2HDM and N2HDM T1 the couplings of the Higgs bosons to the
down-type quarks and leptons are the same
\cite{Muhlleitner:2016mzt,Fontes:2017zfn} the rates in the
$(b\bar{b})(\tau\tau)$ final state are simply reduced by about a
factor 10 without changing the qualitative behaviour so that we do not
show separate plots for this final state. \s

\begin{figure}
  \centering
    \subfigure[C2HDM T1]{\includegraphics[width=0.5\textwidth]{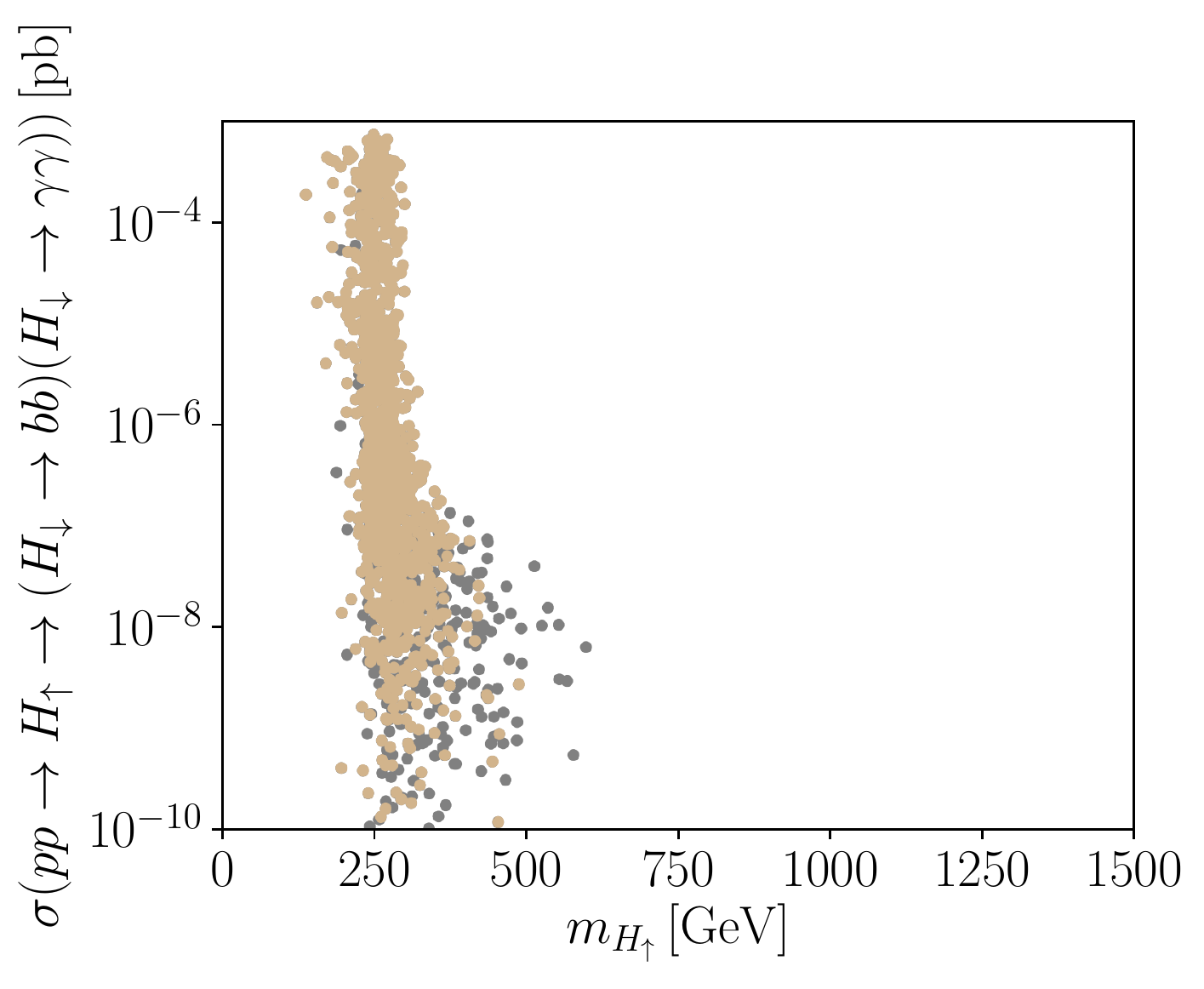}}%
    \subfigure[N2HDM T1]{\includegraphics[width=0.5\textwidth]{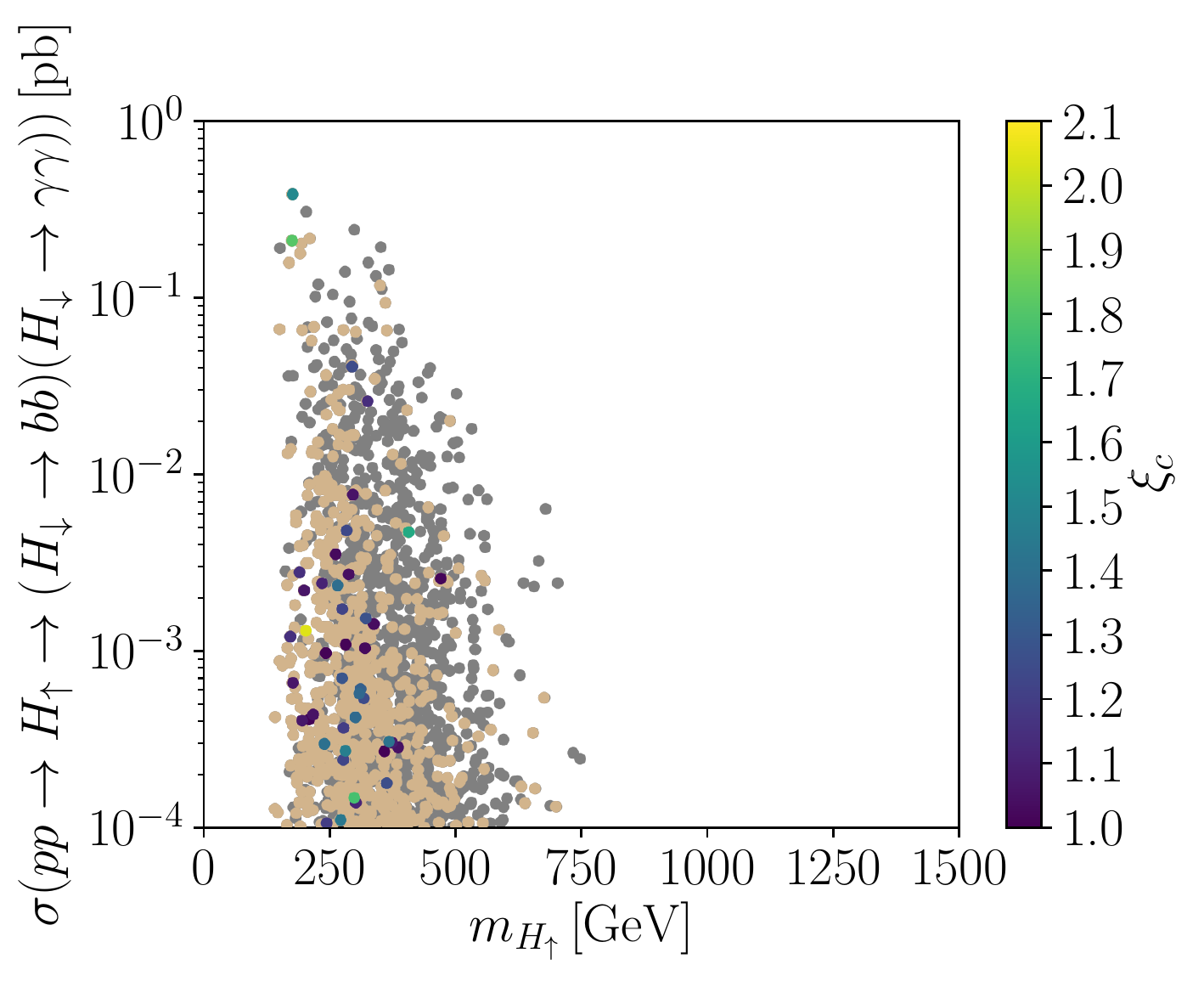}} %
  \caption{T1: Production rate of $H_\uparrow$ with subsequent decay
    into $H_\downarrow H_\downarrow$ in the $(2b)(2\gamma)$ final state versus
    $m_{H_\uparrow}$ for the C2HDM (left) and the N2HDM 
    (right). The color code is the same as in
    Fig.~\ref{N2HDM::massspectrum}. The color bar indicates 
    the strength of the phase transition for $\xi_c \ge 1$. The
    di-Higgs search constraints are included.}
  \label{COMP::HUP::hdownhdown::BBGAMGAM}
\end{figure}

\begin{figure}
  \centering
    \subfigure[C2HDM T1]{\includegraphics[width=0.5\textwidth]{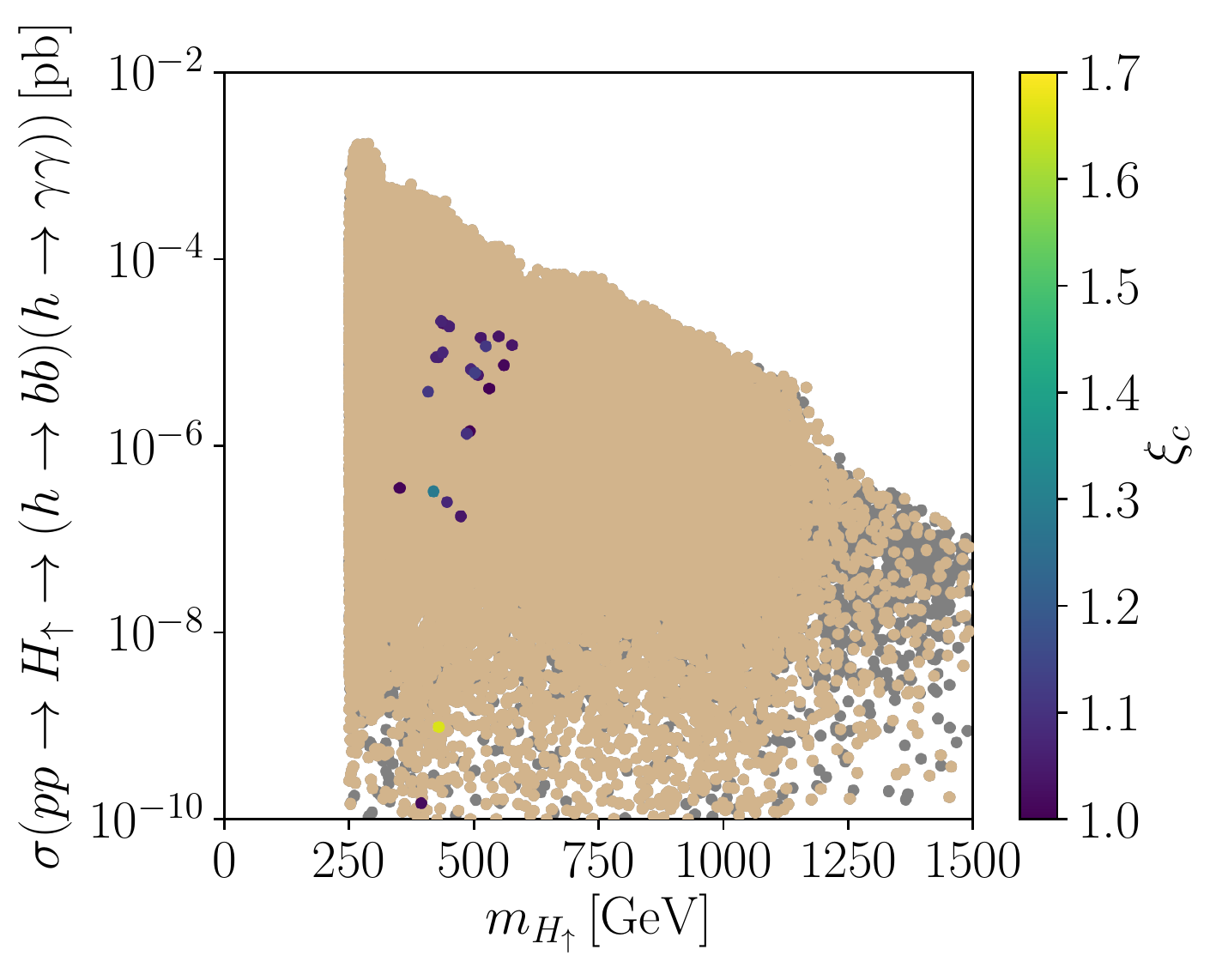}}%
    \subfigure[N2HDM T1]{\includegraphics[width=0.5\textwidth]{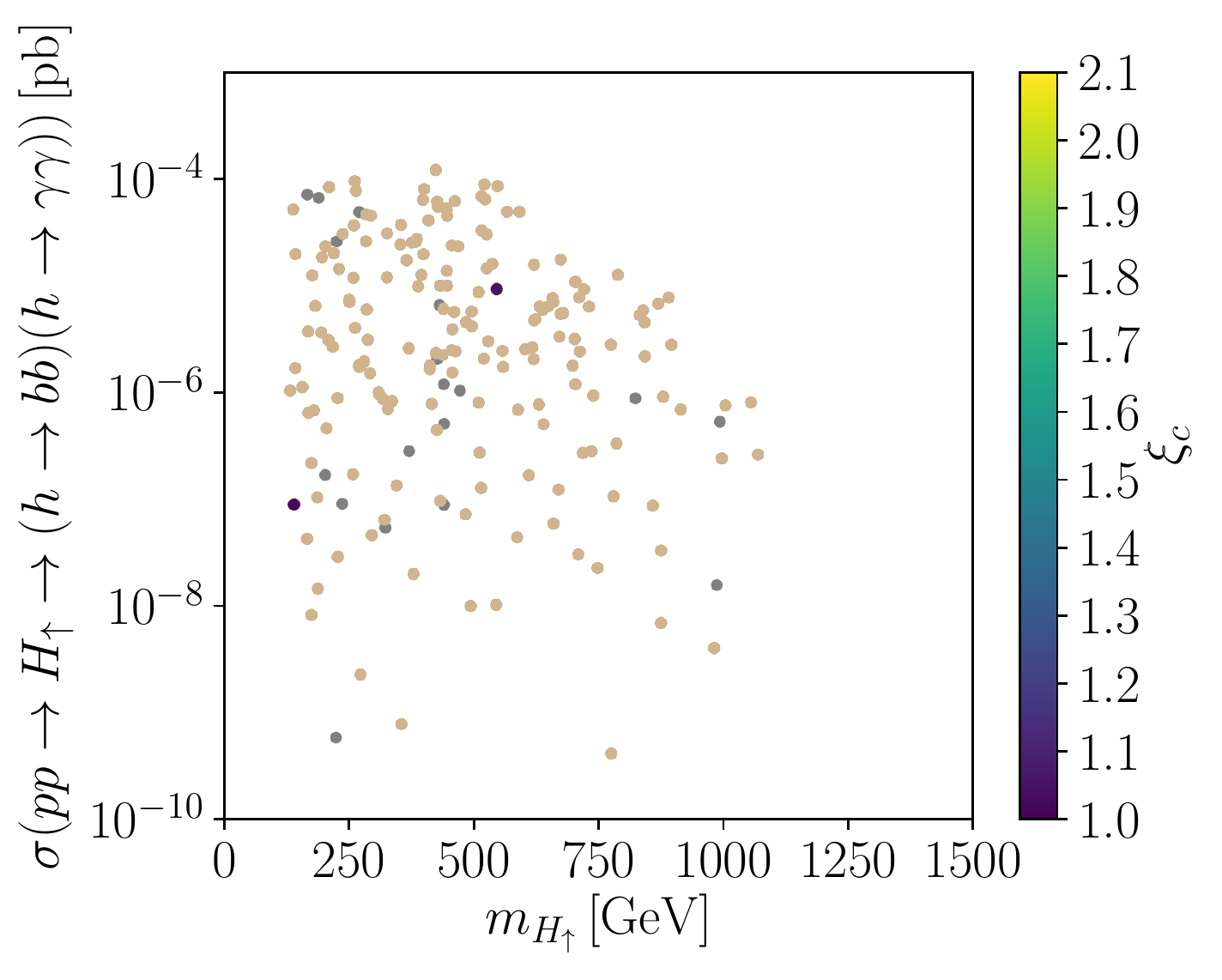}} %
  \caption{T1: Production rate of $H_\uparrow$ with subsequent decay
    into $hh$ in the $(2b)(2\gamma)$ final state versus
    $m_{H_\uparrow}$ for the C2HDM (left) and the N2HDM 
    (right). The color code is the same as in
    Fig.~\ref{N2HDM::massspectrum}. The color bar indicates 
    the strength of the phase transition for $\xi_c \ge 1$. The
    di-Higgs search constraints are included.}
  \label{COMP::HUP::hh::BBGAMGAM}
\end{figure}
\subsubsection*{$\cbrak{\overline b b}\cbrak{\gamma\gamma}$ Final
  State}
The rates for the $(bb)(\gamma\gamma)$ final states from $H_\uparrow$
production with subsequent decay into $H_\downarrow H_\downarrow$ are
shown in Fig.~\ref{COMP::HUP::hdownhdown::BBGAMGAM} and those from
$H_\uparrow \to hh$ in Fig.~\ref{COMP::HUP::hh::BBGAMGAM}. We observe
the same behaviour as in the final state $\cbrak{\overline b
  b}\cbrak{\overline b b }$ with the difference that the overall
signal strength is smaller due to the smaller branching ratio for the
Higgs decays into photons. Again, in the C2HDM the signal rate is
significantly reduced by requiring an SFOEWPT whereas this is not the
case in the N2HDM. The top threshold reduces the overall signal in the
C2HDM while again the N2HDM is not affected due to the possibility of
the singlet admixture. 
\subsubsection{C2HDM and N2HDM T2}
\begin{figure}
  \centering
    \subfigure[C2HDM T2]{\includegraphics[width=0.5\textwidth]{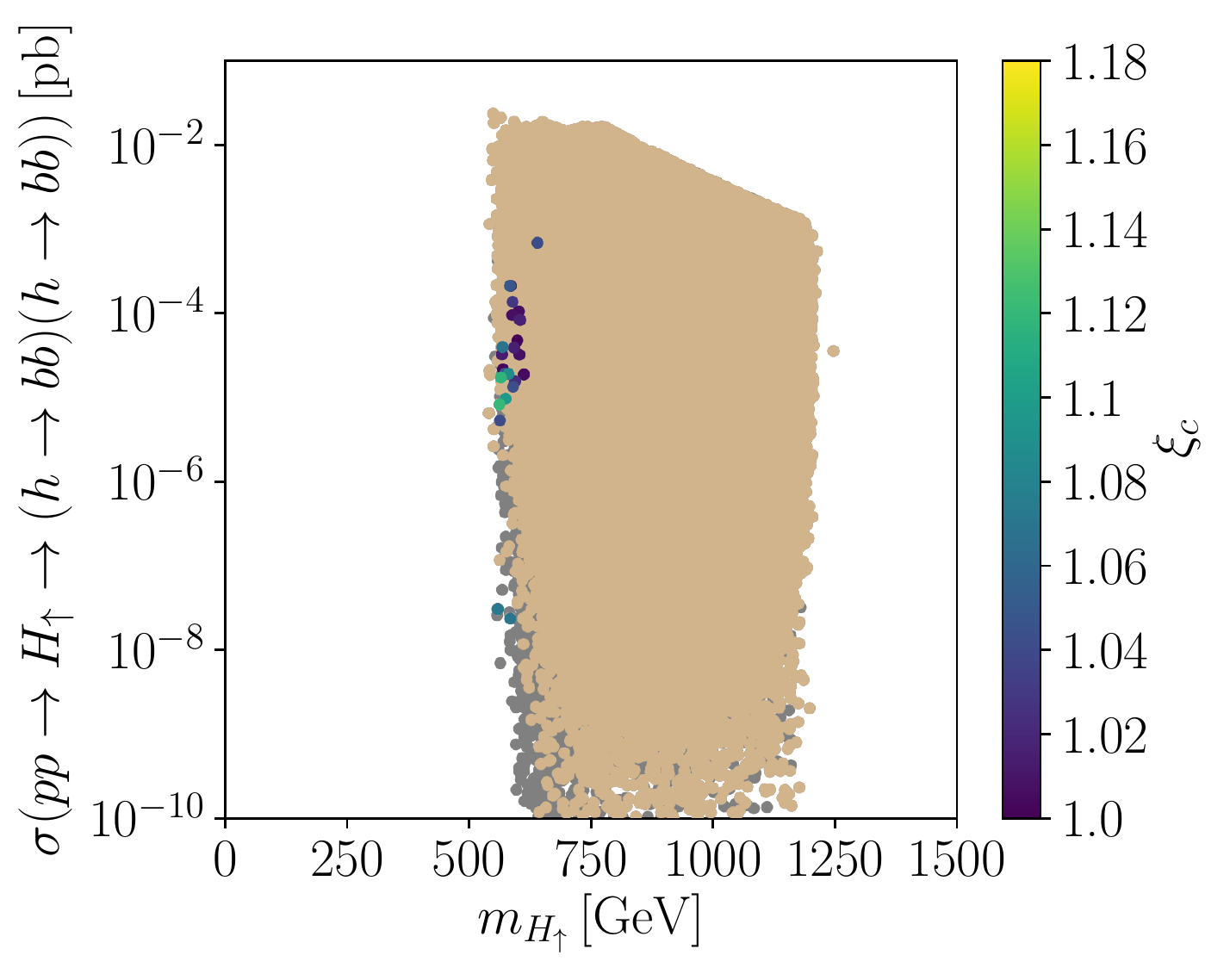}}%
    \subfigure[N2HDM T2]{\includegraphics[width=0.5\textwidth]{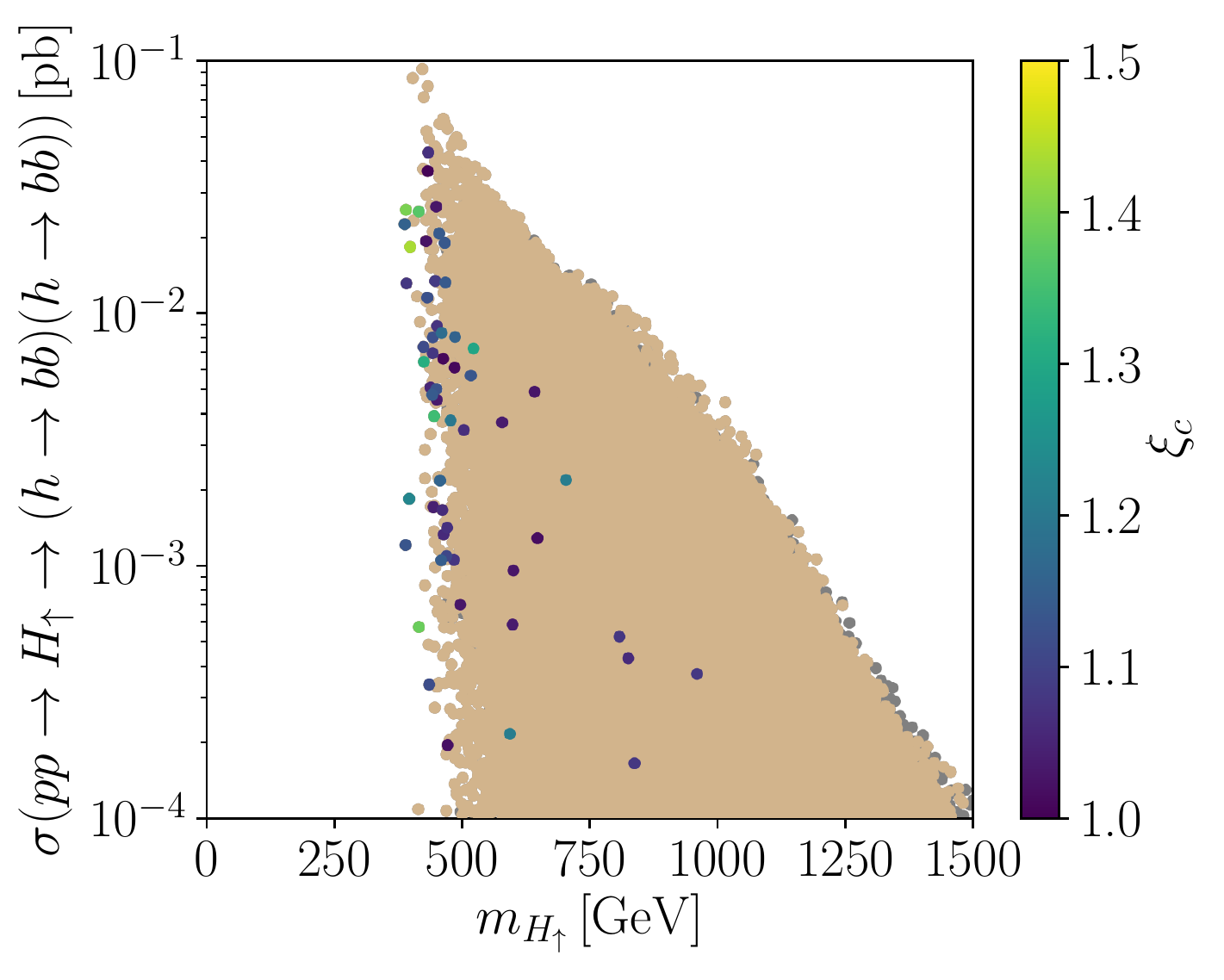}} %
  \caption{T2: Production rate of $H_\uparrow$ with subsequent decay
    into $hh$ in the $(b\bar{b})(b\bar{b})$ final state versus
    $m_{H_\uparrow}$ for the C2HDM (left) and the N2HDM 
    (right). The color code is the same as in
    Fig.~\ref{N2HDM::massspectrum}. The color bar indicates 
    the strength of the phase transition for $\xi_c \ge 1$. The
    di-Higgs search constraints are included.}
  \label{COMPT2::HUP::hdownhdown::BBBB}
\end{figure}
We now turn to the comparison of the C2HDM and N2HDM T2 models. We
display the signal rates
$\sigma\cbrak{pp\rightarrow\hup\rightarrow\cbrak{h\rightarrow
    \overline b b}\cbrak{h\rightarrow \overline bb}}$ versus
$m_{H_\uparrow}$ in \cref{COMPT2::HUP::hdownhdown::BBBB}. In the left plot, the
results for the C2HDM are shown and in the right plot those for the N2HDM. 
The signal rates obtained in the N2HDM and C2HDM T2 are smaller than
in T1 of the models. The maximal signal rates with an SFOEWPT are 
\begin{subequations}
\begin{align}
  &\max\sigma\cbrak{pp\rightarrow\hup\rightarrow\cbrak{h\rightarrow \overline b b}\cbrak{h\rightarrow \overline bb}}\big\vert_{\text{C2HDM}} = 0.67 \fb \\
  &\max\sigma\cbrak{pp\rightarrow\hup\rightarrow\cbrak{h\rightarrow \overline b b}\cbrak{h\rightarrow \overline bb}}\big\vert_{\text{N2HDM}} = 43.24 \fb \,.
\end{align}
\end{subequations}
Although the N2HDM still provides larger signal rates than the C2HDM
the differences are not as significant as in T1. 
Since we do not have parameter points with the mass relation $2 \mdown
\lesssim \mup$ in the C2HDM T2, we cannot discuss the comparison of
the decay channel $\hup\rightarrow\hdown\hdown$ in the
T2. Furthermore, the rates for the decay channel
$\hup\rightarrow\hdown h $ are not sufficiently enhanced to be
interesting for di-Higgs searches so that we do not discuss these
final states further here. The rates for the
$(b\bar{b})(\gamma\gamma)$ final states are below 1~fb in both models
so that we also do not consider these channels here. \s

We conclude by remarking that the signal $\sigma\cbrak{pp\rightarrow
  h\rightarrow\cbrak{\hdown\rightarrow\overline
    bb}\cbrak{\hdown\rightarrow \overline bb}}$ in the N2HDM T2 can
reach
\beq
\max\sigma\cbrak{pp\rightarrow h\rightarrow\cbrak{\hdown\rightarrow
    \overline b b}\cbrak{\hdown\rightarrow \overline
    bb}}\big\vert_{\text{N2HDM}} = 704 \fb \;.
\eeq
To enable such decays a
semi-inverted mass hierarchy is required where $m_h\gtrsim 2 \mdown$. (The
inverted mass hierarchy is not realised in the N2HDM T2.) 
While the signal rate is large, it has to be kept in mind, that this 
mass scenario is already under pressure to fulfill all
required constraints.

\section{Conclusions \label{sec:concl}}
In this paper, we investigated the possibility of an SFOEWPT in models
with non-minimal Higgs sectors, the C2HDM and the N2HDM. For the C2HDM
we updated our analysis \cite{Basler:2016obg} by allowing for heavier
neutral scalar masses up to $1.5\tev$ and including the most recent
collider constraints. Still similar mass regions compatible with
theoretical and experimental constraints and the requirement of an
SFOEWPT were found. The inclusion of the new constraints from the
Higgs data significantly reduced, however, the scenarios compatible
with an SFOEWPT. The strength of the phase transition $\xi_c$ in
both types of the C2HDM is rather small
($\max\xi_c^{\text{C2HDM,T1}}=1.7\,,\max\xi_c^{\text{C2HDM,T2}}=1.18$). In
the N2HDM the inclusion of the
additional real scalar singlet slightly enhances the strength of the phase
transition ($\max\xi_c^{\text{N2HDM,T1}}=
2.1\,,\max\xi_c^{\text{N2HDM,T2}}=1.5$) compared to the C2HDM.
The compatibility with the EW precision constraints requires two of
the Higgs boson masses in the spectrum to be close. The additional requirement of an
SFOEWPT reduces this mass gap further. 
For the N2HDM, we found that not all mass
hierarchies were compatible both with the theoretical and
experimental constraints and an SFOEWPT. Thus, the inverted mass
hierarchy ($\mdown<\mup<m_h$) was not found in the N2HDM T1 and T2.
In the N2HDM T2 the interplay of the requirement of small
mass gaps and a heavy charged Higgs boson mass yields an overall
heavier Higgs boson mass spectrum than in the N2HDM T1. We
showed that the SFOEWPT favors two different mass configurations in
the N2HDM T1. The first region $\mathcal{M}_{\text{deg}}$ features parameter
points with $m_A\approx \mHc$, in the second one, $\mathcal{M}_{\text{sep}}$,
the pseudoscalar Higgs boson $A$ is significantly heavier than the
charged Higgs boson $H^{\pm}$. In particular the mass region $\mathcal
M_{\text{sep}}$ is often neglected in the context of electroweak
baryogenesis. For simplicity, it is often assumed that $m_A\approx
\mHc$. In the N2HDM T2 only points in $\mathcal{M}_{\text{deg}}$ were found to be
compatible with an SFOEWPT. \s

We furthermore investigated in both models the trilinear Higgs
self-couplings between neutral Higgs bosons. An SFOEWPT favors
enhanced trilinear self-couplings. Their values remain, however,
significantly below the values allowed by the experimental and theoretical constraints. 
For the Type I of both models they are significantly constrained by the recent
updates in the di-Higgs searches. Since the SFOEWPT favors Higgs boson
masses in the range $\sim 400-600\gev$, which is also the most
sensitive region in the di-Higgs searches, a large part of our
previous sample of parameter points providing an SFOEWPT is
excluded. Future updates in the di-Higgs searches will therefore have
a significant impact on the valid parameter space of the N2HDM and
C2HDM with respect to an SFOEWPT. 
On the other hand, Type II of both models is strongly
constrained due to the results for $B\rightarrow X_s \gamma$
\cite{Deschamps:2009rh,flavor1,flavor2,flavor3,charged580}. We
observed that the charged Higgs mass tends to smaller mass regions if
an SFOEWPT is required. This cannot be realized in Type II as the
charged Higgs mass must be above $580\gev$ because of the
$B\rightarrow X_s \gamma$ constraints. 
Most of the valid parameter points in the C2HDM T2 in
\cref{C2HDM::mcharged} and in the N2HDM T2 in \cref{N2HDM::T2::mHctbeta}
are located on the left edge in the $\mHc-\tan\beta$ plane. Exactly
this region is sensitive to the $B\rightarrow X_s \gamma$
constraints. We therefore expect the 
strongest constraints for the Type II to come from future updates on
the charged Higgs mass constraint in the flavor sector. 
We finally investigated the phenomenology of both models with respect
to the expected di-Higgs signals at the LHC. Requiring an SFOEWPT,
the expected di-Higgs signals in the C2HDM are significantly suppressed
compared to the N2HDM.  

\subsection*{Acknowledgments}
The research of MM was supported by the Deutsche
Forschungsgemeinschaft (DFG, German Research Foundation) under grant
396021762 - TRR 257.
P.B. acknowledges financial
support by the Graduiertenkolleg GRK 1694: Elementarteilchenphysik bei
h\"ochster Energie und h\"ochster Pr\"azision.





\end{document}